\documentclass{elsart1p}
\usepackage{mathtext}
\usepackage{mathrsfs}
\usepackage{bm}
\usepackage{graphicx}
\usepackage{fancyhdr}

\usepackage{amssymb}

\def\la{\langle}
\def\ra{\rangle}

\def\om{\omega}
\def\Om{\Omega}

\def\d{{\rm d}}

\newcommand{\beq}{\begin{equation}}
\newcommand{\eeq}{\end{equation}}
\newcommand{\beqa}{\begin{eqnarray}}
\newcommand{\eeqa}{\end{eqnarray}}

\begin{document}

\begin{frontmatter}



\title{Quantum transients}

\author{A. del Campo\ead{a.del-campo@imperial.ac.uk}}
\address{Institute for Mathematical Sciences, Imperial College London, SW7 2PG, UK}
\address{QOLS, The Blackett Laboratory, Imperial College London, Prince Consort Rd., SW7 2BW,UK}
\author{G. Garc\'\i a-Calder\'on\ead{gaston@fisica.unam.mx}}
\address{Instituto de F\'{\i}sica,
Universidad Nacional Aut\'onoma de M\'exico,
Apartado Postal {20 364}, 01000 M\'exico, D.F., M\'exico}
\author{J. G. Muga\ead{jg.muga@ehu.es}}
\address{Departamento de Qu\'\i mica-F\'\i sica, UPV-EHU, Apdo. 644, 48080
Bilbao, Spain}
\begin{abstract}
Quantum transients are
temporary features
of matter waves before they reach
a stationary regime. Transients may arise after the preparation of an unstable initial state or
due to a sudden interaction
or a change in the boundary conditions.
Examples are diffraction in time, buildup processes, decay, trapping, forerunners or pulse formation,
as well as other phenomena recently discovered, such as the simultaneous arrival of a wave peak at arbitrarily distant observers. The interest in these transients
is nowadays enhanced by new technological possibilities to control, manipulate and measure matter waves.
%
\end{abstract}
\end{frontmatter}

\pagenumbering{arabic}
\tableofcontents

%
%
%
%

\section{Introduction}
Transients are temporary phenomena before a system attains
a steady-state responding to a sudden change, such as an interaction switch,
an excitation, new boundary conditions,
or the preparation of an unstable initial state. They are ubiquitous in physics, and frequently described by
characteristic times: response times, build-up times, decay times,
or times of emission and arrival.
In quantum mechanics, matter-wave transients are peculiar and generally distinct
from corresponding classical particle phenomena. Typical transients are damped oscillations, forerunners, response to excitations, and trapping or decay processes.

In this work we shall review the theory, results, and experiments
of quantum transients
of single to few-body systems.
We shall leave aside many-body and complex systems, 
such as
electrons in a semiconductor responding to a strong
short optical pulse, since they are treated already in several monographs (see e.g.
\cite{Jauho,Bonitz}) and require specific techniques.
We shall nevertheless discuss many-body systems in which effective
one-particle pictures are still a useful approximation,
e.g. a Bose-Einstein condensate within the mean field Gross-Pitaevskii equation, Tonks-Girardeau gases,
or multi-electron systems treated at a Hartree level.
Many of the concepts and techniques
discussed here are valid for studying exponential-decay processes
and deviations from it,
but we shall not include decay, 
or touch it only in passing, since the amount of work on that field and its applications
deserve separate monographs and many aspects have already been reviewed
\cite{FGR78}.
The field covered, in spite of the above exclusions, is still vast, and we have paid
special attention to work done by the authors, while making an effort to offer a global perspective.

Quantum transients have frequently been studied {\it per se},
because of their interesting dynamics. They are often analytically solvable
and this has also motivated their use to clarify fundamental questions such
as tunneling dynamics, or the Zeno effect.
In recent times, with the advent of technological advances such as laser cooling techniques,
mesoscopic devices, ultrashort laser pulses and nanotechnology, experiments and applications which were
unthinkable one or two decades ago can be implemented or are within reach. This is the case of
operations to prepare and control the dynamics of quantum systems for fundamental studies,
interferometry and metrology, or for quantum information processing. An understanding
of transients is therefore essential: sometimes just to avoid them, as in an atom-laser beam; but also
to take advantage of them, for example to transmit information
simultaneously to receivers at unknown distances;  to optimize gate operations
or transport properties in semiconductors, or as diagnostic tools
for determining momentum distributions, interactions, and other properties
of the system.

A common thread in the mathematical description of many quantum transients is the
Moshinsky function, which is related to the Faddeyeva $w$ function, to Fresnel integrals and to the error function.
Much of this review is devoted to study phenomena where the Moshinsky function
plays a fundamental role.
It was introduced with the method of time-dependent boundary conditions used in the dynamical description of heavy-ion collisions \cite{Moshinsky51}. It was later associated with the ``quantum shutter'' problem, one of the archetypal quantum transient phenomena, which lead to the concept of ``diffraction in time'' (DIT) \cite{Moshinsky52}.
A flurry of experimental and theoretical work has been carried out ever since.
The hallmark of DIT consists in temporal and spatial oscillations of Schr\"odinger matter waves
released in one or several pulses from a preparation region in which the
wave is initially confined. The original setting consisted in a sudden opening
of a shutter to release a semi-infinite beam, and provided a quantum, temporal analogue
of spatial Fresnel-diffraction by a sharp edge \cite{Moshinsky52}.
Later on, more complicated shutter windows \cite{SK88}, initial confinements \cite{GN57,GK76,Godoy02,Godoy03}, time-slit combinations \cite {BZ97,DM05} and periodic gratings \cite{Frank1,Frank2,DMM07} have been considered \cite{Kleber94}. The effects of an external linear potential, e.g. due to gravity, on the DIT phenomenon have also be taken into account \cite{DM06b}.
The basic results on Moshinsky's shutter are reviewed in Section \ref{moshDIT}.

The quantum shutter problem has important applications indeed,
for example the modelization of turning on and off a beam of atoms, as done e.g. in
integrated atom-optical circuits or a planar atom waveguide \cite{SHAPM03}, and   has been used to translate
the principles of spatial diffractive light optics \cite{Fowles68,BW99} to the time domain for matter waves \cite{BAKSZ98}.
It is at the core of different schemes for loading ultracold atoms or Bose-Einstein condensates into traps \cite{MetStra99} and,
besides, it provides a time-energy uncertainty relation \cite{MoshinskyRMF52b,Moshinsky76,Busch02} which has been verified experimentally
for atomic waves in a Young interferometer with temporal slits \cite{SSDD95,ASDS96,SGAD96}.
The Moshinsky shutter has been discussed in phase space \cite{MMS99,Moshinsky04}, for relativistic equations \cite{Moshinsky52,GRV99,DMRGV,BT98,MB00,jv00,DMR04,Kalbermann01},
with dissipation \cite{MS01,DMR04}, in relation to Feynman paths \cite{GCVY03},
or for time dependent barriers \cite{SK88,Kleber94,KM05,KM05b,DMK08}.

Transients for the expansion dynamics from box-like traps have also been
examined both at single \cite{GK76,DMN92,Godoy02} and many-particle level
\cite{Gaudin71,Cazalilla02,Cazalilla04,BGOL05,DM05b}, as well as
from spherical traps similar to actual experimental
magneto-optical traps \cite{Godoy03}.
The experimental buildup and depth control of box-like (square well) traps for ultracold atoms
with an all-optical implementation \cite{MSHCR05}, or in microelectronic chips \cite{HHHR01}
has excited a great deal of attention for these geometries.
The transient effects when the trap is turned off
provide information on the initial state and allow for its reconstruction \cite{BDZ08}.
Once the ballistic dynamics dominates the expansion,  the regime of interactions or the momentum distribution
can be inferred  in a Bose-Einstein condensate \cite{KDS99,PS02}
and effectively one-dimensional gases as well \cite{OS02,DMM08}.
The involved time scales play therefore a key role and are reviewed,
at a single-particle level, in Section \ref{free_expansion}, and with
particle interactions in Section \ref{ucgtw}.

Section \ref{onepulse} deals with
pulse formation, a key process in ultracold neutron interferometry,
and a timely subject due to the
possibilities to control the aperture function of shutters
in atom optics, the ionization by ultrashort laser pulses, 
or the development of atom lasers.
Some of the first mechanisms proposed explicitly for building atom lasers
implied periodically switching off and on the cavity mirrors,
that is, the confining potential of the lasing mode.
As an outcome, a pulsed atom laser is obtained.
Much effort has been devoted to designing a continuous atom laser,
whose principle has been demonstrated using Raman transitions \cite{MHS97,H99}.
With this output coupling mechanism, an atom initially
trapped undergoes a transition to a non-trapped state,
receiving a momentum kick due to the photon emission.
These transitions can be mapped  to the pulse formation, so that
the ``continuous'' nature of the laser arises as a consequence of the
overlap of such pulses.
The fields of coherent control, femto and attosecond 
laser pulses and ultracold matter are also merging rapidly creating  
new opportunities for inducing, studying, and applying quantum
interference of matter pulses and the corresponding transients.  
There is, in summary, a strong motivation for
a thorough understanding of matter-wave
pulse creation, even at an elementary single-particle level.

The shutter model, with adequate interaction potentials added, see Section \ref{epo},
has been used to study and characterize transient dynamics of tunneling matter waves \cite{Stevens,TKF87,JJ89,BM96,GR97,GV01,gcvdm02,DMRGV}, reviewed in
Section \ref{tune}; Section \ref{sape} is devoted to the transient response to abrupt changes of the interaction potential
in semiconductor structures and quantum dots \cite{DCM02,AP,DMCLA05}.
The study of such transients has been further proposed for potential barrier classification \cite{GM06}.

In Section \ref{otp} some transient phenomena different from DIT
which have been recently discovered or studied are reviewed.
They are intriguing effects such as: ultrafast propagation of a peak which arrives simultaneously
at arbitrarily distant observers in an absorbing medium; breakdown of classical conservation of energy in quantum collisions
and momentum-space interferences; transients associated with  classically forbidden momenta, e.g. negative incident momenta that
affect the transmitted part of a wavepacket impinging
on a barrier from the left; and the Zeno effect in a time-of-arrival measurement model.

The experimental work is reviewed in Section \ref{experiments}.
The first field where transient effects and a corresponding time-energy uncertainty relation were investigated was ultracold neutron interferometry 
\cite{GKZ81,GZ91,RWCKW92,HFGGGW98,BAKSZ98,BAKOSZ00}. Experimental observations of transients have been carried out with
ultracold atomic waves where diffraction, phase modulation and interference
were studied using temporal slits \cite{SSDD95,ASDS96,SGAD96}.
More recently, a double temporal slit experiment with
electrons has been performed \cite{LSWBGKMBBP05}.
The applications of atom interferometry in time domain have also been explored
with Bose-Einstein condensates at both theoretical \cite{HSKD94}
and experimental level \cite{CMPL05}. A very recent and relevant result is the
realization of an electromagnetic analogy allowing to study DIT phenomena
by means of the vertical propagation of the lasing modes emitted from the laterally
confined cavities \cite{chinos}. 
Note also that transients do not necessarily involve spatial translation
and release from 
traps or shutters. They may be 
observed in the evolution of internal state populations of an atom due to time-dependent laser excitations \cite{ct2001}.  

The review ends up with some conclusions and technical appendices.
Most sections could be read independently although there are strong links between
Sections \ref{free_expansion} and \ref{ucgtw}, or between Sections \ref{epo}
and \ref{tune}. Multiple use of some symbols has been unavoidable
and the notation is only guaranteed to be consistent within
a given section.
\section{Diffraction in time\label{moshDIT}}
\subsection{The Moshinsky shutter}
\label{mosh52}
In 1952, Moshinsky tackled the quantum dynamics of a suddenly released beam of independent particles of mass $m$.
The corresponding classical problem, that of a beam with a well defined incident velocity $p/m$,
and initially confined to the negative semiaxis ($x<0$), admits a trivial solution:
the density profile is uniform for $x<pt/m$ and vanishes elsewhere.
In non-relativistic quantum mechanics, we shall first consider a quasi-monochromatic
atomic beam of nominal momentum  $\hbar k$ (or velocity $v_k=\hbar k/m$)
impinging on a totally absorbing shutter located at the origin $x=0$,
\begin{equation}
\psi(x,t=0)=e^{ikx}\Theta(-x),
\label{initial}
\end{equation}
where $\Theta(x)$ is the Heaviside step function defined as $\Theta(x)=1$ for $x>0$ and $0$ elsewhere.
Note that, in spite of the prominent role of $k$, the state is not really monochromatic because of the spatial truncation. Such state is clearly not normalized, but it can be considered as an elementary
component of a wave packet truncated (fully absorbed) at the origin.
Since for $t>0$ the shutter has been removed, the dynamics is free and the time evolution can be obtained using
the superposition principle,
\beq
\label{superposition}
\psi(x,t)=\int_{-\infty}^{\infty}dx'
K_0(x,t|x',t'=0)\psi(x',t'=0),
\eeq
with the free propagator
\beqa
\label{freepropagator}
K_0(x,t\vert x',t')=\bigg[\frac{m}{2\pi i\hbar
(t-t')}\bigg]^{\frac{1}{2}}e^{i\frac{m(x-x')^{2}}{2\hbar (t-t')}},
\eeqa
see Appendix \ref{propa} for a brief account of propagators and their role
in quantum transients.
Following customary practice, we shall liberally speak of $|\psi(x,t)|^2$
as a ``density'' even if it is dimensionless.
%
\begin{figure}
\begin{center}
\includegraphics[height=6cm]{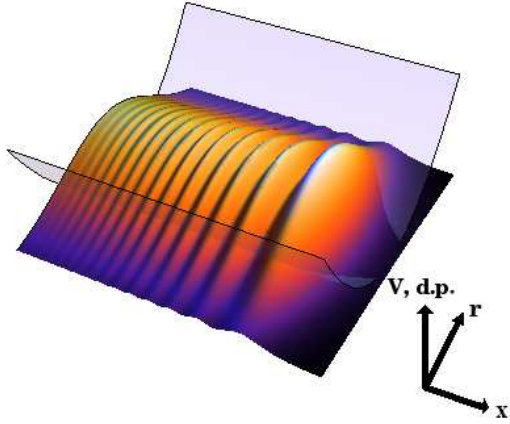}
\caption{\label{artdit}
Snapshot of the density profile of a suddenly released beam, confined transversely to a harmonic trap along a waveguide, and
exhibiting diffraction in time in the longitudinal direction, see Eq. (\ref{moshprofilefresnel}). }
\end{center}
\end{figure}
%

In a seminal paper \cite{Moshinsky52}, Moshinsky  proved that
\beqa
\label{eqDIT}
\psi(x,t)=M(x,k,t).
\eeqa
This is a simple compact result, which constitutes the main reference case for more complex setups.
$M$ is the so-called Moshinsky function \cite{Moshinsky51,Moshinsky52,Moshinsky76},
\beq\label{mou}
M(x,k,t):=\frac{e^{i\frac{mx^{2}}{2 \hbar t}}}{2}w(-u),\qquad
u=\frac{1+i}{2}\sqrt{\frac{\hbar t}{m}}\left(k-\frac{mx}{\hbar t}\right).
\eeq
Many of its properties relevant to the study of quantum transients are listed in Appendix \ref{app_moshfunction}. Here we just note that it can be related to the Faddeyeva function $w(z):=e^{-z^{2}}{\rm{erfc}}(-i z)$ \cite{FT61,AS65}, see
Appendix \ref{app_wfunction}.
Such solution entails the diffraction in time phenomenon, a set of oscillations in the beam
profile (see Fig. \ref{artdit}) in dramatic contrast with a
classical monochromatic and homogeneous beam density
$\Theta(p t/m-x)$.

\begin{figure}
\begin{center}
 \includegraphics[height=6cm]{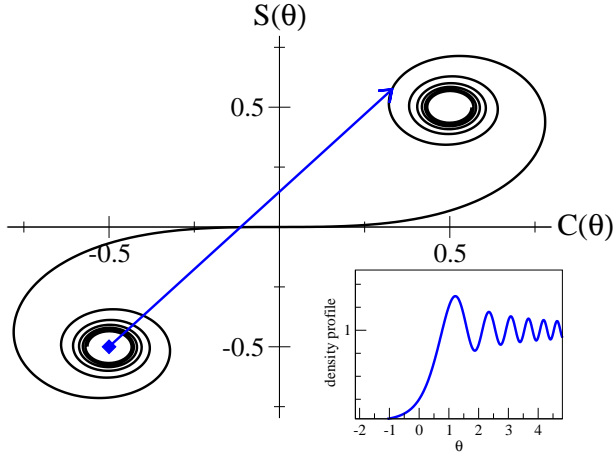}
\caption{\label{fig_cornu}
Cornu spiral. The density profile characteristic of the diffraction in
time, Eq. (\ref{moshprofilefresnel}),
is given by half the distance from the point $(-1/2,-1/2)$ to any
point of the spiral.
}
\end{center}
\end{figure}

The ``diffraction in time'' (DIT) term was introduced because the quantum density profile,
$|M(x,k,t)|^2$, admits a simple geometric interpretation
in terms of the Cornu spiral or clothoid, which is the curve that
results from a parametric representation of the Fresnel integrals $C$ and $S$,
see Fig. \ref{fig_cornu} and Eq. (\ref{A1fresnel}).
%
%
As shown also by Moshinsky \cite{Moshinsky52}, it follows that
\beq
\label{moshprofilefresnel}
\vert M(x, k,t)\vert^{2}=\frac{1}{2}
\Bigg\{\left[\frac{1}{2}+C(\theta)\right]^{2}
+\left[\frac{1}{2}+S(\theta)\right]^{2}\Bigg\},
\eeq
with
\beq
\theta=\sqrt{\frac{\hbar t}{m\pi}}\left(k-\frac{m x}{ \hbar t}\right).
\eeq
This is precisely the form of the intensity profile of a light beam
diffracted by a semi-infinite plane \cite{Fowles68}, which prompted the choice of the DIT term.
The probability density is half the distance from the point
$(-1/2,-1/2)$ to any other point of the Cornu spiral. In such representation
the origin corresponds to the classical particle with momentum $p$ released at time $t=0$ from the shutter position.
Thanks to it, the width of the main fringe, $\delta x$, can be estimated
from the intersection between the classical and quantum probability densities \cite{Moshinsky52,DM05},
leading to a dependence of the form
\beqa
\label{fringewidth}
\delta x\propto(\pi\hbar t/m)^{1/2}.
\eeqa
Moreover, any point of constant probability deviates from the classical spacetime path.
In particular, the principal maxima and minima obey
\beqa
x_{max}(t) &=& p t/m-\sqrt{\frac{\pi\hbar t}{m}}\theta_{max},\nonumber\\
x_{min}(t) &=& p t/m-\sqrt{\frac{\pi\hbar t}{m}}\theta_{min},\nonumber\\
\eeqa
where  $\theta_{max}=1.217$, $\theta_{min}=1.872$ are specific values for the totally absorbing shutter.
Therefore, the quantum signal is slowed down for $t\lesssim({m\pi\hbar}\theta_{min/max}^{2}/p^{2})^{1/4}$ with respect to the classical case, $x_{cl}(t)=pt/m$.

A more general type of initial state was considered in \cite{Moshinsky76,DM05},
\beqa
\label{Rbeam}
\psi(x,t=0)=e^{ikx}+Re^{-ikx},
\eeqa
with $R=\exp(i\alpha \pi)$ corresponding to a shutter with reflectivity $|R|^2=1$.
Under free evolution, $\psi(x,t)=M(x,k,t)+R M(x,-k,t)$.
%
\begin{figure}
\begin{center}
\includegraphics[height=5cm]{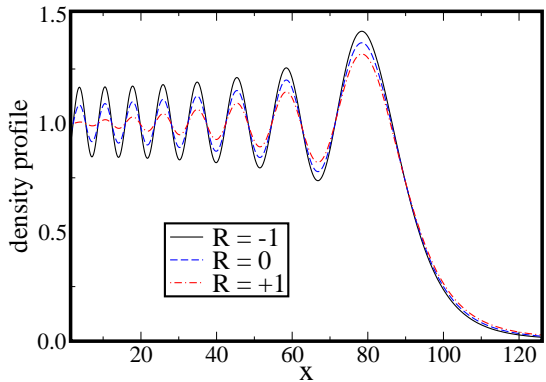}
\caption{\label{rmoshwf} Enhancement effect of the reflectivity of the shutter
on the visibility of the diffraction-in-time fringes of a suddenly released
quasi-monochromatic matter-wave beam (in $\hbar=m=1$ units).}
\end{center}
\end{figure}
%
For such a beam, the visibility of the fringes increases monotonically from $\alpha=0$ (cosine initial condition, see  \ref{mbsources} below) to $\alpha=1$ for which the shutter is equivalent to a totally-reflecting hard-wall potential, see Fig. \ref{rmoshwf}
and Appendix \ref{apis}.

The possibility of closing the shutter after a time $\tau$ to form a pulse was studied by Moshinsky himself \cite{MoshinskyRMF52b,Moshinsky76}. This is the essential step of a chopper in an ultracold neutron interferometer
used for velocity selection. For small values of $\tau$  a time-energy uncertainty relation comes into play, broadening the energy distribution of the resulting pulse. For the initial condition (\ref{Rbeam}), with $R=+1$, the explicit calculation of the energy spread $\Delta E$ in the resulting wavepacket leads to
\beqa
\Delta E\, \tau\simeq \hbar.
\eeqa
Using the full width at half-maximum (FWHM) of the energy distribution, it was found that \cite{DM05}
\beqa
{\rm FWHM}(E)\, \tau \gtrsim 2\hbar.
\eeqa
The time-energy uncertainty relation lacks the simple status of that holding for position and momentum observables, having many different non-equivalent versions reviewed in  \cite{Busch02}. The one for chopped beams as discussed here \cite{MoshinskyRMF52b,Moshinsky76} was experimentally observed and
will be further discussed in Section \ref{experiments}.
\subsection{Time-dependent shutter \label{tidesh}}
%
%
%
%
%
\begin{figure}
\begin{center}
\includegraphics[height=5cm]{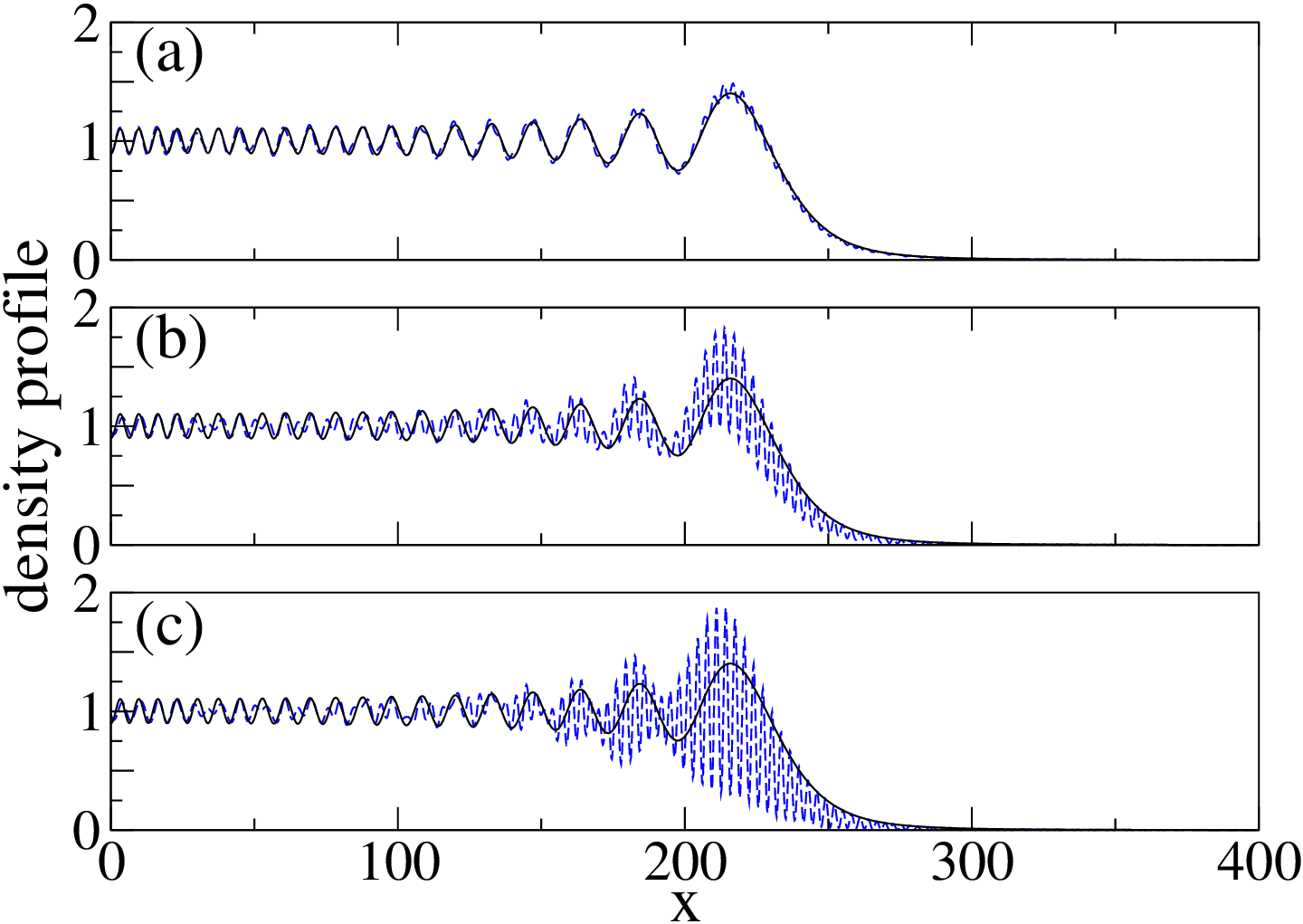}
\caption{\label{kleber88} Snapshots of the density profile of a chopped beam ($\hbar=m=1$) with
$\gamma=0.5$ (a), $5$ (b), and $20$ (c) taken in all cases at $t=250$ with $k=1$.
The solid line corresponds to the sudden totally reflecting shutter
($\gamma=0$), while the dashed line is for the time-dependent $\delta$-potential.}
\end{center}
\end{figure}
The paradigmatic result by Moshinsky rests on the sudden approximation for the removal of the shutter.
The effect of a more realistic time dependence has also been looked into.
In \cite{SK88,Kleber94}, the shutter potential $V(x,t)=\gamma\delta(x)/t$ was assumed,
which reduces to an infinite wall at $t=0$ and vanishes asymptotically for $t\rightarrow\infty$.
Using the Laplace transform method, the associated propagator can be related to the free one \cite{SK88},
%
%
from which the time-dependent wavefunction can be calculated by numerical integration.
Scheitler and Kleber analyzed the flux at the origin $x=0$ and found that the effect of
the slow removal of the shutter is to suppress the DIT fringes in the current.
It turns out that the Massey parameter $\beta=\gamma k/\hbar$ identifies qualitatively two different regimes:
for $\beta\lesssim1$, one recovers the sudden approximation; but for $\beta\gtrsim 5$ the particle
will experience a quasi-static barrier, no DIT-like transients in the current are observed,
and the flux is indeed well-described by the equilibrium current
\beqa
J(x=0,t)=\frac{\hbar k}{m}T\left(E=\frac{\hbar^2k^2}{2m}\right)
\qquad {\rm with}\qquad T(E)=\frac{E}{E-\frac{m\gamma^ 2}{2\hbar^ 2t^ 2}},
\eeqa
which is the transmission probability for a $\delta$-barrier of strength $\gamma/t$.
Nonetheless it is noteworthy that the dynamics of the density profile can be dramatically distorted
with respect to the reference case (see Eq. (\ref{eqDIT})) as shown in Fig. \ref{kleber88},
due to the partial reflection from the shutter at short times.
\begin{figure}
\begin{center}
\includegraphics[width=7.5cm,angle=0]{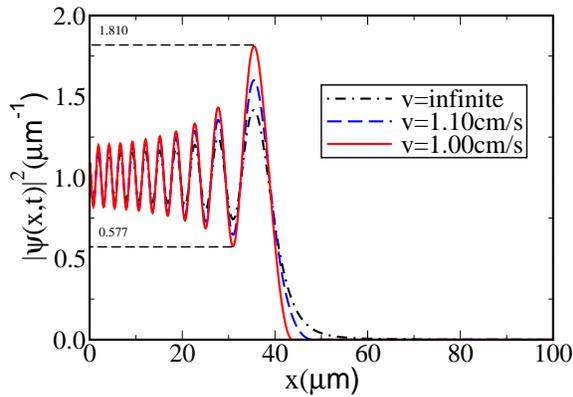}
  \caption{
Characteristic modulation of the density profile of a matter-wave
beam of $^{87}$Rb atoms induced by a totally reflecting mirror which
is displaced at constant velocity $v>v_k$ ($v_k=1.0$ cm/s). The
amplitude of the oscillations is maximized as $v$ approaches the velocity of the
beam $v_k$, when the density exceeds by more than $80\%$ the classical case. In the limit  $v\rightarrow\infty$ the dynamics reduces to the diffraction in time setup with the lowest fringe visibility.} \label{fig_enhancement}
\end{center}
\end{figure}

DIT is actually a delicate quantum phenomenon, easily suppressed or averaged out.
An exception is the effect of an alternative removal of the shutter consisting of its  displacement along the direction of the propagation of the beam. For a mirror moving with velocity ${\rm v}$ \cite{DMK08}, the constructive interference with the reflected components leads to an enhancement of DIT. In particular, the sharpness of the fringes in the density profile and their visibility are maximized whenever ${\rm v}\rightarrow pt/m$, as illustrated in Fig. $\ref{fig_enhancement}$.
\subsection{Dissipation}
It is clear that DIT requires quantum coherence. From this point of view it should come as no surprise that the presence of dissipation leads to its suppression. This was shown by Schuch and Moshinsky \cite{MS01} using the phenomenological Caldirola-Kanai model \cite{Caldirola41,Kanai48}, which describes dissipation without explicitly considering the degrees of freedom of the environment, though it is roughly tantamount to the Caldeira-Leggett model in the Ohmic regime \cite{SY95}. Within this approach, the fringes in the DIT analytical solution are washed out as the time of evolution goes by.
\subsection{The wave equations}
One may wonder if DIT could be observed in other systems, such as in the electromagnetic field \cite{Moshinsky52}.
Let us consider the ordinary wave equation
\beq
\frac{\partial^2 E(x,t)}{\partial x^2}=\frac{1}{c^2}\frac{\partial^2 E(x,t)}{\partial t^2},
\eeq
where the speed of light $c$.
Moshinsky showed that no DIT phenomenon arises, the solution being
\beq
E(x,t)=\Theta(t-x/c)\Big[e^{i(kx-\om t)}-\frac{1}{2}\Big],
\eeq
with $\om=ck$, which oscillates uniformly for $t<x/c$ and vanishes elsewhere, in  contrast to the DIT solution.
Therefore, the DIT is absent in the free propagation of light. DIT may however be seen in constrained geometries, such as waveguides.
Moreover, a relativistic approach to the diffraction in time using the Klein-Gordon equation \cite{Moshinsky52}
shows that the probability density is restricted to the accessible region $x<ct$, at variance with the non-relativistic solution $\psi(x,t)=M(x,k,t)$ which is non-zero everywhere already for $t=0^+$, see also \ref{relativi}.
A similar discussion of DIT in the Dirac equation can be found in \cite{moshinskyRMF52}.
Finally, a mapping from space to time serves also to observe DIT from the 
vertical emission of a lasing mode in a laterally confined cavity \cite{chinos}.
More on this in Section \ref{experiments}.  
%
%
%
%
%
%
\section{Transients in free expansion}
\label{free_expansion}
The free dynamics of a suddenly released matter-wave is frequently used to probe the properties of the initial state and its relaxation quantum dynamics \cite{BDZ08}. The first studies generalizing the Moshinsky shutter setup in the context of neutron interferometry focused on rectangular wavepackets of the form $e^{ikx}\chi_{[0,L]}$, where the characteristic function in the interval $[a,b]$ is $\chi_{[a,b]}=\Theta(b-x)-\Theta(a-x)$ \cite{GN57,GK76}.
We shall discuss the more realistic case of particle-in-a-box eigenstates,
\beq
\phi_{n}(x,t=0)=\sqrt{\frac{2}{L}}\sin(n\pi x/L)\chi_{[0,L]}(x),
\eeq
with $n\in\mathbb{N}$. Box-like traps have been implemented in the laboratory
both in an all-optical way \cite{MSHCR05} and with microchip technology \cite{HHHR01}.
The expansion after suddenly switching off the confining potential has been
analyzed
at short \cite{Godoy02} and arbitrary times \cite{DM05}.
Using the superposition principle and introducing
$k_{n}=n\pi/L$, the time evolved wavefunction is
\beqa
\label{3phit}
\phi_{n}(x,t) & = & \int_{-\infty}^{\infty}dx'
K_0(x,t\vert x',t'=0)\psi(x',t'=0)\nonumber\\
& = & \frac{1}{2i}\sqrt{\frac{2}{L}}\sum_{\nu=\pm1}\nu
\big[e^{i\nu k_{n}L}M(x-L,\nu k_{n},t)
-M(x,\nu k_{n},t)\big].
\eeqa
%
\begin{figure}
\begin{center}
\includegraphics[width=4cm,angle=0]{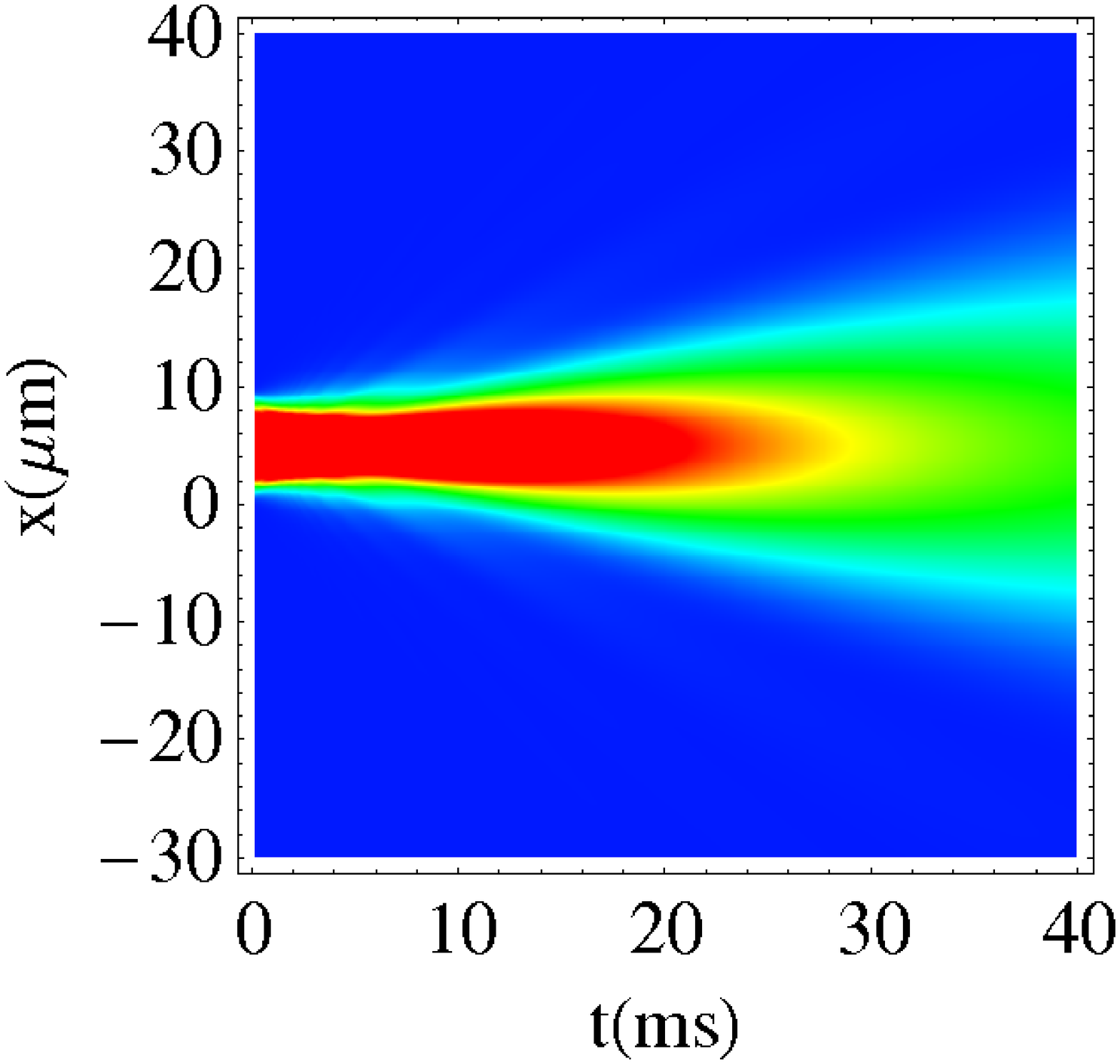}
\includegraphics[width=4cm,angle=0]{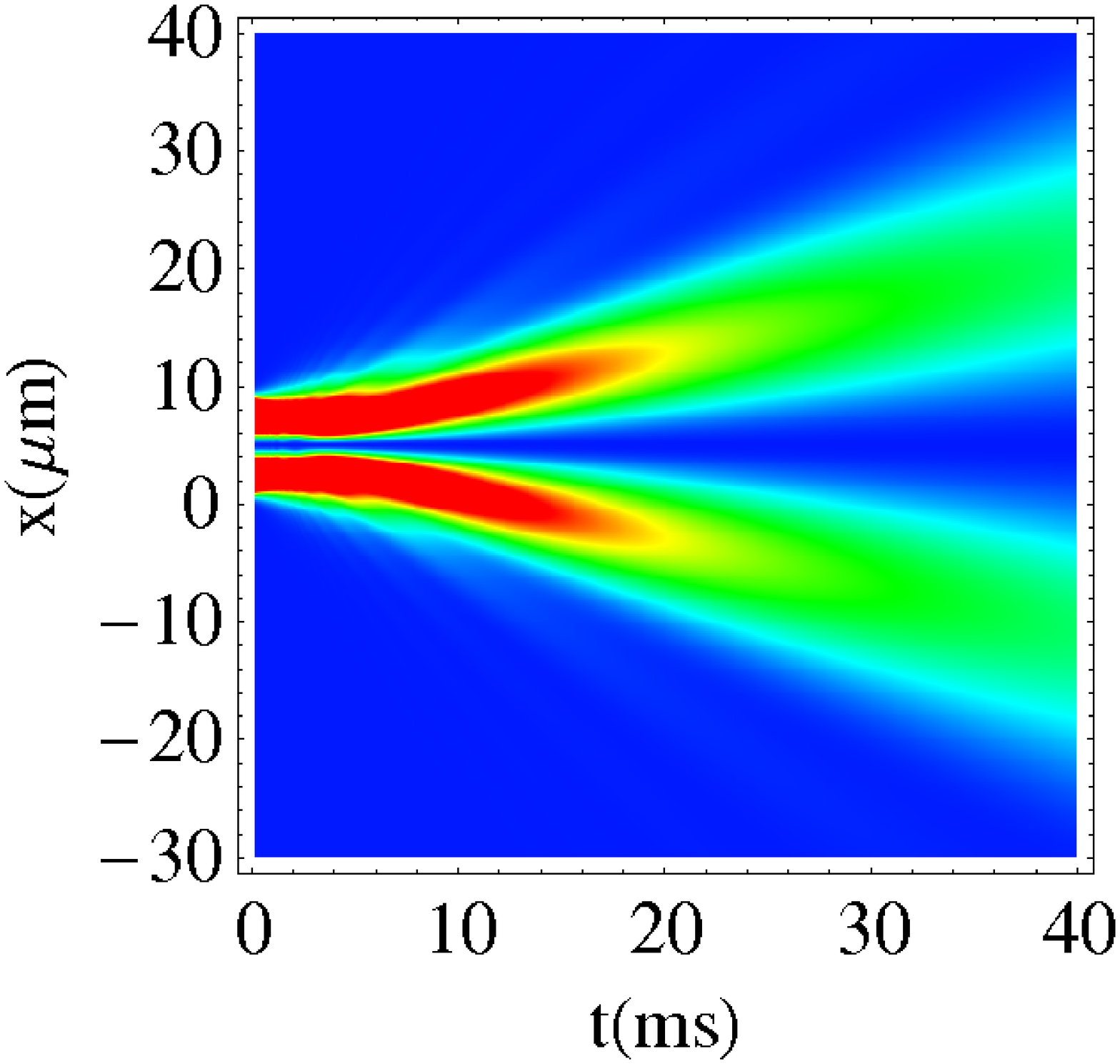}
\includegraphics[width=4cm,angle=0]{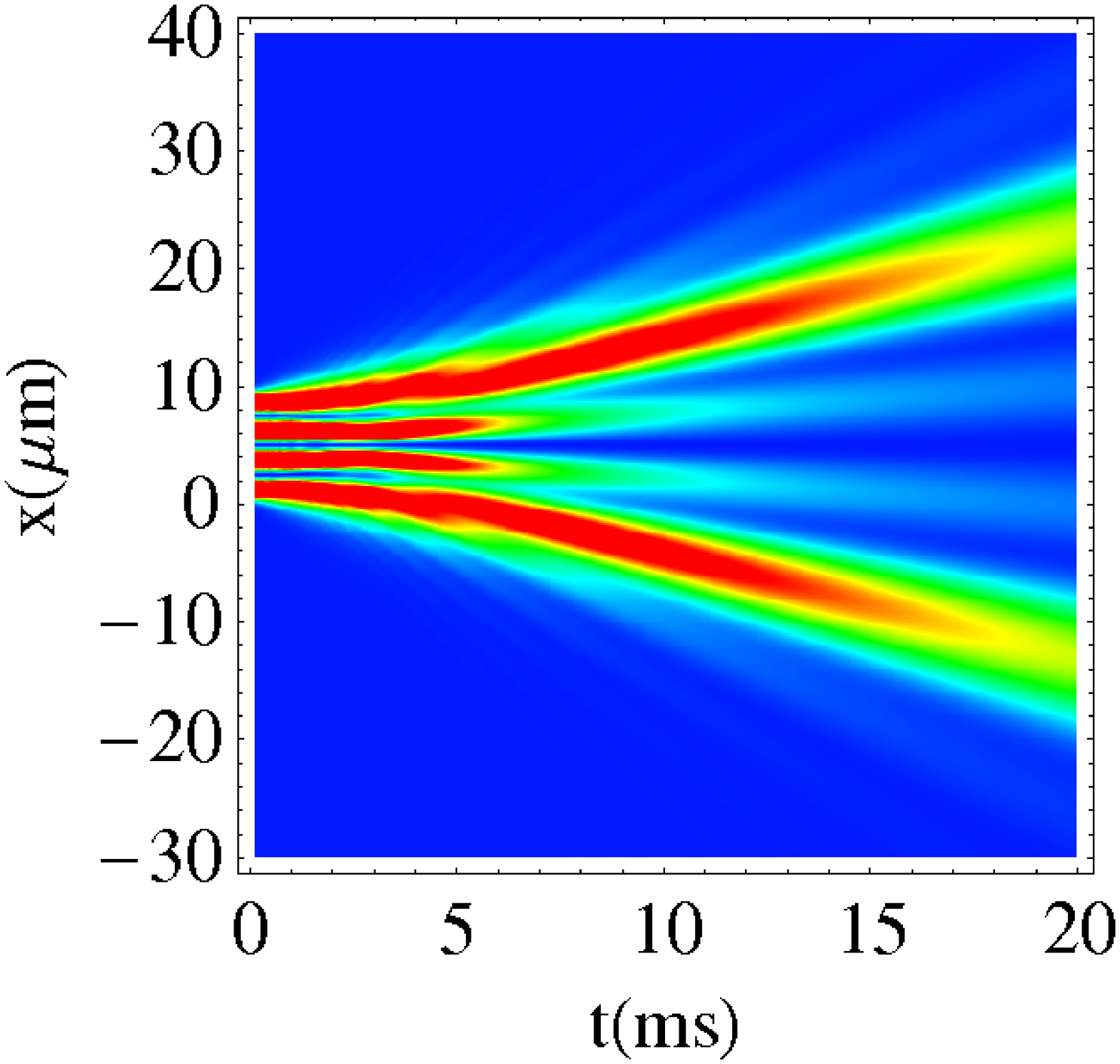}
\vspace*{.2cm}
\caption{\label{hwn}
(Color online) Free expansion of the $n=1,2,4$ eigenstates
for a $^{87}$Rb atom released from a hard-wall trap of $10\mu$m, see Eq. (\ref{3phit}).
}
\end{center}
\end{figure}
Registration of the probability density at a given point as a function of time, allows to distinguish the following two regimes. Whenever the de Broglie wavelength is of the order of the box trap, which is to say $n\sim 1$,
the particle expands following a bell-shaped profile with few transients as sidelobes. This is the {\it Fraunhoffer} limit of diffraction by a narrow slit in time. The DIT shows up for excited states $n\gg 1$, which corresponds to the so-called {\it Fresnel} regime, in which the effect of the two edges of the trap can be identified \cite{Godoy02}. The space-time density profile is actually richer in structure as shown in Fig. \ref{hwn}.

In what follows we shall compare the dynamics with the expansion from a harmonic trap.
Let us first recall the spectrum of the harmonic oscillator (HO) trap,
$Sp(H_{HO})=\{E_{l}^{HO}=\hbar\om(l+1/2)|l=0,1,2,\dots\}$,
and that of the hard-wall (HW) trap,
$Sp(H_{HW})=\{E_{n}^{HW}=(\hbar\pi n/L)^{2}/(2m)|n=1,2,3,\dots\}$.
The condition for the $n$-th eigenstate of the HW trap to have the same energy than
the $(n-1)$-th eigenstate of the HO leads to
\beq
\om=\frac{\hbar n^{2}\pi^{2}}{(2n-1)mL^{2}}.
\eeq
The eigenstates of the harmonic oscillator trap are well-known to be given by
\beq
\phi_{n}(x,t=0)=\left(\frac{m^{2}\om^{2}}{\pi\hbar^{2}}\right)^{1/4}
e^{-\frac{m\om x^{2}}{2\hbar}}
H_{n}\left(\sqrt{\frac{m\om}{\hbar}}x\right),
\eeq
where $H_{n}(x)$ are the Hermite polynomials \cite{AS65}.
Although the time-evolution of such eigenstates can be
calculated for a broad variety of functional dependencies $\om(t)$ \cite{PZ98}, in the
following we shall restrict ourselves
to $\om(t)=\om\Theta(-t)$ which is tantamount to
the time-evolution under the free propagator $K_0(x,t\vert x',0)$ in (\ref{freepropagator}) for $t>0$.
The evolution is given in terms of a scaling law of coordinates up
to an additional phase which will play no role in our discussion.
Indeed, it reads \cite{PZ98}
\beq
\phi_{n}(x,t)=\frac{1}{\sqrt{b}}\phi_{n}\left(\frac{x}{b},t=0\right)
e^{i\frac{mx^{2}}{2\hbar}\frac{\dot{b}}{b}-i\tau\om(n+1/2)},
\eeq
where $\tau:=\int_{0}^{t}dt'/b^{2}(t')$ and $b$ must be a solution of
$\ddot{b}+\om(t)b=0$ with $b(0)=1$ and $\dot{b}(0)=0$.
An instantaneous turn-off of the harmonic potential at $t=0$ leads to $b(t)=\sqrt{1+\om^{2}t^{2}}$.
%
\begin{figure}
\begin{center}
\includegraphics[height=5cm]{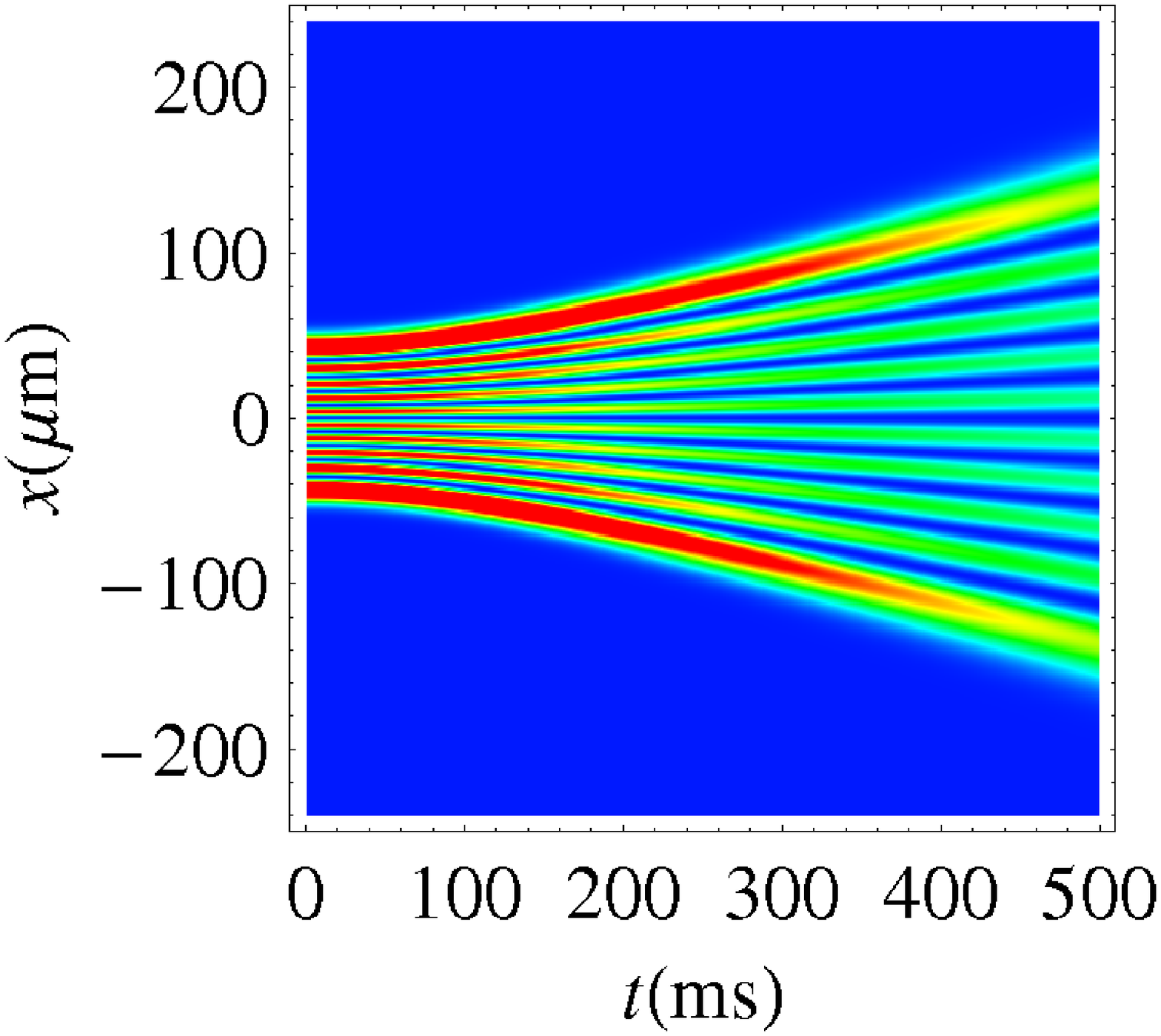}
\includegraphics[height=5cm]{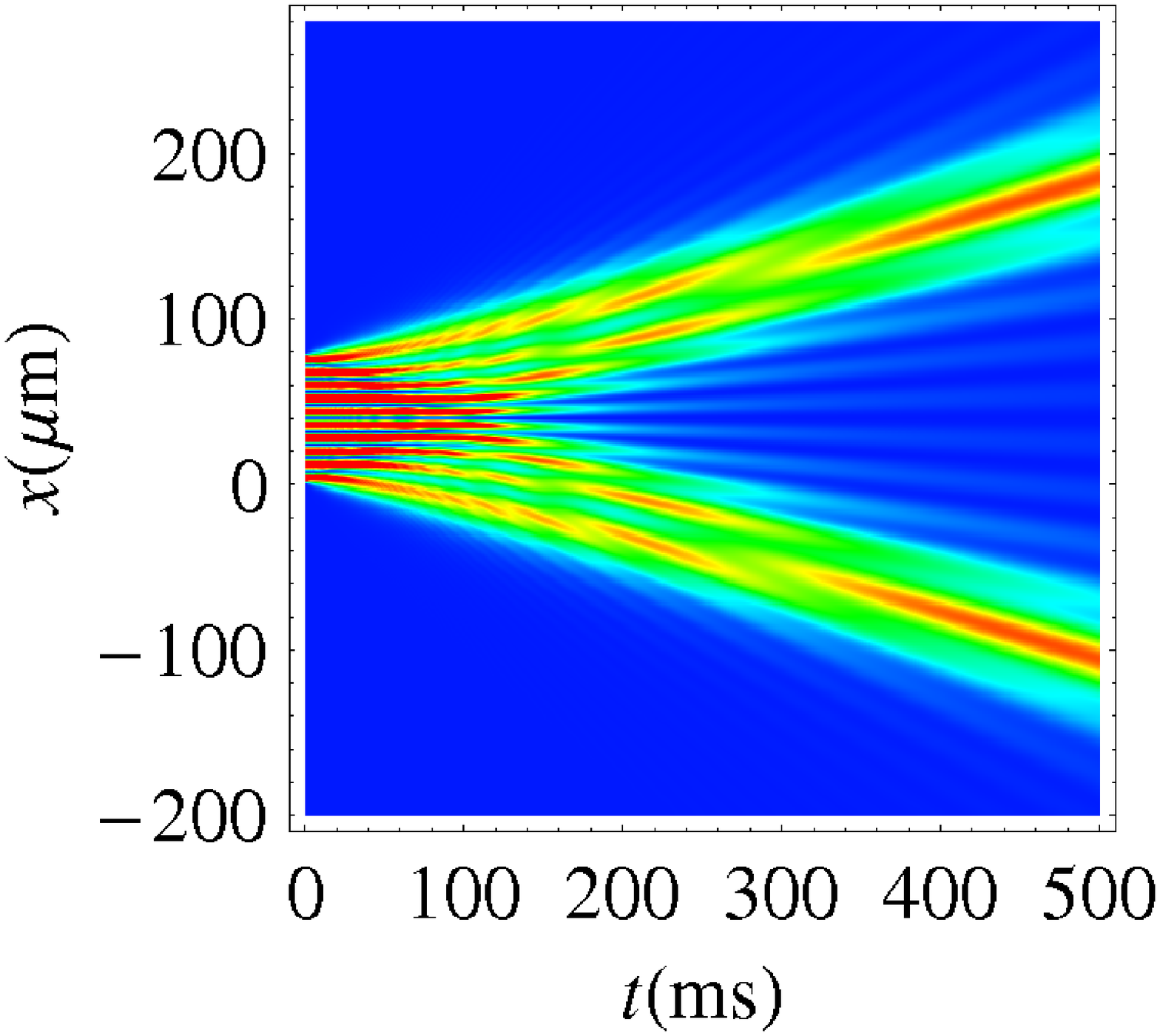}
\caption{\label{3sphw} Density plots of the probability
density for the dynamical evolution of a ${}^{87}$Rb atom in the tenth eigenstate released from
a harmonic (left) and box (right) traps. $L=80$ $\mu$m and $\om$ is chosen so that the energies
of the eigenstate in both traps coincide.}
\end{center}
\end{figure}
Figure \ref{3sphw} shows the expansion of a single particle wavefunction suddenly released from a HO and a HW trap.
If the particles are in the ground state of the trap the evolution is alike.
However, for any excited state a remarkable difference appears.
For the former case, the scaling law gives rise to a uniform spreading without affecting
the initial quantum structure. As can be seen in the $x$-$t$ density plot,
the full with at half maximum (FWHM) becomes linear for any $t\gg\om^{-1}$.
By contrast, whenever the initial trap is box-like, the density profile exhibits a bifurcation
into two main branches after the  semiclassical time
\beqa
t_{n}=\frac{m L}{2 p_{n}}=\frac{mL^{2}}{2n\pi\hbar}
\eeqa
of expansion \cite{DM05}.
This is the result of the mapping of the underlying momentum distribution to the density profile expected asymptotically.
Indeed, the probability density in $k$-space for a box eigenstate is $\vert\widetilde{\phi}_n(k)\vert^{2}=
\frac{2\pi n^2}{L^3}[1-(-1)^n\cos(kL)]/(k^2-k_{n}^{2})^2$, and has
two main peaks at $k=\pm k_n$
except for the ground state: $\vert\widetilde{\phi}_1(k)\vert^{2}$, is a bell-shaped distribution centered at $k=0$.

The more general dynamics in an expanding box has been described in \cite{DMN92,DKN93}.
The transient dynamics of particles released from a cylindrical trap has also been discussed in \cite{Godoy05} and  from  spherical traps in \cite{MoshinskyRMF52b,Godoy03,Godoy07}.
We recall that the 3D s-wave case is tantamount to the 1D problem in half space, with a hard-wall at $x=0$.
One can exploit the method of images \cite{Kleber94} to write down the propagator,
\beq
K_{s}^{3D}(r,t|r',0)=K_{0}(r,t|r',0)-K_{0}(-r,t|r',0),
\eeq
and the wavefunction in the radial coordinate as
\beq
\psi_{s}^{3D}(r,t)=\psi_{0}(r,t)-\psi_{0}(-r,t), \qquad r>0,
\eeq
where $\psi_{0}(r,t)=\int_{-\infty}^{\infty}dr'K_{0}(r,t|r',t'=0)\psi_{0}(r',t'=0)$,
$\psi_{0}(r',t'=0)$ being the radial wavefunction of the initial state.
The effect of interactions during the expansion will be considered in Section \ref{ucgtw}.
\subsection{Phase space representation}
\subsubsection{The Wigner function}
The Wigner function \cite{Wigner32},
\beq
\label{wigner}
{W}(x,p) := \frac{1}{\pi\hbar}\int_{-\infty}^{\infty}dy\,
\psi(x+y)^{*}\psi(x-y)e^{2ipy/\hbar},
\eeq
is the best known quasi joint probability distribution for position and momentum.
Nowadays, it is possible to study it experimentally as has been
demonstrated in a series of works \cite{wigexp1,wigexp2,wigexp3,wigexp4,wigexp5,wigexp6}. We recall that
the marginals of the Wigner function
are precisely the momentum and coordinate
probability densities, this is,
$\int dx {W}(x,p)=\vert\widetilde{\psi}(p)\vert^2$ and
$\int dp {W}(x,p)=\vert\psi(x)\vert^2$.

In the context of the Moshinsky shutter, its time evolution was derived for the
free case problem \cite{MMS99} and in the presence of a linear potential \cite{DM06a}.
The Wigner function for a cut-off plane wave initial condition
is\footnote{Since $\psi(x,t=0)=e^{ip_0x/\hbar}\Theta(-x)$
is dimensionless, $W(x,p)$ here has dimensions $[p]^{-1}$.}
\beq
\label{wignermosh}
{W}(x,p;p_{0},t=0)=\frac{1}{\pi}
\frac{\sin[-2x(p_{0}-p)/\hbar]}{p_{0}-p}\Theta(-x),
\eeq
which clearly assumes negative values. If the potential is linear or quadratic,
the time evolved Wigner function follows classical trajectories,
\beq
{W}(x,p;t)=\int\!\!\!\int dx_{i}dp_{i}\delta[x-x_{cl}(x_{i},p_{i},t)]
\delta[p-p_{cl}(x_{i},p_{i},t)]{W}(x_i,p_i;t=0).
\eeq
For (\ref{wignermosh}) the result is
%
limited to the classically
accessible region of phase space.
Moreover, we may take the limit $\hbar\rightarrow 0$ to find
the classical distribution \cite{MMS99,MS00,SM05,DM06a},
%
%
in which no hint
about the quantum oscillations -hallmark feature of DIT- remains.

Similarly, one can describe the time evolution of the eigenstates of a hard-wall trap.
At the beginning of the experiment the Wigner function is given by \cite{Almeida98}
\beqa
{W}_{\phi_n}(x,p)=\mathcal{P}_{n}(x,p)\chi_{[0,L/2]}(x)+
\mathcal{P}_{n}(L-x,p)\chi_{[L/2,L]}(x),
\eeqa
where
\beqa
\mathcal{P}_{n}(x,p)= \frac{2}{\pi\hbar L}
\bigg\{\sum_{\nu=\pm 1}\frac{\sin\left[2(p/\hbar+\nu n\pi/L)x\right]}
{4(p/\hbar+\nu n\pi/L)}-\cos\left(\frac{2n\pi x}{L}\right)
\frac{\sin(2 p x/\hbar)}{2p/\hbar}\bigg\}.\nonumber
\eeqa
%
\begin{figure}
\begin{center}
\includegraphics[width=4cm,angle=0]{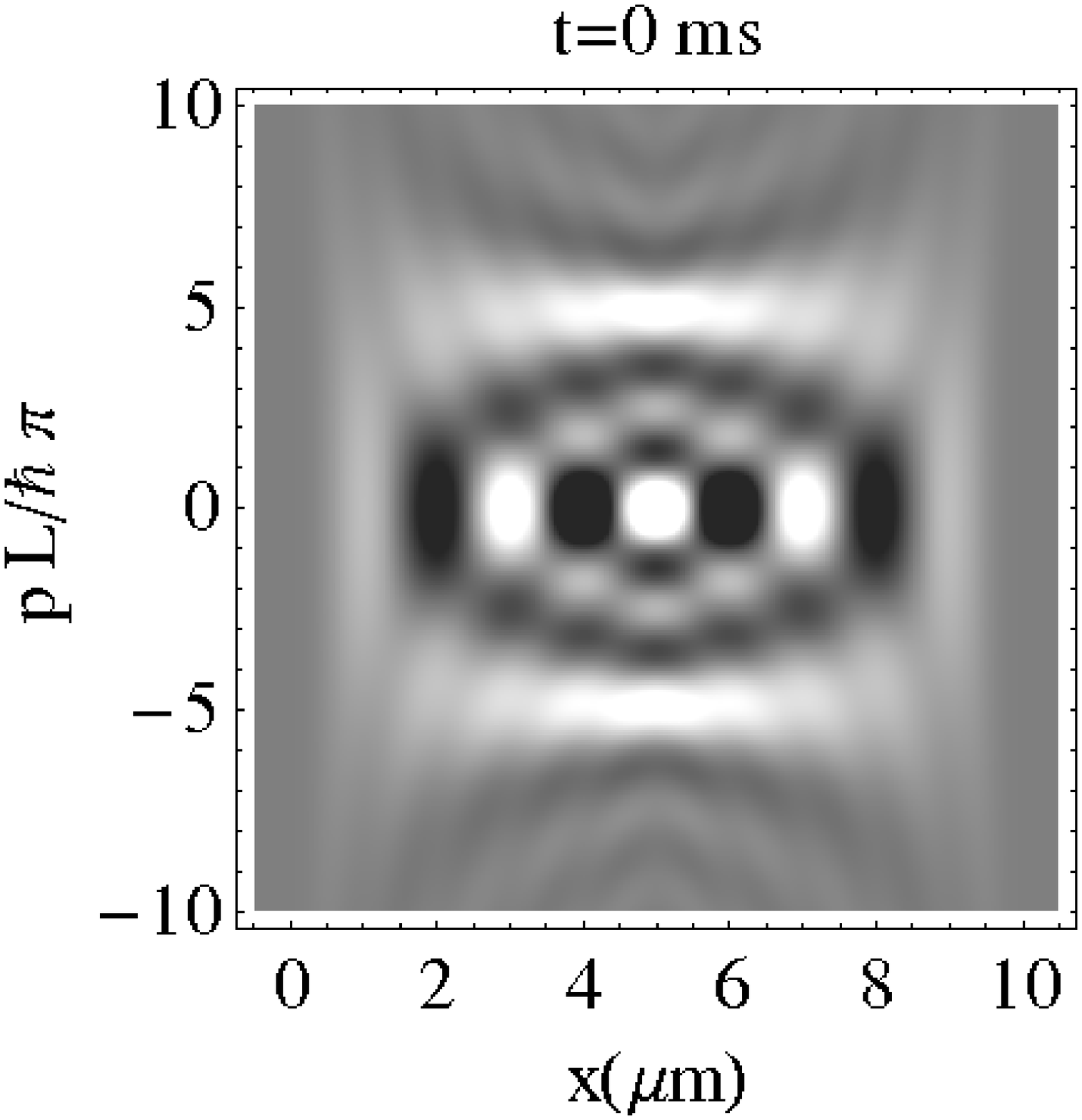}
\includegraphics[width=4cm,angle=0]{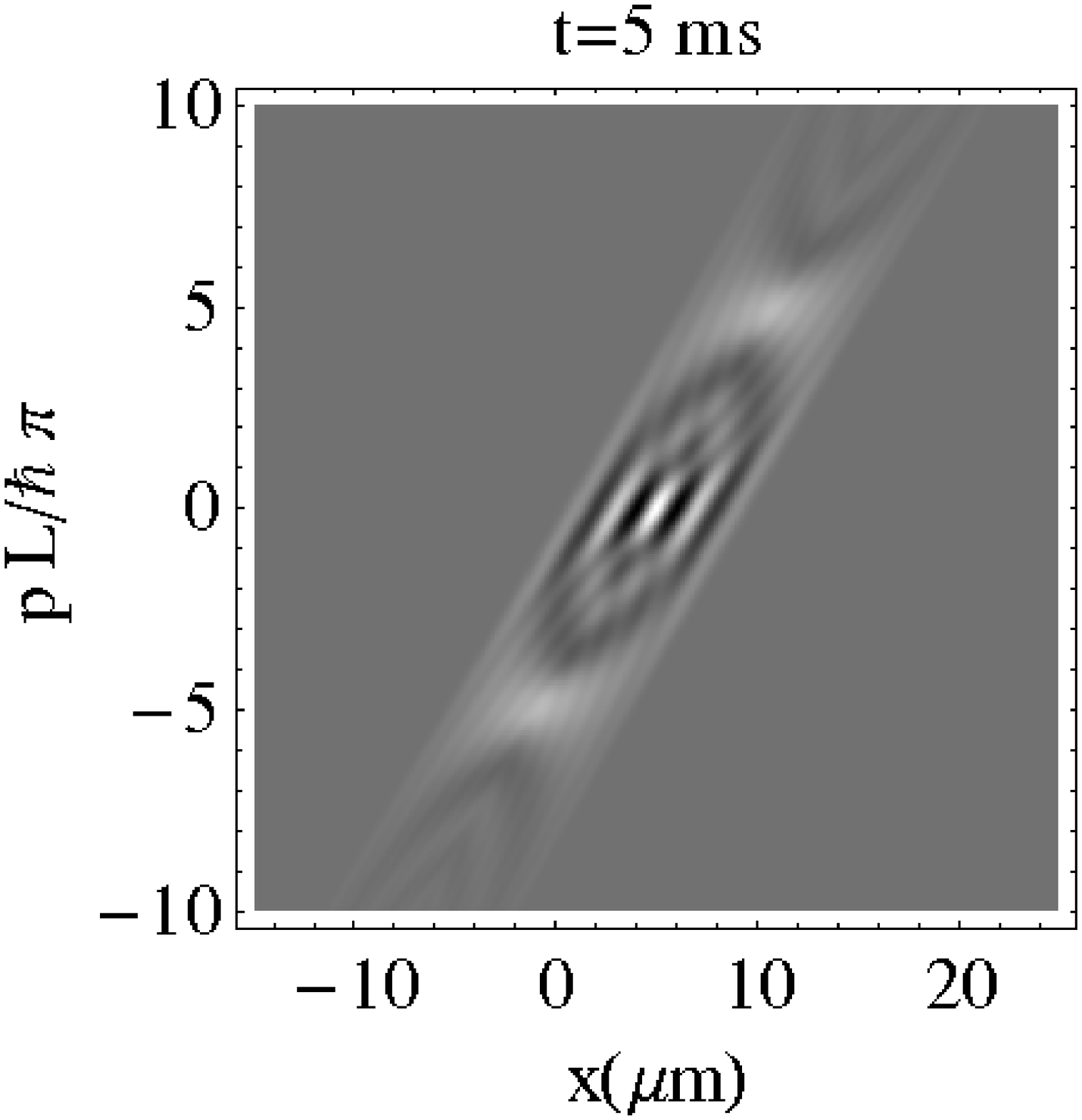}
\includegraphics[width=4cm,angle=0]{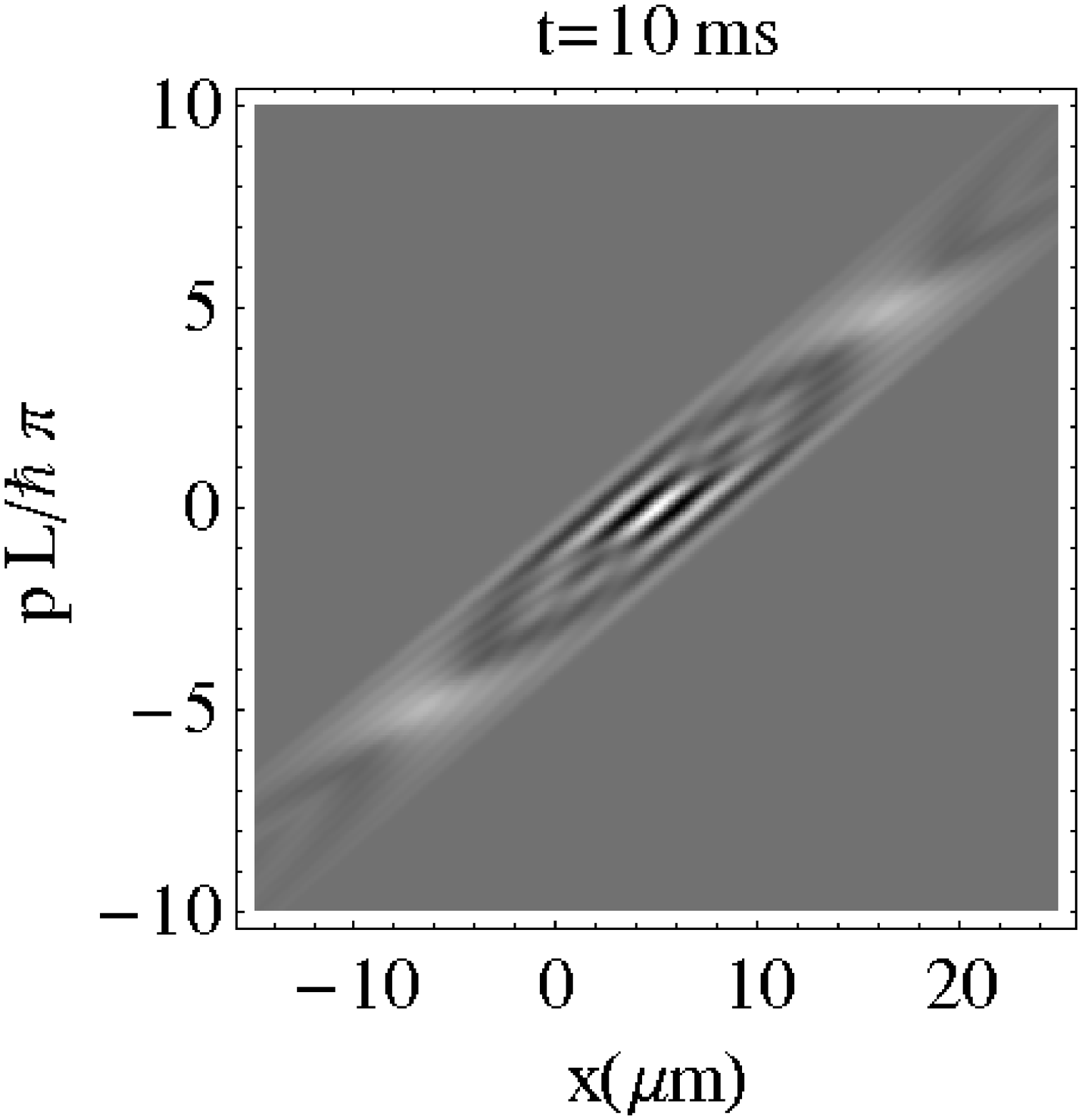}
\vspace*{.2cm}
\caption{\label{wigner5} Time evolution of the Wigner function for the fifth eigenstate of the box trap
($L=10\mu$m, for a $^{87}$Rb atom). As time goes by, the the quasi-probability distribution is stretched along the $x$-axis.
The gray scale varies from black to white as the function value increases,
corresponding the background gray to zero value. Note in particular the presence of negative values
(in black), telltale sign of the non-classicality of the state.
}
\end{center}
\end{figure}

Figure \ref{wigner5} shows the stretching of the Wigner function along the $x$-axis as time goes by.
The momentum distribution, being bimodal for all $n>1$, is responsible for the splitting in
real space for $t>t_n$ already noticeable at $t=5$ ms.
For a general Hamiltonian the propagation of a Wigner function obeys the quantum Liouville equation \cite{Wigner32,Wyatt05}
\beq
\partial_{t}{W}(x,p;t)+\frac{p}{m}\partial_{x}{W}(x,p;t)-\sum_{l=0}^{\infty}
\frac{(-1)^l(\hbar/2)^{2l}}{(2l+1)!}\partial_{x}^{2l+1}V(x)\partial_{p}^{2l+1}{W}(x,p;t)=0.\nonumber
\eeq
%
In such a way, the phase space description of transient effects can be extended to tunneling problems as well \cite{Moshinsky04,LSA04}.

The Wigner function has been also proposed to study the
quantum transient response to changes in an externally applied
voltage in semiconductor structures \cite{Frensley87}, see also
Section \ref{sape},
dynamics of polaron formation in compound semiconductors \cite{Jacoboni},
or for transients in scattering processes \cite{MuSn92} and classical-quantum comparisons \cite{MuSn93}.
\subsubsection{Symplectic tomography}
A related representation in phase space is symplectic tomography \cite{MMT96}. The wave function $\psi(x)$
or the density matrix $\rho(x,x')$ can be mapped onto the
standard positive distribution ${\cal W}(X,\mu,\nu)$
of the random variable $X$ depending on two real
extra parameters, $\mu$ and $\nu$. The symplectic tomogram can be related to the
Wigner function $W(x,p)$ \cite{Wigner32},
\begin{equation}\label{wignertomo}
{\cal W}(X,\mu,\nu)= \int W(x,p)\delta(X-\mu x - \nu p)\frac{d x d p}{2\pi\hbar}
\end{equation}
as its Radon transform.
The parameters $\mu$ and $\nu$
can be expressed in the form $s\cos\theta$, $\nu=s^{-1}\sin\theta$,
so that $s>0$ is a real squeezing parameter and $\theta$ is a rotation
angle. Then the variable $X$ is identical to the position measured
in the new reference frame in the phase-space which has new scaled axes $sz$ and $s^{-1}p$
and rotated by an angle $\theta$.
The Moshinsky solution in the tomographic representation reads \cite{MMS99}
\beqa
{\cal W}(X,\mu,\nu)=\frac{1}{2|\mu|}
\Bigg\{\left[\frac{1}{2}+C(\xi)\right]^{2}
+\left[\frac{1}{2}+S(\xi)\right]^{2}\Bigg\},
\eeqa
where
\beqa
\xi=\frac{k(\mu \tau+\nu)-X}{\sqrt{2\mu(\mu \tau+\nu)}}
\eeqa
with $\tau=\hbar t/m$. In the reference frame with $(\mu,\nu)=(1,0)$, the quantum tomogram reduces to the density profile in Eq. (\ref{moshprofilefresnel}).
\subsection{Propagation of wavepackets with sharp boundaries.}
The sharp boundaries of the initial wavefunction considered so far, which
also arise naturally in the Zeno dynamics under periodic spatial projections \cite{FPSS02},
might cast some doubt on the realizability of DIT.
The study of the generic time propagation of initially confined wave-packets has shed new light on the
effects of the boundaries on quantum transients at short times \cite{GM05,GM06b,MaS05}.
In the spirit of the Moshinsky shutter, let us consider a state $\psi(x,t=0)$ with support in the negative semiaxis ($x<0$),
and denote its Fourier transform by
\beqa
g(k)=\frac{1}{(2\pi)^{1/2}}\int \psi(x,t=0)e^{-ikx}dx.
\eeqa
For any $t>0$ it follows that
\beqa
\label{sharpwp}
\psi(x,t)=\frac{1}{(2\pi)^{1/2}}\int g(k)M(x,k,t) dk.
\eeqa
An expansion of the Moshinsky function with respect to $k$, allows to perform the integral in (\ref{sharpwp}),
and to express the solution in terms of derivatives of the initial state $\psi(x,t=0)$ at the boundary $x=0$,
\beqa
\psi(x,t)=\frac{1}{(2\pi)^{1/2}}M\left(x,-i\frac{\partial}{\partial y},t\right)\psi(y,t=0)\bigg|_{y=0}.
\eeqa
Expanding in a power series of $(\hbar t/2m)^{1/2}x=\eta$, one finds for $\hbar t/2m x^2\ll1$,
\beqa
\sqrt{2\pi}\psi(x,t)&=&\sqrt{\frac{i}{\pi}}e^{\frac{imx^2}{2\hbar t}}\{\eta-2i\eta^3(1+x\partial_y)\nonumber\\
& & -4\eta^5[3(1+x\partial_y)+x^2\partial_{y} ^2+\dots]\}\psi(y,t=0)|_{y=0}.
\eeqa
So, for short times, the value of the wavefunction depends exclusively on $\psi(x,t=0)$ and has a $t^{1/2}$
dependence, whereas should $\psi(x,t=0)$ vanish, higher order terms are to be considered.
The effect of the exact wave packet boundary may however be washed out
at longer times as demonstrated by the revival effect, see 
\ref{switching}.
\section{Beam chopping and formation of pulses\label{onepulse}}
\subsection{Quantum transients in beam chopping}
Diffraction in time is generally associated with the release of a matter wave initially confined in a given region of space.
Nonetheless, the opposite problem is of great physical interest in interferometry, namely, the chopping of a beam  by a shutter, or mirror.
Consider a monochromatic beam described by a travelling plane-wave $\psi(x,t=0)=e^{ikx}$ and the action of a totally reflecting mirror suddenly turned on at $x=0$ and $t=0$. For all $t>0$ the beam is effectively split into two different parts. Noticing that the corresponding propagator is that of a hard-wall at the origin, $K_{wall}(x,t|x',0)=K_0(x,t|x',0)-K_0(-x,t|x',0)$, the evolution of the beam is
\begin{equation}
\psi (x,t)=\left\{
\begin{array}{cc}
M(x,k,t)-M(-x,k,t), & x<0 \\ [.3cm]
M(-x,-k,t)-M(x,-k,t),  &  x \geq 0,
\end{array}
\right.
\label{beamchopwf}
\end{equation}
where we have neglected the width of the mirror for simplicity. From
the asymptotic properties of the Moshinsky function (see Appendix \ref{app_moshfunction})
it is clear that the matter wave
$\psi(x)=2i\sin(kx)e^{-i\frac{\hbar k^2t}{2m}}$ is eventually formed at the left of the mirror.
%
\begin{figure}
\begin{center}
\includegraphics[height=5cm]{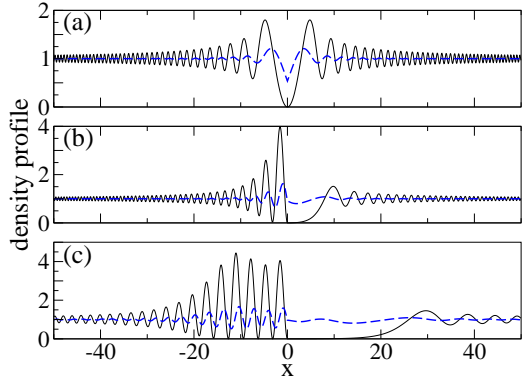}
\caption{\label{beamchop} Snapshots of the density profile of a chopped beam ($\hbar=m=1$) with
a) $k=0$ at $t=5$, b) $k=1$ at $t=5$, and c) $k=1$ at $t=20$.
The solid line corresponds to the totally reflecting shutter
($\nu\rightarrow\infty$) in Eq. (\ref{beamchopwf}),
while the dashed line is for a finite $\delta$-potential with $\nu=0.25$, see Eq. (\ref{deltabeamchopwf}).}
\end{center}
\end{figure}
The effect of finite reflectivity of the shutter can be taken into account considering a delta mirror $V(x)=\eta\delta(x)$ which represents a weak and narrow potential. The explicit dynamics is slightly more involved, for the propagator is \cite{EK88}
\beqa
K_{\delta}(x,t|x',0)=K_{0}(x,t|x',0) +\varkappa M(|x|+|x'|,-i\varkappa,\hbar t/m),
\eeqa
with $K_{0}(x,t|x',0)$ as in (\ref{freepropagator}) and $\varkappa=m\eta/\hbar^2$.
Using the integrals in Appendix \ref{app_moshfunction}, it turns out that
\beqa
\label{deltabeamchopwf}
\psi(x,t)=e^{ikx-i\hbar k^2 t/2m}
-\sum_{\nu=\pm 1}\frac{\varkappa}{\varkappa-i\nu k}[M(|x|,\nu k,t)
-M(|x|,-i\varkappa,t)],
\eeqa
which asymptotically, at long times,  tends to
\beqa
\psi(x,t)\sim e^{-i\hbar k^2 t/2m}\bigg[\frac{k}{k+i\varkappa}e^{ikx}
-\frac{\varkappa}{k-i\varkappa}e^{-ikx}\bigg].
\eeqa
This is nothing but the stationary scattering state of the mirror with the energy of the initial beam
(notice that $\frac{k}{k+i\varkappa}$ and $-\frac{\varkappa}{k-i\varkappa}$ are
the corresponding transmission and reflection probability amplitudes, respectively).
The dynamical buildup of this standing matter-wave after the action of the shutter is illustrated in Fig. \ref{beamchop}.

Further work on beam chopping and pulse formation has almost invariably
been carried out using the ``quantum source approach'',
to which we now turn out attention.
\subsection{Formation of single pulses}
The quantum mechanical effects of beam chopping have been studied in time-dependent
neutron optics \cite{GG84,FGG88}. The energy distribution of the initial beam is expected to be modified in this process in agreement with a time-energy uncertainty relation \cite{MoshinskyRMF52b,Moshinsky76}.
In this section we examine the formation of a single-pulse of duration $\tau$ from a
quasi-monochromatic quantum source $\psi(x=0,t)\sim e^{-i\om_0t}$
modulated according to a given aperture function $\chi^{(1)}(t)$.\footnote{Baute, Egusquiza and Muga \cite{BEM01} have made explicit  the relation between a source
boundary condition where $\psi(x=0,t)$ is specified for $t>0$ and the standard
initial value problem in which $\psi(x,t=0)$ is specified.
The simple ``Kirchhoff boundary condition'' $\psi(x=0,t)=\Theta(t)e^{i\om_0t}$ is discussed
further in  \ref{mbsources}, see also \cite{GOD07}.}
Here, the superscript
denotes the creation of a single pulse \cite{MoshinskyRMF52b,Moshinsky76,GG84,FGG88,FMGG90,BZ97,DM05,DMM07}, and the free dispersion relation holds, $\om_0=\hbar k_0^2/2m$.
Many works have been carried out within this approximation in neutron interferometry \cite{GG84,FGG88,BSCK92},
considering also a triangular  \cite{GG84,FMGG90} or more general apodized aperture functions \cite{DM05}, atom-wave
diffraction \cite{BZ97}, tunneling dynamics \cite{Stevens,Ranfa90,Ranfa91,Mor92,BT98,MB00,DMRGV},
and absorbing media \cite{DMR04}.
After the experimental realization of a guided atom laser \cite{Guerin06}, the same formalism has been employed to describe the
dynamics in such system \cite{DMM07}.

For the sake of concreteness, let us focus on the family of aperture functions
\beqa
\chi_{n}^{(1)}(t)=\sin^{n}(\Om t)\Theta(t)\Theta(\tau-t),
\eeqa
with $\Om=\pi/\tau$ \cite{DM05,DMM07}. The solution for an arbitrary aperture function is described in Section \ref{generapo}.
We first note that the energy distribution varies for increasing $n$ and $T$.
The energy distribution of the associated wavefunction
is proportional to $\om^{1/2}|\widehat{\psi}_n^{(1)}(x=0,\om)|^2$ \cite{BEM01,DMM07},
where $\widehat{\psi}_n^{(1)}(x=0,\om)=(2\pi)^{-1/2}\int dt \exp[i(\omega-\omega_0)t)
\chi_n^{(1)}$.
For increasing $n$, the aperture function becomes smoother broadening the energy distribution.
Similarly, the smaller the time for pulse formation $\tau$,  the broader is its associated
energy distribution. Indeed, whenever $\hbar\om_0\tau<1$, the distribution is shifted to higher energy components.

As an example, let us consider the case $\chi_{0}^{(1)}(t)=\Theta(t)\Theta(\tau-t)$
corresponding to a rectangular single-slit in time which is switched on and off at infinite velocity.
The time evolution of the pulse is given by
\beqa
\label{recpulse}
\psi_{0}^{(1)}(x,k_0,t;\tau)&=&[M(x,k_0,t)+M(x,-k_0,t)]\nonumber\\
& &-
\Theta(t-\tau)e^{-i\om_0 \tau}[M(x,k_0,t-\tau)+M(x,-k_0,t-\tau)].
\eeqa
If $t<\tau$, before the pulse has been fully formed, the problem reduces to that of
a suddenly turned-on source, $\psi_{0}^{(1)}(x,k_0,t<\tau)=M(x,k_0,t)+M(x,-k_0,t)$, discussed in \ref{mosh52}.
An explicit calculation of the energy distribution of the resulting pulse, taking the overlap of
the wavefunction with the free particle eigenstates which vanish at the closed shutter \cite{MoshinskyRMF52b,Moshinsky76}, shows that
$$
\mathcal{P}(E;E_0,\tau)=\mathcal{N}\sqrt{E}\,
\frac{\sin^{2}[(E-E_0)\tau/2\hbar]}{(E-E_0)^{2}},
$$
$\mathcal{N}$ being a normalization constant.
Such distribution is peaked at the energy of the initial beam $E_0=\hbar\om_0$ if $E_0\tau\gtrsim\hbar$ and shifted to higher energies otherwise \cite{DMM07}. Indeed, for some measure of the width $\Delta E$ it leads to the time-energy uncertainty relation $\Delta E\tau\simeq\hbar$  \cite{MoshinskyRMF52b,Moshinsky76,Busch02,DM05}.

It is illuminating to consider a slightly more complicated aperture function, such
as the sine-square (Hanning) function,
\beqa\label{Hanning}
\chi_2^{(1)}(t)&=&\sin^{2}({\Om t})\,\Theta(t)\,\Theta(\tau-t)
=\frac{1}{2}\left[1-\frac{1}{2}\cos({2\Omega t})\right]\,\Theta(t)\,\Theta(\tau-t).
\eeqa
Defining $k_{\beta}=\sqrt{2m(\om_{0}\pm2\Om)/\hbar}$
with $\beta=\pm1$, the resulting pulse is described by the time-dependent wavefunction
\beqa
\psi_2^{(1)}(x,k_0,t;\tau)=\frac{1}{2}\left[\psi_{0}^{(1)}(x,k_0,t;\tau)
-\frac{1}{2}\sum_{\beta=\pm1}\psi_{0}^{(1)}(x,k_{\beta},t;\tau)\right],
\eeqa
where the effect of the apodization is to subtract to the pulse
with the source momentum two other matter-wave trains associated with $k_{\beta}$,
all of them formed with rectangular aperture functions.
The effect of creating the pulse by a smooth modulation as opposed to the sudden rectangular aperture function,
 is to suppress the diffraction sidelobes in the density profile, at the expense of broadening
the corresponding energy distribution. These two features are the key elements of apodization
in classical optics and Fourier analysis, and hence the effect was dubbed {\it apodization in time} \cite{DM05,DMM07}.
\subsection{Arbitrary modulation of a matter-wave source}
\label{generapo}
Though the dynamics of matter-wave sources associated with different aperture functions
has been worked out in detail \cite{Moshinsky76,GG84,FGG88,FMGG90,BZ97,DM05}, it is desirable
to tackle the general case. In order to do so, we consider an
arbitrary aperture function $\chi(t)$ which is zero outside the interval
$[0,\tau]$ \cite{DMM07}. An analytical result can be obtained in such case by means of Fourier series.
Noting that
\beq
\chi(t)=\sum_{r=-\infty}^{\infty}c_{r}e^{i\frac{2\pi rt}{\tau}}\Theta(t)\Theta(\tau-t)
\quad {\rm with}\qquad
c_{r}=\frac{1}{\tau}\int_{0}^{\tau}\chi(t)e^{-i\frac{2\pi rt}{\tau}}dt,
\eeq
the  wavefunction at the origin can be written as the coherent superposition of its Fourier components
modulated with a rectangular aperture function
\beqa \label{apBeq:sbc}
\psi_{\chi}(x=0,k_0,t) &=& \sum_{r=-\infty}^{\infty}c_{r}e^{-i \omega_{r} t}\Theta(t)\Theta(\tau-t),
\eeqa
with $\omega_{r}=\omega_{0}+2\pi r/\tau$. Clearly, the subsequent dynamics for $x,t>0$ reads
\begin{eqnarray}
\psi_{\chi}(x,k_0,t;\tau)
& = & \sum_{r=-\infty}^{\infty}c_{r}\psi_{0}^{(1)}(x,k_r,t;\tau),
\end{eqnarray}
where $\psi_{0}^{(1)}(x,k,t;\tau)$ is defined in Eq. (\ref{recpulse}),
and $\hbar k_{r}=\sqrt{2m\hbar\om_r}$, with the branch cut taken along the negative
imaginary axis of $\om_{r}$.
The upshot is that the dynamics of a source with an arbitrary
aperture function can be described in terms of the coherent superposition of
rectangular pulses associated with each of the Fourier components.
\subsection{Diffraction-in-time revival after slow switching}
\label{switching}
The dynamics of a matter-wave source can be tamed by switching it slowly.
For the sake of concreteness, we introduce the continuous switching function
\begin{equation}
\label{switch}
\chi^{switch}(t)=\cases{0 &if $\,\, t<0$\,,\cr
\chi(t), &if $\,\, 0\leq t<\tau$\,,\cr
1, &if $\,\, t\geq\tau$,\cr}
\end{equation}
with a characteristic time scale $\tau$, which we shall refer to as the switching time.
The family for which $\chi(t)=\sin^n\Om_s t$ with $n=0,1,2$ and $\Om_s=\pi/2\tau$ was considered in \cite{DMM07,DLPMM08},
where it was shown that for fixed $\tau$, the apodization of the beam increases with the smoothness of $\chi(t)$,
this is, with $n$. Note that for $n=0$, one recovers the sudden aperture $\chi_0(t)=\Theta(t)$ which
maximizes the diffraction-in-time fringes. We shall consider a sine-square modulation in what follows ($n=2$),
and concentrate on the dependence on $\tau$.

%
\begin{figure}
\begin{center}
\includegraphics[width=7.5cm,angle=0]{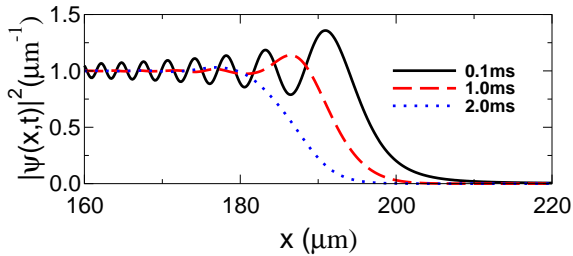}
\caption{\label{apodization}
Apodization in time. Removal of the sudden approximation in the switch of a matter wave
source leads to the suppression of oscillations in the beam profile characteristic of the diffraction in time.
The $^{87}$Rb beam profile is shown at $t=20$ms, and $\hbar k_0/m=1$cm/s.
For increasing $\tau$ the amplitude of the oscillations diminishes and the signal is delayed.
}
\end{center}
\end{figure}

Increasing the switching time $\tau$  leads to an apodization of the oscillatory pattern. As a result
the fringes of the beam profile are washed out, see Fig. \ref{apodization}.
In this sense, the effect is tantamount to that of a finite band-width source \cite{MB00},
and as already remarked, this is the essence of apodization (in time), the suppression of diffraction by means of broadening
the energy distribution \cite{BT59,Fowles68}.

%
\begin{figure}
\begin{center}
\includegraphics[width=7.5cm,angle=0]{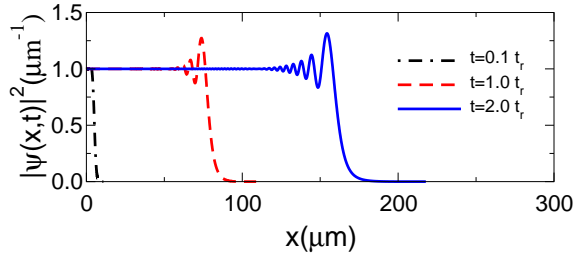}
\caption{\label{revival}
Revival of the diffraction in time.
During the time evolution of an apodized source, there is a revival of the diffraction
in time in the time scale $t_r=\om_0\tau^2$ ($\tau=1$ms, $\hbar k_0/m=0.5$cm/s,
and $t_r=16.8$ms for a $^{87}$Rb source which is switched on following a sine-square function).
}
\end{center}
\end{figure}
%
However, the apodization lasts only for a finite time after which a {\it revival}
of the diffraction in time occurs, as shown in Fig. \ref{revival}.
The intuitive explanation is that the intensity of the signal from the apodization ``cap''
associated with the switching process during the interval $[0,\tau]$, decays with time, whereas the intensity of the main
signal (coming from the step excitation for $t>\tau$) remains constant.
For sufficiently large times, the main signal, carrying its
diffraction-in-time phenomenon, overwhelms the effect of the
small cap.
Indeed, from the linearity of the Schr\"odinger equation,
it follows that the wavefunction is the sum of a ``half-pulse'' term
associated with the switch released during the interval $[0,\tau]$ and a semi-infinite beam
suddenly turned on at time $t=\tau$.
The revival time can be estimated to be \cite{DMM07,DLPMM08}
\beq
t_r\approx \om_0\tau^2,
\eeq
and plays a similar role to the Rayleigh distance of classical diffraction theory,
but in the time domain \cite{Brooker03}.
The smoothing effect of the apodizing cap
cannot hold for times much longer than $t_r$.
This is shown in Fig. \ref{revival}, where the initially
apodized beam profile eventually develops spatial fringes, which reach the maximum value associated
with the sudden switching at $t\approx 2t_r$.
\subsection{Multiple pulses}
\label{mp}
\subsubsection{Application to atom lasers}

Knowledge of the dynamics  associated with a single-pulse $\psi^{(1)}(x,t)$
can be used to tackle matter wave sources periodically modulated in time.
Frank and Nosov first described the periodic action of a quantum chopper in ultracold neutron interferometry leading to multiple  rectangular aperture functions \cite{Frank1,Frank2}. More recently, del Campo, Muga and Moshinsky \cite{DMM07}
found the general dynamics of an atom source under an arbitrary aperture function as described in \ref{generapo}, and applied it to the description of an atom laser in the non-interaction limit.
The experimental prescription for an atom laser
depends on the ``out-coupling'' mechanism.
Here we shall focus on the scheme described in \cite{Phillips99,Phillips00}.
The pulses are extracted from a condensate trapped in a magneto-optical trap (MOT), and a well-defined momentum is
imparted on each of them at the instant of their creation through a
stimulated Raman process. The result is that each of the pulses  $\psi^{(1)}$
does not have a memory phase, the wavefunction describing the coherent atom laser being then
\beq
\phi^{(N)}(x,k_0,t)=\sum_{j=0}^{N-1}\psi^{(1)}(x,k_0,t-jT;\tau)
\Theta(t-jT).
\eeq
Such modulation describes the formation of $N$ consecutive $\chi^{(1)}$-pulses,
each of them of duration $\tau$ and with an ``emission rate''
(number of pulses per unit time) $1/T$.
The Heaviside function $\Theta(t-jT)$ implies that the $j$-th
pulse starts to emerge only after $jT$. We can impose a
relation between the out-coupling period $T$ and
the kinetic energy imparted to each pulse $\hbar\om_0$, such that $\om_0 T=l\pi$:
if $l$ is chosen an even (odd) integer
the interference in the beam profile $|\phi^{(N)}|^2$
is constructive (destructive).

The diffraction in time is a coherent quantum dynamical effect,
so we may expect it to be affected by noise in the environment.
If the phases between different pulses are allowed to fluctuate, the
interference pattern is blurred \cite{VVDHR06}.
Moreover, if there is no phase coherence between different pulses,
the resulting density profile becomes the incoherent sum
of the single-pulse densities,
$\rho(x,t)=\sum_{j}\vert\psi^{(1)}(x,k_0,t-jT;\tau)\vert^{2}\Theta(t-jT)$,
which we shall use as a reference case.

\begin{figure}
\begin{center}
\includegraphics[height=6cm,angle=0]{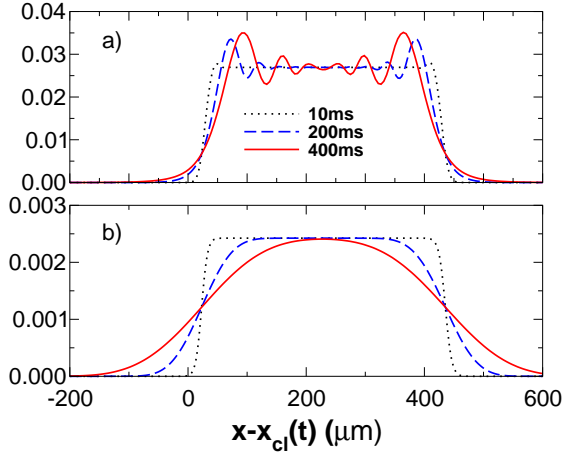}
\hspace*{.2cm}
\caption{
\label{atltime}
Dynamics of the density profile of an atom laser  a) coherently constructive, b) incoherent case.
Note that $x_{cl}=\hbar k_0 t/m$ is the classical trajectory. In all cases the norm is relative to the
incoherent case, as explained in the text. ($^{23}$Na source, modulated by a sine-square function
with $\tau=0.833$ms, $T=0.05$ms, moving at $3.73$cm/s, for which the revival time is $174.4$ ms.)
}
\end{center}
\end{figure}

One advantage of employing stimulated Raman pulses is that any desired fraction of atoms
can be extracted from the condensate. Let us define $s$ as the number of atoms out-coupled
in a single pulse, $\psi^{(1)}$. The total norm for the $N$-pulse incoherent atom laser is simply $\mathcal{N}_{inc}=sN$.
A remarkable fact for coherent sources is that the total number of atoms $\mathcal{N}_c$
out-coupled from the Bose-Einstein condensate reservoir depends on the nature of the interference \cite{Felipe}.
Therefore, in Fig.  \ref{atltime} we shall consider the signal relative to the incoherent case,
namely, $|\phi^{N}(x,t)|^2/\mathcal{N}_{inc}$.
%

%
Note the difference in the time evolution between coherently
constructive and incoherent beams in Fig. \ref{atltime}.
At short times both present a desirable saturation in the probability density of the beam.
However, in the former case a revival of the diffraction in time phenomena takes place in a
time scale $\om_0\tau^{2}$, similar to the smooth switching considered in \ref{switching},
whereas the latter develops a bell-shape profile.
This revival time can actually be related to the frequency of the BEC reservoir trap $\om_{trap}$,
which determines the width of the out-coupled pulses \cite{DLPMM08}.
Assume that the atoms are out-coupled from the $n$-th longitudinal mode of the reservoir trap, whose
spatial width is $\xi_n=[(n+1/2)\hbar/m\om_{trap}]^{1/2}$. Using $\om_0=\hbar k_{0}^2/2m$,
and the semiclassical relation $\tau_n=m\xi_n/\hbar k_0$,
the revival time becomes
\beq
t_r^{(n)}=\left(n+\frac{1}{2}\right)\frac{1}{2\om_{trap}}.
\eeq
For a non-interacting BEC, one can take $n=0$.
Therefore, when the trap of the BEC reservoir is very tight, DIT will arise ever since the creation of the atom laser beam,
while if the trap is wide, so will be the out-coupled pulses and apodization will suppress the oscillations
on the density profile for times smaller than the revival time.
Conversely, if the initial front of the beam profile is known to be smoothed out in a given length scale $\xi$,
it can be shown that density modulations arise in the time scale $m\xi^2/\hbar^2$.

\subsubsection{Coherent control}

As an outlook, 
the techniques of ``coherent control'' \cite{cc1,cc2,cc3}, could merge with the 
atom laser to design matter-wave pulses for specific aims. A remarkable example is the deterministic generation of arbitrary de Broglie wave-fronts in atom lithography \cite{ODHP00}. 
Coherent control has been applied in recent years to a wide range of systems in atomic, molecular, solid-state physics, semiconductor devices, biology or quantum information. 
The basic idea is to make use of
quantum interference to manipulate the dynamics with the aid of a control field,  typically formed by laser light. Examples of objectives so far are selective bond breaking, choosing a reaction pathway, or performing a quantum information operation. The experimental output
may be used in the optimization procedure with closed-loop learning
control techniques. Instead of a control field made of light, we envision 
the formation of control matter-fields by combining matter-wave
pulses with tunable 
intensity, energy, timing and phase. An elementary example is the formation 
of a continuous atom laser by constructive interference, as discussed above, 
and other applications may follow in which the elementary pulses and their transients will be helpful building blocks for analyzing the resulting matter field.        

\subsubsection{Ultra-short laser pulses \label{uslp}}

Ultrashort laser pulses in femtosecond or attosecond time domains are a basic tool
in coherent control research and applications. These nowadays manipulable  
pulses are able to induce 
many transient phenomena for nuclear or electronic motion. Using pump and probe 
or streaking methods, the transient dynamics following a system excitation may be 
recorded and/or steered.  
A detailed account is out of the scope of this review and
impossible to summarize here, but  
we shall mention briefly some recent examples to show that there exists a link with
work on elementary quantum transients which has not been fully exploited and is worth 
pursuing.  

Strong field ionization of atoms via tunnelling can be nowadays followed
with attosecond resolution      
so that some aspects of electron tunnelling (see others in Section \ref{tune}) 
can be observed \cite{tun1,tun2}. Impressive ``streaking techniques'' are used for this 
purpose, in which emission times and momenta of electrons are mapped, thanks to the action of 
the short pulses on the ejected electron, into final ion and electronic momenta which are readily observable with time-of-flight detectors. The emission may also be controlled so as to 
produce double or multiple time slits with few-cycle pulses and the corresponding interference-in-time phenomena
\cite{doubleslit}. 
Simple models for apodized pulses or time-dependent shutters
(see e.g. Section \ref{tidesh}) could be contrasted with these experimental procedures 
and may help to interpret the observed results and design new experiments.      
\section{Dynamics of ultracold gases in tight-waveguides\label{ucgtw}}
The models described so far do not explicitly consider quantum statistics
and rigorously only apply to single-particles.
Therefore, even though the use of semi-infinite beams and quantum sources
is suitable to describe neutron optics and ideal atom waves,
a more rigorous approach is desirable.
As already pointed out by Stevens, a consistent
interpretation for several -non-interacting- particles requires the appropriate
quantum symmetrization \cite{Stevens84}.
Moreover, there is nowadays a flurry of theoretical and experimental work
dealing with ultracold atomic gases in the non-linear and strongly
interacting regimes.

Ultracold atoms in waveguides, tight enough so that the transverse degrees of freedom are frozen out \cite{Olshanii98,DLO01},
are accurately described by the Lieb-Liniger (LL) model,
\beq\label{LiLi}
\hat{\mathcal{H}}_{LL}=-\frac{\hbar^{2}}{2m}\sum_{i=1}^{N}
\frac{\partial^{2}}{\partial x_{i}^{2}}
+g_{1D}\sum_{1\le i< j\le N}\delta(x_{i}-x_{j}),
\eeq
exactly solvable by coordinate Bethe ansatz \cite{LL63}. This model describes $N$ bosons in one-dimension
interacting with each other through a short-range $\delta$-function.
The use of such pseudo-potential is justified provided that the scattering
is restricted to $s$-wave type in the zero-energy limit of relevance to ultracold temperatures,
when the details of the true interparticle potential become unimportant \cite{Huang87}.

The cloud can then be characterized by the interaction parameter $\gamma
=mg_{1D}L/\hbar^2N$, where $g_{1D}$ is the one-dimensional coupling strength,
$L$ the size of the system, $m$ and $N$ the mass and number of atoms respectively;
$\gamma$ can be varied \cite{Olshanii98} allowing to realize the mean-field regime ($\gamma\ll1$)
or the Tonks-Girardeau regime ($\gamma\gg 1$) \cite{PSW00}.
\subsection{Dynamics in the Tonks-Girardeau regime}
We shall next consider the dynamics
of strongly interacting bosons confined in a tight-waveguide,
the so-called  Tonks-Girardeau (TG) gas.
In the TG regime, bosons mimic an effective Pauli exclusion principle as a result of the repulsive interactions rather than the quantum statistics itself. The Fermi-Bose (FB) duality \cite{Girardeau60,YG05,CS99,GW00b} provides
the many-body wavefunction of $N$ strongly interacting bosons from that of a
free Fermi gas with all spins frozen in the same direction.
The Fermi wavefunction, antisymmetric under permutation of particles,
 is built as a Slater determinant, with one particle in each eigenstate $\phi_{n}$ of the trap,
\beq
\psi_{F}(x_{1},\dots,x_{N})=\frac{1}{\sqrt{N!}}{\rm det}_{n,k=1}^{N}\phi_{n}(x_{k}).
\eeq
Whenever the position of two particles coincide, the wavefunction $\psi_{F}$ vanishes as
a consequence of the Pauli exclusion principle. This contact condition
is the desirable one in the TG gas due to the effectively infinite repulsive interactions.
Symmetrization can then be carried out ``by hand'', applying the  so-called ``antisymmetric unit function''
\beq
\mathcal{A}=\prod_{1\leq j<k\leq N}{\rm sgn}(x_{k}-x_{j}),
\eeq
so that
\beq
\psi_{B}(x_{1},\dots,x_{N})=
\mathcal{A}(x_{1},\dots,x_{N})\psi_{F}(x_{1},\dots,x_{N}).
\eeq
This is the Bose-Fermi map introduced by Girardeau in 1960 \cite{Girardeau60}.
A remarkable advantage of this mapping is that it holds for time dependent
processes (governed by one-body external potentials),
since the $\mathcal{A}$ operator does not include time explicitly
\cite{GW00b,YG05}.
The glaring upshot is that as far as local
correlation functions are concerned,
to deal with the many-body TG gas it suffices to work out the single
particle problem, since
\beq
\vert\psi_{B}(x_{1},\dots,x_{N};t)\vert^{2}
=\vert\psi_{F}(x_{1},\dots,x_{N};t)\vert^{2}.
\eeq
In particular, from the involutivity of the $\mathcal{A}$ operator ($\mathcal{A}^{2}=1$)
and the fact that $\la\phi_{n}\vert U^{\dag}U\vert\phi_{m}\ra=\delta_{nm}$, where $U$ is
the time-evolution operator,
it follows that the time-dependent density profile can be calculated as \cite{GW00b}
\beq
\label{3dp}
\rho(x,t)=N\int\vert\psi_{B}(x,x_{2},\dots,x_{N};t)\vert^{2}dx_{2}
\cdots dx_{N}
=\sum_{n=1}^{N}\vert\phi_{n}(x,t)\vert^{2}.
\eeq
Once the single-particle time evolution is known, we are
ready to study the spreading of the TG gas through its density profile, Eq. (\ref{3dp}).

\begin{figure}
\begin{center}
\includegraphics[height=5cm]{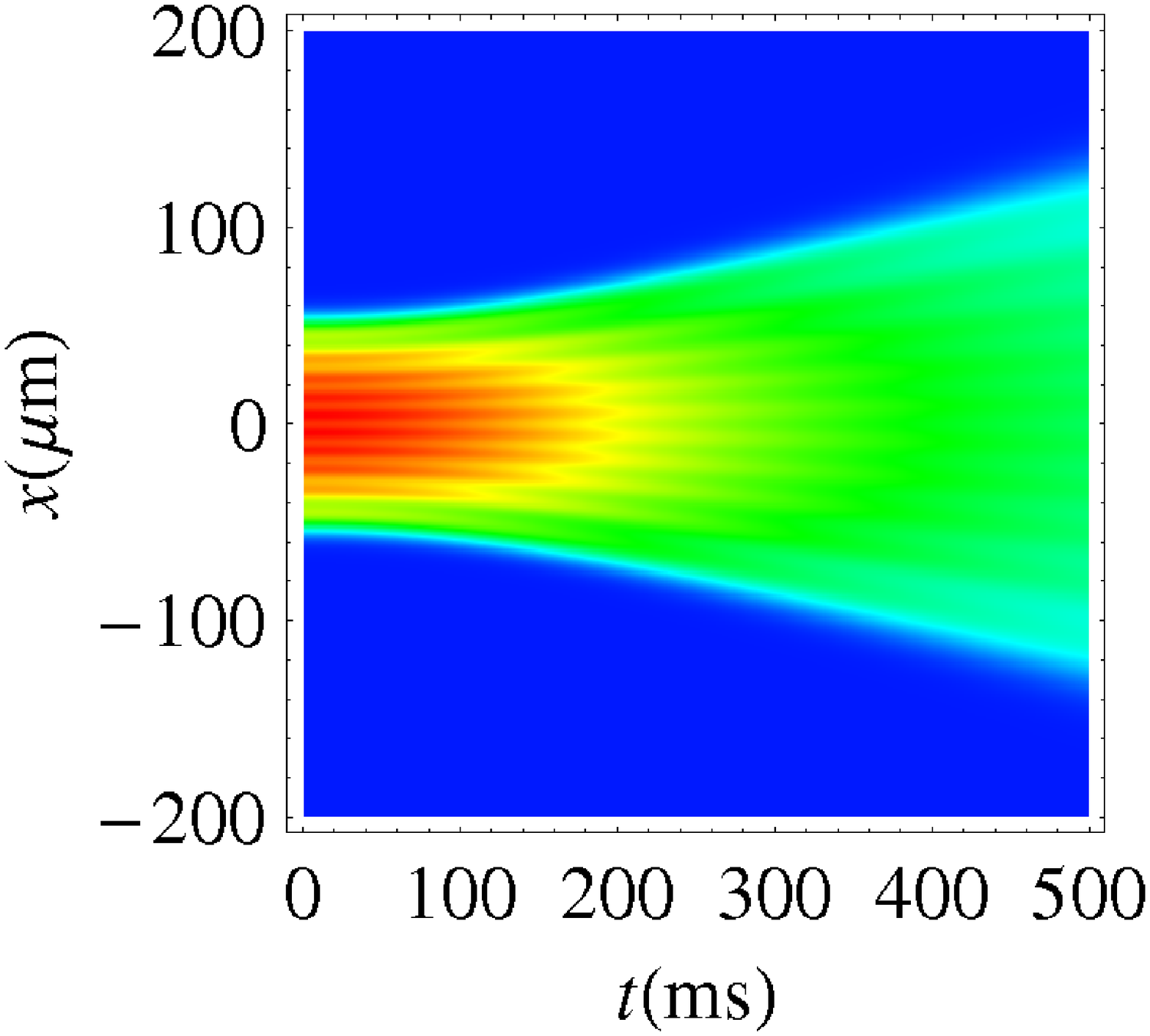}
\includegraphics[height=5cm]{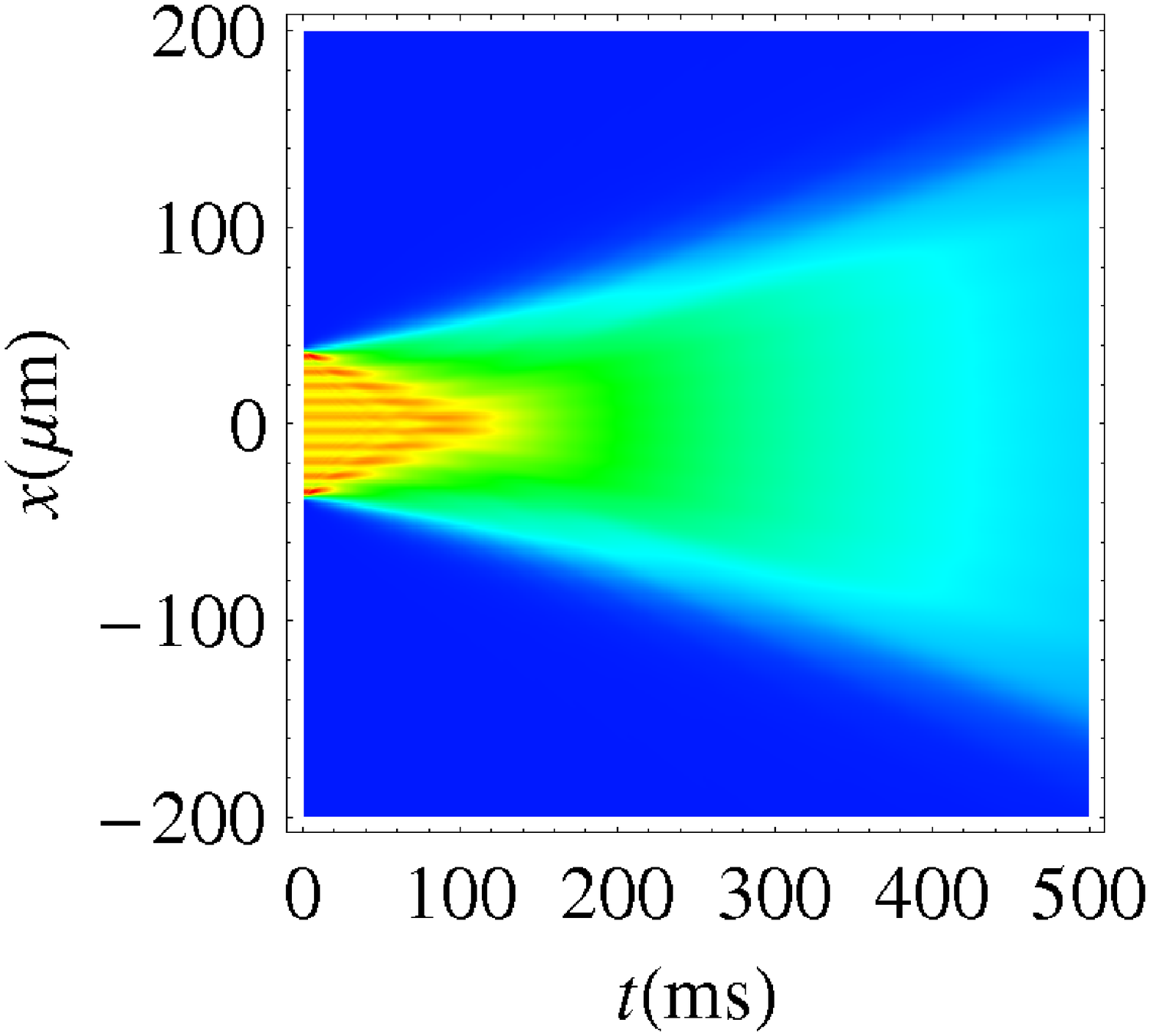}
\caption{\label{3figdp} Density plots of the probability
density $\rho$ for a TG gas composed of $N=10$ atoms of ${}^{87}$Rb released from a harmonic (left) and box-like (right) trap.
$L=80$ $\mu$m, and the frequency $\om$ is chosen according to Eq. (\ref{3isoen}).}
\end{center}
\end{figure}

For the HO trap, several studies have been carried out describing
the scaling law governing the expansion of the TG gas \cite{OS02,PSOS03} as well as the dynamical
fermionization of the system exhibited in the momentum distribution \cite{RM05,MG05}.
Here, we shall focus on the transition to the ballistic regime in the expansion from both
HO and HW traps. Moreover, we shall be interested in low number of particles $N$,
where the mean-field approach is not accurate, missing the spatial anti-bunching in the density profile \cite{KNSQ00,GW00b}.
If we are to compare the expansion from both traps, it seems natural to consider that, for the same number of particles, the total energy of the TG gas is to be the same.
Then, the following relation must hold between the harmonic frequency and the length of the box,
\beq
\label{3isoen}
\om=\frac{\hbar \pi^{2}(N+1)(2N+1)}{6mL^{2}N}.
\eeq
For the HO trap, the similarity transformation
\beq
\rho(x,t)=\frac{1}{\sqrt{1+\om^{2}t^{2}}}
\rho\left(\frac{x}{\sqrt{1+\om^{2}t^{2}}},0\right)
\eeq
has been shown to hold, where the ballistic regime sets for $t\gg\om^{-1}$ \cite{OS02}.
However, as we have already discussed, no such simple expression can be obtained for the hard-wall trap,
exhibiting the dynamics a lack of self-similarity (which should be clear once the transient features entailed in
Eq. (\ref{3phit}) are recognized), see Fig. \ref{3figdp}. The $\pm p_{N}=\pm\hbar N\pi/L$ components govern in this case the width of the expanding cloud for $t\gtrsim t_{N}=mL^2/2N\pi\hbar$.

\begin{figure}
\begin{center}
\includegraphics[height=7cm,angle=-90]{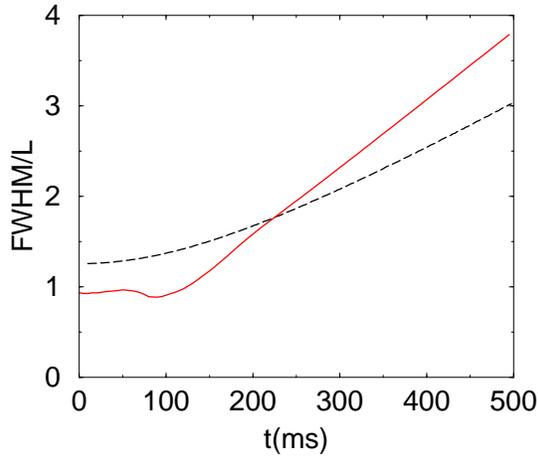}
\caption{\label{3tonksw000} Time dependence of the FWHM of a cloud of $N=10$ ${}^{87}$Rb atoms in the Tonks-Girardeau regime,
released from a box-like ($L=80$ $\mu$m, solid line) and harmonic (dashed line) traps. Note that the transition to the ballistic regime is sudden when the gas is released from a box-like trap, but smooth if initially confined in a harmonic trap (parameters as in Fig. \ref{3figdp}).}
\end{center}
\end{figure}

%
Figure \ref{3tonksw000} shows the variation in time of the full width at
half maximum (FWHM) for a TG gas expanding from both types of traps.
The transition to the ballistic regime is sharp for the HW trap at $t_{N}$,
whereas for the HO happens gradually, and only sets for $t\gg\om^{-1}$.
This fact, together with transient versus self-similar dynamics,
points out the relevance of the confining geometry.

The density profile we have described so far in the TG regime, is actually the same that for spin-polarized non-interacting fermions.
We note that for the latter the momentum distribution
\beq
\label{nk}
n(k)=\frac{1}{2\pi}\int dxdye^{ik(x-y)}\rho(x,y)
\eeq
remains invariant under time evolution provided
that the momentum operator commutes with the purely kinetic Hamiltonian.
Since the reduced single-particle density matrix (RSPDM) of spin-polarized
fermions is
\beqa
\rho(x,y)&=&N\int dx_2\dots dx_N \psi_{F}(x,x_{2},\dots,x_{N})^*\psi_{F}(y,x_{2},\dots,x_{N}),\nonumber\\
&=&\sum_{n=1}^N\phi_{n}^*(x)\phi_n(y),
\eeqa
the momentum distribution is the sum
\beq
n_F(k)=\sum_{n=1}^N|\widetilde{\phi}_n(k)|^2,
\eeq
$\widetilde{\phi}_n$ denoting the Fourier transform of $\phi_n$.
\footnote{Note the errata  of the published version (Physics Reports 476, 1-50, 2009),  where in Eq. (60) $\psi_n^*$ should read $\phi_n^*$, and in Eqs. (62) and (64)  $\psi_l^*$ should read $\phi_l^*$, as corrected here.}
However, the momentum distribution of an expanding Tonks-Girardeau gas exhibits a transient dynamics.
An efficient way to compute the RSPDM has been recently derived by Pezer and Buljan \cite{PB07} explicitly using the Bose-Fermi map
and the Laplace expansion of the determinant, see also \cite{BLPJ07}.
Thanks to the orthonormality of the single-particle eigenstates, the RSPDM is given by
\beqa
\label{RSPDMTG}
\rho_{TG}(x,y)=\sum_{l,n=1}^N\phi_l^*(x){\rm A}_{ln}(x,y)\phi_n(y),
\eeqa
where
\beqa
{\rm \bf A}(x,y)=({\bf P}^{-1})^T {\rm det}{\bf P},
\eeqa
and the elements of the matrix ${\bf P}$ are
\beqa
P_{ln}=\delta_{ln}-2\int_{x}^{y}\d z \phi_{l}^*(z)\phi_n(z)
\eeqa
for $x<y$ with no loss of generality. A generalization of this result for hard-core anyons can be found in \cite{delcampo08}.

\begin{figure}
\begin{center}
\includegraphics[height=5cm,angle=0]{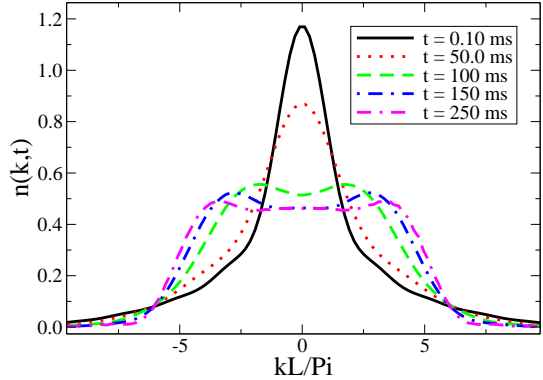}
\caption{\label{dynfer} Dynamical fermionization of  $N=5$ ${}^{87}$Rb atoms in the Tonks-Girardeau regime,
released from a box ($L=80$ $\mu$m). The asymptotic momentum distribution is that of a spin-polarized ideal Fermi gas.}
\end{center}
\end{figure}

From (\ref{nk}) and (\ref{RSPDMTG}), the time-evolving $n(k,t)$ can then be found.
For $t=0$  it presents a prominent peak at $k=0$, whereas as the cloud expands and the particles cease to interact, the interaction energy is gradually transformed into kinetic energy, and the asymptotic momentum distribution is the quasi-momentum distribution, namely, the flat one typical of the dual Fermi system. This is the so-called dynamical fermionization \cite{RM05,MG05},
illustrated in Fig. \ref{dynfer} for a TG gas expanding from a box-like trap.

In conclusion, strongly repulsive interactions tend to
suppress quantum transients in the density profile,
while enriching the dynamical picture in the momentum space.
Conversely, attractive interactions have been shown to lead to an
enhancement of density modulations in both expanding clouds and propagating beams \cite{KGAK03}.

\subsection{Finite interactions: dynamics of Lieb-Liniger gases}
\label{sec:LL}
Whenever the interactions are finite, the exact dynamics of ultracold gases becomes a highly non-trivial subject.
The first steps to cope with it made use of a hydrodynamical approach \cite{DLO01,OS02,PSOS03} and numerical simulations, but only recently have the required analytical techniques been developed \cite{Girardeau03,BPG08,JPGB08}. Within the Lieb-Liniger model the wavefunction of the gas  $\psi_B(x_1,\ldots,x_N,t)$ describes $N$ identical $\delta$-interacting bosons in one-dimension governed by the Hamiltonian (\ref{LiLi}) \cite{LL63}. It is customary to express the corresponding time-dependent and many-body Schr\"odinger equation (TDMBSE) with ``Lieb-Liniger'' units ($2m=1, \hbar=1$) in the form
\begin{equation}
i \frac{\partial \psi_B}{\partial t}=
\sum_{i=1}^{N}\bigg[-\frac{\partial^2 }{\partial x_i^2}+V(x_i,t)\bigg]\psi_B+
\sum_{1\leq i < j \leq N} 2c\,\delta(x_i-x_j)\psi_B.
\label{LLmodel}
\end{equation}
%
We assume that the gas is initially confined in an external potential $V(x)$, which is suddenly turned off at $t=0$,
so that the cloud starts expanding. Therefore for $t>0$ the domain of each coordinate is the whole real line $x_j \in\mathbb{R}$.

The wavefunction $\psi_B$ is totally symmetric under permutations of particles,
and hence it suffices to focus on the fundamental sector $R_1:x_1<x_2<\ldots<x_N$, where the TDMBSE reduces to
\begin{equation}
i \frac{\partial \psi_B}{\partial t}=
-\sum_{i=1}^{N}\frac{\partial^2 \psi_B}{\partial x_i^2}.
\label{free}
\end{equation}
The Lieb-Linger contact conditions are responsible for creating a cusp in
the wave function when two particles touch. This can be expressed as a boundary
condition at the borders of $R_1$ \cite{LL63},
\begin{equation}
\left [
1-\frac{1}{c}
\left (
\frac{\partial}{\partial x_{j+1}}-\frac{\partial}{\partial x_j}
\right)
\right]_{x_{j+1}=x_j}\psi_B=0.
\label{interactions}
\end{equation}
These boundary conditions can easily be rewritten for any
permutation sector. In the TG limit ($c\rightarrow \infty$), the cusp
condition implies an effective exclusion principle whereby the wavefunction identically vanishes when
two particles are in contact \cite{Girardeau60}.
Exact solutions of the time-dependent Schr\"odinger equation (\ref{LLmodel})
can be obtained by using a Fermi-Bose mapping operator \cite{Gaudin83,BPG08,JPGB08} acting on
fermionic wave functions. Assume $\psi_F(x_1,\ldots,x_N,t)$ is an antisymmetric (fermionic)
wave function, which obeys the Schr\" odinger equation
for a non-interacting Fermi gas, $i \frac{\partial \psi_F}{\partial t}=
-\sum_{i=1}^{N}\frac{\partial^2 \psi_F}{\partial x_i^2}$. Then, the wavefunction of the Lieb-Liniger
gas can be written as
\begin{equation}
\psi_{B}= {\mathcal N}_{c} \hat O_c \psi_F,
\label{ansatz}
\end{equation}
using the generalized Fermi-Bose mapping operator
\begin{equation}
\hat O_c=\prod_{1\leq i < j \leq N}
\left[
\mbox{sgn}(x_j-x_i)+\frac{1}{c}
\left(
\frac{\partial}{\partial x_{j}}-
\frac{\partial}{\partial x_{i}}
\right)
\right],
\label{oO}
\end{equation}
up to the normalization constant ${\mathcal N}_{c}$ \cite{Gaudin83}.
This surprising result states that the time evolution of a Lieb-Liniger gas at a given time $t$ can be computed
from that of non-interacting fermions at the same time, and only then take into account the interactions through the generalized Fermi-Bose map. The applicability of this map is not limited to the expansion dynamics of an excited Bose gas initially confined in an external potential $V(x)$, for it actually describes the ground state whenever $[\hat O_c,V(x)]\approx0$. Nonetheless, as the gas expands in 1D, the linear density decreases while the effective coupling constant becomes larger, prompting the system to enter the strongly interacting regime. Indeed, the asymptotic
form of the wavefunction can be probed to possess the Tonks-Girardeau
structure \cite{JPGB08}. However, the properties of this asymptotic state can
considerably differ from the properties of the Tonks-Girardeau gas in the
ground state of an external potential \cite{JPGB08}.
We note that such dynamics is restricted to free expansion, while the inclusion of specific boundary conditions
(periodic, Dirichlet, Neumann) and its extension to other quantum statistics (interacting fermionic or anyonic systems) and multi-component, multi-channel problems remain as interesting open problems.
%
%
%
%
%
\subsection{Finite interactions: mean-field approach}
\label{sec:TDGPE}
By now, it should be clear that suddenly released matter-waves initially confined in a compact support exhibit quantum transients. Nonetheless, the effect of strongly repulsive interactions tends to smooth out the density profile in the resulting dynamics, as in the Tonks-Girardeau gas.
In this section we discuss the effect of the interactions on quantum transients within the mean-field approach.

Weakly interacting ultracold gases in 1D are then described by the Gross-Pitaevskii equation, as a result of approximating the wavefunction of the Bose-Einstein condensate by the Hartree-Fock ansatz $\Psi(x_1,\dots,x_N)=\prod_{i=1}^N\phi(x_i)$.
For an effectively 1D Bose-Einstein condensate, in the presence of some external potential $V(x)$ (i.e. the trap)
the condensate wavefunction $\Phi(x)=\sqrt{N}\phi(x)$ obeys the 1D time-dependent Gross-Pitaevskii equation
\beqa
i\hbar\frac{\Phi(x,t)}{\partial t}
=- \frac{\hbar^2}{2m}\frac{\partial^2\Phi(x)}{\partial x^2}  + \bigg[V(x)+g_{1D}|\Phi (x)|^2\bigg] \Phi(x),
\eeqa
where $m$ is the atomic mass, and $g_{1D}=-2\hbar^2/ma_{1D}$ the effective 1D coupling parameter,
$a_{1D}$ being a known function of the three-dimensional scattering length \cite{Olshanii98}.
In the mean-field regime, for low enough temperatures, the phase fluctuations can be suppressed \cite{GanShl03}.
Making use of time-dependent shutters a way to enhance the diffraction in time was shown in \cite{DMK08}. An alternative
way is to make the interactions in the system attractive, (say, by means of a Feschbach resonance \cite{BDZ08}) as discussed in \cite{KGAK03}. Quantum transients arising in suddenly released matter-waves can then be enhanced in systems governed by attractive interactions and suppressed in the repulsive case.

Indeed, it is worthy to consider the case in which the mean-field interaction dominates over the kinetic energy (but at boundaries). In such a case, the so-called Thomas-Fermi approximation assumes that the kinetic energy can be neglected
in the Hamiltonian so that the time-independent Gross-Pitaevskii equation reads
$\mu\Phi_{TF}(x)=\big[V(x)+g_{1D}|\Phi_{TF}(x)|^2\big]\Phi_{TF}(x)$, where
$\mu$ is the chemical potential.
The Thomas-Fermi wavefunction is then given by
$\Phi_{TF}(x)=[(\mu-V(x))/g_{1D}]^{1/2}$ whenever $\mu>V(x)$ and zero elsewhere.
The dynamics of the Thomas-Fermi density profile $n_{TF}(x)=|\Phi_{TF}(x,t)|^2$ after switching off the trap
obeys well-known scaling laws with no transient behavior whatsoever \cite{CD96,KSS96}.
For instance, for the case of a harmonic trap $V(x,t)=m\om(t)^2x^2/2$,
\beqa
\label{ntf}
n_{TF}(x,t)=\frac{1}{b(t)}n\bigg[\frac{x}{b(t)},t=0\bigg],
\eeqa
where the scaling coefficient satisfies the differential equation $\ddot{b}=\om^2/b^2$ subjected to the initial conditions $b(0)=1$, and $\dot{b}(0)=0$. Clearly, the dynamics entailed in Eq. (\ref{ntf}) lacks any DIT-related transient.

\subsection{Turning interactions on and off in a Bose-Einstein Condensate}

The world of cold atoms provides many interesting transients
due to the possibility to turn on and off interatomic
and/or external interactions.
Ruschhaupt {\it et al}. \cite{RCM06} have examined the short-time behavior
of a Bose-Einstein condensate when the interatomic interaction
is negligible for the preparation in the harmonic
trap, and strongly increased when the potential trapping
is removed. A quantum interference effect
in momentum space is then found in the Thomas-Fermi regime:
the momentum distribution expands
due to the release of mean field energy and the number of
peaks increases with time one by one because of the interference of
two positions in coordinate space contributing to the same
momentum. The effect is stable in a parameter range and
could be observed with current technology.
Interestingly enough, similar abrupt changes in time of the coupling constant $g_{1D}$ \cite{LMBKP07} and density perturbations \cite{Damski} can also induce self-modulations in the density profile and shock waves. The appearance of shock waves is a phenomenon also present in classical non-linear equations and is out of the scope of this review. We refer the reader to the good existing works on the topic \cite{shockwaves1,shockwaves2,shockwaves3}. More generally, the dynamics after a quantum quench in the Hamiltonian in a many-body system exhibits non-trivial quantum transients, i.e. see \cite{IC09} and reference therein. Nonetheless, one has to bear in mind that the nature of this type of transient is dramatically different of those related to the DIT, for the non-linear interactions entail a dispersion relation different from the free one. 
\subsection{State reconstruction}
The advances in the dynamical description of ultracold atoms in tight-waveguides have paved the way to the study of transients of Lieb-Liniger gases \cite{Girardeau03,BPG08,JPGB08}.
Nevertheless, the sudden quench of interactions using  Feschbach resonances \cite{BDZ08} combined with a switch of the confining potential stands for its applications in tomography of trapped ultracold gases \cite{DMM08}. The evolution of
the essentially free single-particle density profile, $n_0(x,t)$, allows us
to obtain the reduced density matrix at $t=0$,
\beqa
\label{rhokk}
\rho(k,k')=\frac{\hbar}{m}\int \tilde{n}_0(k'-k,t)|k'-k|e^{i\frac{\hbar(k'^2-k^2)t}{2m}} d t,
\eeqa
where the Fourier transform of the density profile reads
\beqa
\tilde{n}_0(k,t)=\frac{1}{2\pi}\int n_0(x,t)e^{-ikx}d x.
\eeqa
Alternatively, the RSPDM can be diagonalized \cite{DG03} as $\rho(x,x';t=0)=
\sum_j\lambda_j\varphi_j^{*}(x)\varphi_j(x')$ in terms of the
orthonormal natural orbitals $\varphi_j(x)$ with occupation numbers $\lambda_j>0$
satisfying $\int\rho(x,x')\varphi_j(x)dx=\lambda_j\varphi_j(x')$ and $\sum_{j=1}^N\lambda_j=N$.
Under free evolution, having set up $c(t>0)=0$, the density profile reads
\beqa
n_0(x,t)=\rho(x,x;t)=\sum_j\lambda_j|\varphi_j(x,t)|^2,
\eeqa
with $\varphi_j(x,t)=(2\pi)^{-1/2}\int dk
\widetilde{\varphi}_j(k)e^{ikx-i\hbar k^2 t/2m}$, which, using Eq. (\ref{rhokk}),
leads to the density matrix $\rho(k,k')=\sum_j\lambda_j\widetilde{\varphi}_j^{*}(k)\widetilde{\varphi}_j(k')$.
Experimental measurements are restricted to $t>0$, limiting the possibility of state reconstruction by means of Eq. (\ref{rhokk}), but assuming time-reversal invariance or using spatial symmetries of the system, the integral in time can be extended over the whole real line \cite{wigexp4}. More generally, once the interactions in the system are negligible, the reconstruction of the RSPDM of the initial quantum state from  the cloud dynamics in a potential $V(x)$ has been successfully addressed by Leonhardt and coworkers within the optical tomography \cite{LR96, LS97}. Remarkably, the Wigner function of non-interacting Helium atoms has been experimentally measured by looking at the time-evolution of the density profile \cite{wigexp4}, from which higher order correlations of the trapped gas can also be inferred \cite{ADM04}. 

\section{Transient effects with external potentials\label{epo}}
Transient effects have been considered so far mostly for motion in free space but new and interesting features arise in the presence of external potentials. Most work has been carried out within the sudden approximation for the shutter, namely, its instantaneous release. The quantum transients with short-range external potentials often admit an exact solution,
as is the case for  a $\delta$-barrier \cite{TKF87,EK88,AD02,DA02,hgc03,Moshinsky04,GM06,GM06b}, the  time-dependent $\delta$-barrier \cite{SK88,KM05}, and the Kroning-Penney lattice used to mimic the behavior of electrons in crystals \cite{MML96}. Similarly, it is possible to tackle multichannel problems such as the excitation dynamics of atomic wavepackets interacting with narrow laser beams \cite{DM06b}. Much work on these potential-dependent transients has been carried out to understand time dependent aspects of tunneling and will be reviewed in Section \ref{tune}.

\subsection{Extended quasi-monochromatic initial states}\label{quasim}

Here we review an approach, first considered by Garc\'\i a-Calder\'on and Rubio \cite{GR97},  that leads to an exact analytical solution to the time-dependent Schr\"odinger equation with cutoff wave initial conditions for arbitrary potentials of finite range. The approach involves the complex poles and residues (resonant states) of the outgoing Green's function of the problem and renders the time-dependent solution both along the internal and the transmitted regions of the potential. These authors solve the time-dependent Schr\"odinger equation
\begin{equation}
\left(i \hbar \frac{\partial}{\partial t}- H\right)\psi(x,t)=0,
\label{te1a}
\end{equation}
where  $H=-(\hbar^2/2 m)d^{\,2}/dx^2+V(x)$, and $V(x)$ describes a potential of arbitrary shape extending from $x=0$ to $x=L$, with
the plane wave initial condition
\begin{equation}
\psi(x,t=0)=e^{ikx}\Theta(-x),
\label{te1p}
\end{equation}
which is the same initial condition considered by Moshinsky for the free case, see Eq. (\ref{initial}).
Garc\'\i a-Calder\'on and Rubio \cite{GR97} used Laplace transform techniques
to obtain along the internal region of the potential the expression \cite{GR97}
\begin{eqnarray}
\psi(x,k,t) = &&\phi(x,k)M(0,k,t)- \nonumber \\[.3cm]
&&\sum_{n=-\infty}^{\infty} \phi_n(x) M(0,k_n,t); \,\,\,(0 \leq x \leq L),
\label{te1}
\end{eqnarray}
where $\phi(x,k)$ refers to the stationary solution, and $\phi_n(x)=iu_n(0)u_n(x)/(k-k_n)$
is given in terms of the poles $\{k_n\}$ and resonant states $\{u_n(x)\}$ of the problem, discussed in Appendix \ref{ressta}.
Notice that along the internal region of the potential, the Moshinsky function does not depend on $x$ and hence
it does not correspond to a propagating solution. Similarly, along the transmitted region, the above authors provided the solution
\begin{eqnarray}
\psi(x,k,t) = &&T(k)M(x,k,t)-\sum_{n=-\infty}^{\infty} T_n M(x,k_n,t);\,\,\, (x \geq L),
\label{te2}
\end{eqnarray}
where $T(k)$ is the transmission amplitude and $T_n=iu_n(0)u_n(L)e^{-ik_nL}/(k-k_n)$.
Further notice that in the above two equations the function $M(x,k_n,t)$ corresponds to the Moshinsky function for the complex value $k=k_n$.

\begin{figure}[!tbp]
\begin{center}
\rotatebox{0}{\includegraphics[width=3.3in]{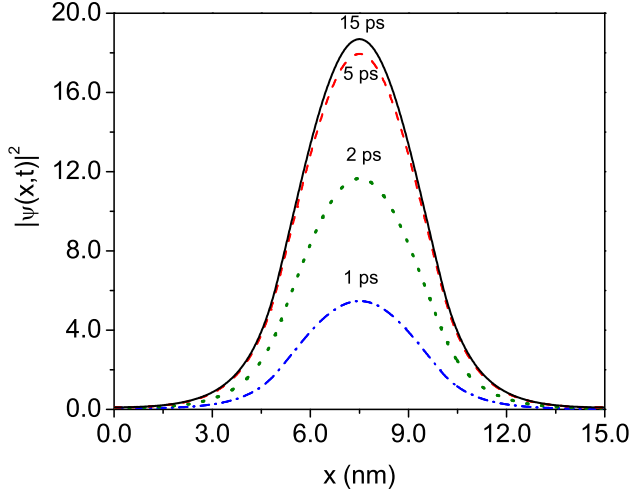}}
\caption{Plot of the building up of $|\psi(x,t)|^2$  along the internal region of a double-barrier potential for several times
with the parameters given in the text. At  $t=15$ ps, $|\psi(x,t)|^2$ becomes indistinguishable from the stationary solution.}
\label{fig1te}
\end{center}
\end{figure}

\begin{figure}[!tbp]
\begin{center}
\rotatebox{0}{\includegraphics[width=3.3in]{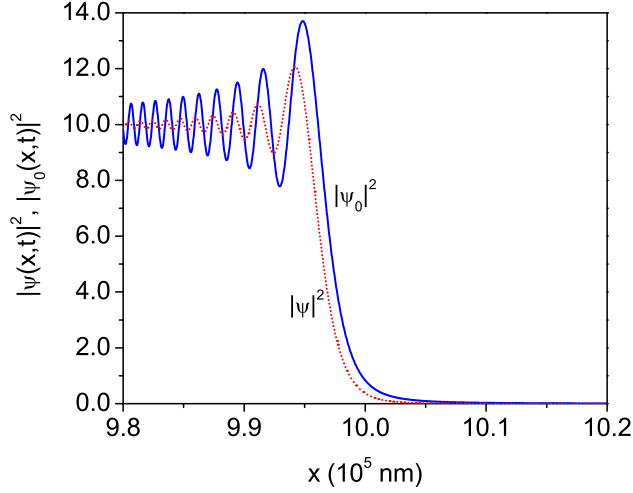}}
\caption{Comparison of the wavefronts of the free-wave evolving solution  $|\psi_0(x,t)|^2$ and the transmitted solution at resonance energy $|\psi(x,t)|^2$ through a double-barrier system  at a long time $t$= 1540 ps,  to show that the delay time is of the order of $2\hbar/\Gamma_1$, with $\Gamma_1$, the corresponding resonance width.}
\label{fig2te}
\end{center}
\end{figure}

In the absence of a potential, the solution given by Eq. (\ref{te1}) vanishes exactly and in
Eq. (\ref{te2}), $T(k)=1$ and the resonant sum vanishes as well, so that the solution becomes the free solution.
In \cite{GR97}, it is shown that the exact solutions given by Eqs.\ (\ref{te1}) and (\ref{te2}) satisfy the corresponding
initial conditions, i.e., they vanish exactly for $t= 0$. Similarly, it is also shown that at asymptotically long times
the terms $M(x,k_n,t)$ that appear in the above equations, tend to a vanishing value,
while $M(x,k,t)$ tends to the stationary solution, as first shown by Moshinsky \cite{Moshinsky52}.  Along the internal region
\begin{equation}
\psi(x,t)\sim\phi(x,k){\rm exp}(-iEt/\hbar),
\label{te5}
\end{equation}
and along the external (transmitted) region,
\begin{equation}
\psi(x,t)\sim T(k){\rm exp}(ikx){\rm exp}(-iEt/\hbar).
\label{te6}
\end{equation}
The above solutions allow us to study the dynamics along the full time span from zero to infinity.
In \cite{GR97} these solutions were used to examine the transient behavior of a double barrier resonant structure possessing an isolated  resonance level and  an initial plane wave with energy equal to the resonance energy of the level. The calculations showed that for times roughly of a fraction of a lifetime onwards the single-resonance approximation to the time-dependent solutions given above was in good quantitative agreement. Figure \ref{fig1te} exhibits the building up of the probability density $|\psi(x,t)|^2$ along the internal region of the double-barrier system (DB) for several times until it reaches the stationary solution. The parameters of the  DB system, are typical of semiconductor resonant structures \cite{ferry}:  barrier heights $V=0.23$ eV, barrier widths $b=5.0$ nm,  well width $w=5.0$ nm and effective electron mass $m=0.067m_e$, with $m_e$ the electron mass. The isolated resonance level is characterized by the resonance parameters ${\mathcal E}_1=0.08$ eV and $\Gamma_1=1.0278$ meV. Figure \ref{fig2te}  provides  a comparison of the corresponding transmitted probability density with the free evolving solution at long times, which exhibits a delay time of the order of the theoretical estimate $\tau= 2\hbar/\Gamma_1$.

The fact that the plane wave initial state jumps from unity to zero at $x=0$  implies
that the resonant sums in Eqs. (\ref{te1}) and (\ref{te2}) have a slow convergence.
%
 For this reason, most subsequent work has considered the initial state \cite{rvprb99,GV01,rprb02,gcv02,RVG02,gcv03,GCVY03,vgc04,YGV05}
\begin{equation}
\psi(x,t=0)= (e^{ikx}-e^{-ikx})\Theta(-x),
\label{te6p}
\end{equation}
which continuously vanishes at $x=0$. It follows then, that the time-dependent solutions along the internal and transmitted regions of the potential become, respectively,
\begin{eqnarray}
\psi(x,t) &=& \phi(x,k)M(0,k,t)-\phi(x,-k)M(0,-k,t)
\nonumber\\
&-&  \sum_{n=-\infty}^{\infty} \phi_n(x) M(0,k_n,t); \,\,\,\,\,(0 \leq x \leq L),
\label{te7}
\end{eqnarray}
and
\begin{eqnarray}
\psi(x,t) &=& T(k)M(x,k,t)-T(-k)M(x,-k,t)
\nonumber \\[.3cm]
&-& \sum_{n=-\infty}^{\infty} T_n M(x,k_n,t);\,\,\,\,\, (x \geq L),
\label{te8}
\end{eqnarray}
where the terms $\phi_n(x)$ and $T_n$ are now given by  $\phi_n(x)=2iku_n(0)u_n(x)/(k^2-k_n^2)$ and $T_n=2iku_n(0)u_n(L)e^{-ik_nL}/(k^2-k_n^2)$. It may be shown that at very long times $M(x,-k,t)$ goes to zero \cite{GR97} and hence the above solutions tend also to the stationary values given by Eqs. (\ref{te5}) and (\ref{te6}).

In \cite{GCVY03} an alternative procedure was derived to obtain the time-dependent transmitted solution. This procedure starts from the expression for the time-evolved wave function along the transmitted region
\begin{equation}
\psi(x,t)=\int_{-\infty}^\infty \frac{dk'}{\sqrt{2\pi}}\, \phi(k')T(k')\,
e^{ik'x-i\hbar k'^2 t/2m},
\label{te9}
\end{equation}
where $\phi(k')$ is the $k'$-space wave function (i.e., the Fourier transform of the initial wave
function) defined by
\begin{equation}
\phi(k')=\int_{-\infty}^\infty \frac{dx}{\sqrt{2\pi}} e^{-ik'x} \psi(x,0).
\label{te10}
\end{equation}
$T(k)$ for $k<0$ is the analytical continuation in that domain
of the transmission amplitude for left incidence, $T(k>0)$, not to be confused with the transmission
amplitude for right incidence \cite{BM96}.

For the initial state  given by Eq. (\ref{te6p}), it follows from Eq. (\ref{te10}) that
\begin{eqnarray}
\phi(k')&=&\int_{-\infty}^\infty \frac{dx}{\sqrt{2\pi}} e^{-ik'x} \psi(x,0)\nonumber\\
&=&\frac{i}{\sqrt{2\pi}}\left(\frac{1}{k'-k+i\epsilon}-\frac{1}{k'+k+i\epsilon}\right),
\label{te11}
\end{eqnarray}
where $\epsilon$ is an infinitesimal positive number. The other ingredient to evaluate Eq. (\ref{te9}) is to make use of the
Cauchy expansion for the transmission amplitude derived in Appendix \ref{ressta}, \textit{i.e.}, Eq. (\ref{e21}), written as
\begin{equation}
T(k')=\sum_{n=-\infty}^{\infty} \left(\frac{r_n}{k'-k_n}+\frac{r_n}{k_n}\right),
\label{te12}
\end{equation}
where $r_n=iu_n(0)u_n(L)\exp(-ik_nL)$.
Substitution of Eqs. (\ref{te11}) and (\ref{te12}) into Eq. (\ref{te9}) leads, after some algebraic manipulation, to exactly the same expression given by Eq. (\ref{te8}). This procedure allows one to obtain more easily the time-dependent solution than the Laplace transform approach since it facilitates the consideration of other initial states. We shall refer to this approach as the Fourier transform approach. Its extension to deal with the internal region is straightforward \cite{CG08}.
\begin{figure}[!tbp]
\begin{center}
\includegraphics[width=2.6in,height=1.6in]{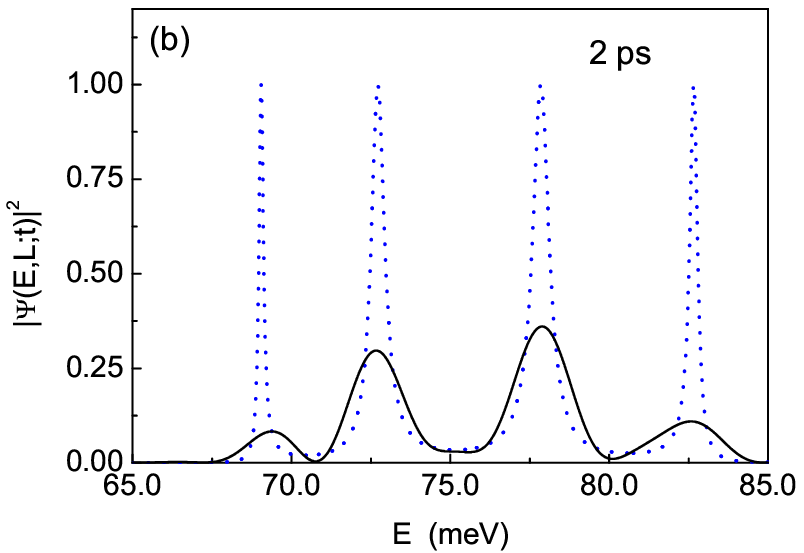}
\includegraphics[width=2.6in,height=1.6in]{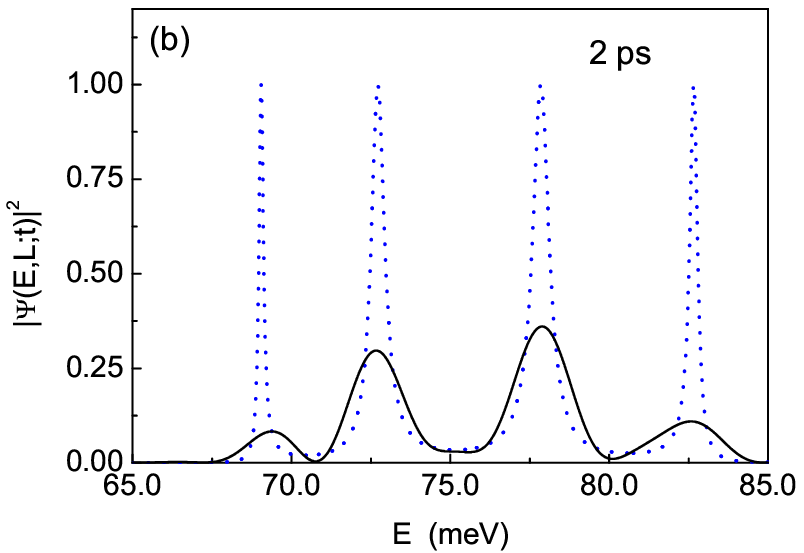}
\includegraphics[width=2.6in,height=1.6in]{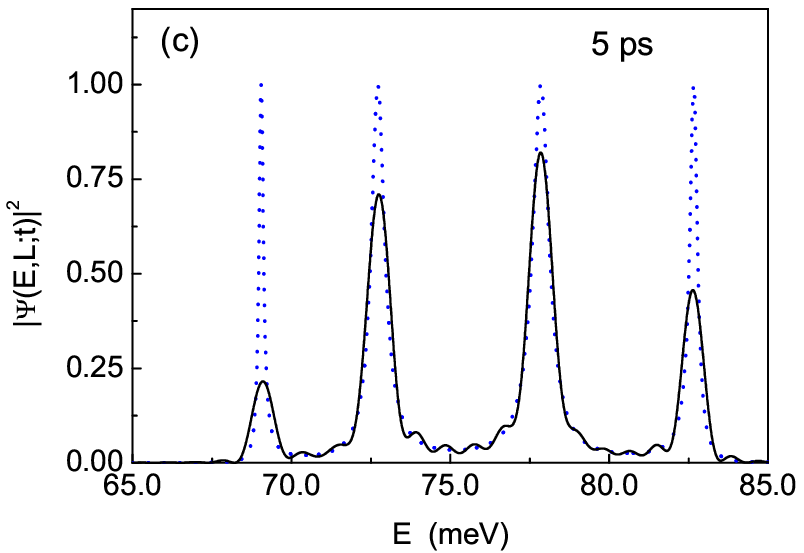}
\includegraphics[width=2.6in,height=1.6in]{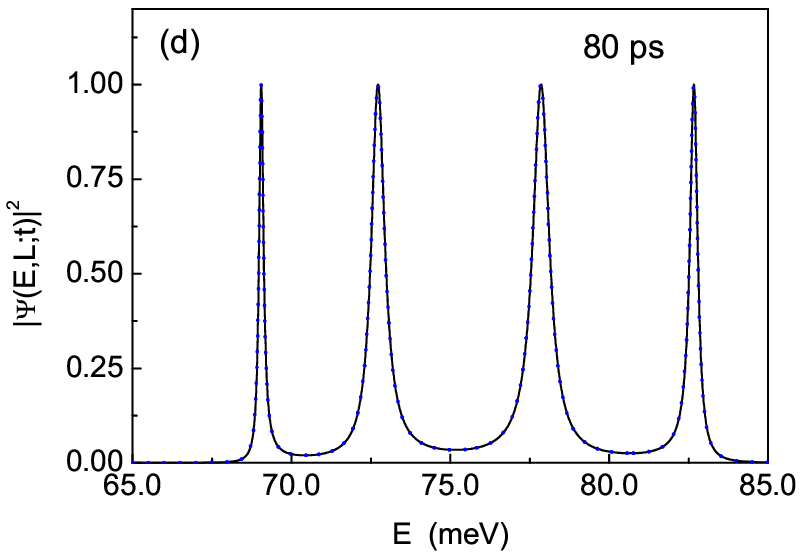}
\caption{Snapshots of the time-dependent probability density for a periodic superlattice of five barriers, with parameters given in the text,
reproducing Fig. 4 of Ref. \cite{rprb02}. The values of $\left| \Psi(E,L;t) \right| ^2$ \textit{vs} $E$ are calculated Eq. (7) of that paper at different fixed times (solid lines): (a) t=0.6 ps; (b) t=2.0 ps; (c) t=5.0 ps; (d) t=80.0 ps. The transmission coefficient is also included for comparison (dotted lines).}
\label{buildup}
\end{center}
\end{figure}

There have been two main lines of research where the above exact time-dependent solutions
for the wave function have been used in the investigation of transient phenomena. One line has addressed detailed studies on the dynamics of resonant tunneling on and off resonance energy in double \cite{GR97,rvprb99,vrap00,rvap01}, triple \cite{RVG02} and multiple \cite{rprb02,vrprb03} resonant tunneling structures, whereas the other line of research, considered in the next section, has addressed the issue of transient effects in relation to the tunneling time problem \cite{GV01,gcv02,gcvdm02,gcv03,GCVY03,hgc03,vgc04,YGV05}.

\subsubsection{Buildup dynamics of the  transmission resonance spectra}

Most studies involving the quantum shutter setup consider a given value  for the energy of the initial state, and evaluate the time-dependent solution for the probability density $|\psi(x,t)|^2$ either as a function of time for a fixed value of the distance or as a function of the distance for a fixed value of the time. In Ref. \cite{rprb02}, however, Romo addresses the issue of the building up of the transmission resonances in a superlattice of finite range by considering the time-dependent solution for fixed values of both the distance and the time and varying instead the energy of the initial state in a relevant energy range.  He considered the initial state given by Eq. (\ref{te6p}) which leads to the time-dependent solution given by Eq. (\ref{te7}) to study this transient behavior for different superlattice potential profiles.
In a finite superlattice involving $N+1$ barriers the resonances group themselves in minibands involving $N$ resonance levels isolated from each other. Romo studied the building up dynamics for a given miniband and found that the solution for $|\psi(x,t)|^2$ using Eq. (\ref{te7}) may
be written as a simple analytic expression which involves only the $N$ resonance levels of the miniband.
Figure \ref{buildup} exemplifies his findings. It consists of a superlattice formed by
five rectangular barriers with heights $V_{0}=0.2$ eV and widths $b_{0}=5.0$ nm and well widths $w_{0}=5.0$ nm, and it shows a series of plots of $\left| \Psi ^{N}(E,L;t)\right| ^{2}$ as a function of the energy $E$ at different fixed times.
In the snapshots depicted in Figures \ref{buildup}(a) to \ref{buildup}(d) one
appreciates the ``birth'' of the transmission resonances and their subsequent
evolution towards the stationary regime. At the very beginning of the tunneling process, one appreciates a smooth curve
with no peaks, see Fig. \ref{buildup}(a); that is, no evidence of the resonances have appeared at
this early stage. However, as time elapses some peaks in the curve gradually begin to appear as can be appreciated in Figs. \ref{buildup}(b) to \ref{buildup}(d). The height of the resonance peaks increases at different rates towards their corresponding asymptotic values.
Notice that the buildup of the peaks is faster for the wider resonances ($n=2$ and $3$), and slower for the thinner ones ($n=1$ and $4)$.
It is also obtained that the buildup of the transmission peaks is governed by the analytic expression $T_n^{peak}[1-\exp(-t/t_b)]$, where $T_n^{peak}$ is the height of the corresponding transmission peak and $t_b=2\hbar/\Gamma_n$, namely twice the lifetime $\hbar/\Gamma_n$ of the
resonance level.

\subsection{Wavepacket initial states}\label{GWpacket}

The transmission of wavepackets through one-dimensional potentials is a model that has been of great relevance both from a pedagogical point of view, as discussed in many quantum mechanics textbooks, and in research, particularly since the advent of artificial semiconductor quantum structures \cite{ferry,mizuta}.
In \cite{YGV05},  Yamada, Garc\'\i a-Calder\'on and Villavicencio extended the quantum shutter initial condition to certain type of wave packets. They considered the following initial condition,
\begin{equation}
\psi(x,0)=A\int_{-\infty}^\infty dk
\left(\frac{e^{ikx}}{k-k_0+i\Delta}+\hbox{c.c.}\right),
\label{te7p}
\end{equation}
where $A=\sqrt{\Delta\left\{1+(\Delta/k_0)^2\right\}}/2\pi$ with $\Delta >0$, and $c.c.$
stands for complex conjugate. An important feature of this initial state is that it automatically
vanishes for $x > 0$. This is immediately seen from the fact that the integrand $e^{ikx}/(k-k_0+i\Delta)$,
which corresponds to a Lorentzian momentum distribution centered at $\hbar k_0$ with width
$\hbar\Delta$, has a simple pole only in the lower-half of the complex $k$-plane. An explicit
expression for $\psi(x,0)$ can be easily obtained by the method of residues. It is found that
\begin{equation}
\psi (x;t=0)=\left\{
\begin{array}{cc}
4\pi A\,e^{\Delta x}\sin k_0 x, & x<0 \\ [.3cm]
0,  &  x \geq 0,
\end{array}
\right..
\label{te7pp}
\end{equation}
A measure of the packet width is $1/\Delta$. It can be easily proved that the wave packet is normalized, i.e., $\int dx |\psi(x,0)|^2=1$.
The Fourier transform of the above expression is
\begin{equation}
\phi(k)=\sqrt{2\pi}A\left(\frac{1}{k-k_0+i\Delta}-\frac{1}{k+k_0+i\Delta}\right),
\label{te15}
\end{equation}
and then, following the same procedure of the previous example, these authors arrive at the solution
\begin{eqnarray}
\psi(x,t)&=&-i\sqrt{\Delta\{1+(\Delta/k_0)^2\}}
\Bigg[T(k_0-i\Delta)M(x,k_0-i\Delta;t)\nonumber\\
&-&T(-k_0-i\Delta)M(x,-k_0-i\Delta;t)\nonumber\\
&-&2k_0\sum_{n=-\infty}^{\infty}\frac{r_n}{k_0^2-(k_n+i\Delta)^2}M(x,k_n;t)\Bigg],
\label{te16}
\end{eqnarray}
where, we recall that $r_n=iu_n(0)u_n(L)\exp(-ik_nL)$.

Most time-dependent numerical studies consider  Gaussian wavepackets as initial states \cite{mizuta,konsek,harada,hauge91}, though  in some recent work, the formation of a quasistationary state in the scattering of wavepackets on finite one-dimensional periodic structures also involves also some analytical considerations \cite{peisakhovich}.
As discussed above, analytical approaches have been mainly concerned with  quasi-monochromatic initial states in a quantum shutter setup. In some recent work, however, analytical solutions to the time-dependent wave function have been discussed using initial  Gaussian wavepackets for square barriers \cite{PBM03}, delta potentials \cite{AD04} and resonant tunneling systems near a single resonance \cite{wulf}.

In Ref. \cite{vrcpra07}, Villavicencio, Romo and Cruz  derive an analytical solution to the time-dependent wavefunction for an initial cutoff Gaussian wavepacket, for the free evolving case and two barrier potentials: the $\delta$-potential and a very thin and very high  rectangular barrier. They consider the Fourier transform approach and the analytical solution follows provided the center of the Gaussian wavepacket is far from the interaction region, as is usually assumed on physical grounds. In such a case, the tail of the Gaussian is small near the interaction region, a regime they refer to as the \textit{small truncation regime}.

The cutoff Gaussian wave packet is given by the expression
\begin{equation}
\psi_0(x) = \left\{ \begin{array}{l l} A_0e^{-(x-x_0)^2/4\sigma^2}
e^{ik_0x}, & x<0 \\ 0, & x>0
\end{array} \right.,
\label{te16a}
\end{equation}
where $A_0$ is the normalization constant, and $x_0$ and $\sigma$ stand, respectively, for the center and
effective width of the wavepacket.  The interesting point of this approach is that substitution of Eq. (\ref{te16a}) into Eq. (\ref{te10})
gives an exact expression for the corresponding Fourier transform,
\begin{equation}
\phi_0(k) = A_0 w(iz),
\label{te17}
\end{equation}
where $A_0= (1/\sqrt{2\pi})[(2\pi\sigma^2)^{1/4}/\sqrt{\omega(iz_0)}]$, $z = x_0/(2\sigma) - i (k-k_0)\sigma$,
$z_0 = x_0/(\sqrt{2}\,\sigma)$,  and $w(y)$ is the $w$-function \cite{FT61} (see Appendix \ref{app_wfunction}).  Along  $x_0 < 0$ and $z \gg 1$ using the relationship
$w(iz) = 2e^{z^2}-1/(\pi^{1/2}z)-1/(2\pi^{1/2}z^3)+...$, allows to write $w(iz) \simeq 2 \exp(z^2)$.
Substituting this last expression into Eq. (\ref{te17}) and the resulting expression into Eq. (\ref{te9}) yields
\begin{equation}
\psi(x,t)=2A_0\int_{-\infty}^\infty \frac{dk'}{\sqrt{2\pi}}\, e^{z^2}T(k')\,e^{ik'x-i\hbar k'^2 t/2m}.
\label{te17a}
\end{equation}
For $T=1$, the above expression leads to an analytical expression for the free evolving Gaussian wavepacket \cite{vrcpra07},
\begin{eqnarray}
\psi^f_a(x,t)&=&\frac{1}{(2\pi)^{1/4}}\frac{1}{\sigma^{1/2}} \frac{e^{i(k_0x-\hbar k^2t/2m)}}{\sqrt{1+it/\tau}}
\times \nonumber \\ [.3truecm]
&&\exp \left \{- \frac{[x-x_0-(\hbar k_0/m) t]^2}{4\sigma^2 \left [ 1 +it/\tau \right ]} \right\},
\label{free0}
\end{eqnarray}
where $\tau=2m\sigma^2/\hbar$, that is identical to the analytical solution for the extended free Gaussian wavepacket,
\textit{i.e.}, no cutoff Gaussian wavepacket. The above authors considered the quantum shutter setup with a cutoff Gaussian wavepacket initial state for a delta-barrier potential of intensity $\lambda$, \textit{i.e.}, $V(x)=\lambda \delta(x)$. The corresponding transmission amplitude reads
\begin{equation}
T(k)=\frac{k}{k+i(m\lambda/\hbar^2)}.
\label{te10a}
\end{equation}
Notice that the transmission amplitude given above has just one pole, an antibound pole at $k_a=-i(m\lambda/\hbar^2)$.
Substitution of Eq. (\ref{te10}) into Eq. (\ref{te9}) leads to an expression that acquires a simple analytical form in the
\textit{small truncation regime}, namely,
\begin{equation}
\psi(x,t)=\psi^f_a(x,t) - Ce^{ik_0x-i\hbar k_0^2t/2m}M(x',q',t),
\label{te11a}
\end{equation}
where $\psi^f_a(x,t)$ stands for the free solution, $C=(\pi/2)^{1/4}(2m\sigma^{1/2}\lambda)/\hbar^2$ and
$M(x',q',t)$ stands for a Moshinsky function where $x'=x-x_0-v_0t$, $q'=-(k_0+im\lambda/\hbar^2)$, and $t'=t-i\tau$.
The solution given by Eq. (\ref{te11a}) holds also for a rectangular barrier of height $v_0$ and width $L$ in the limit of small opacity
$[2mV_0]^{1/2}/\hbar \ll 1$, by letting $\lambda=V_0L$ and performing the translation  $x \rightarrow (x-L)$.

\begin{figure}[!tbp]
\begin{center}
\includegraphics[width = 2.6in]{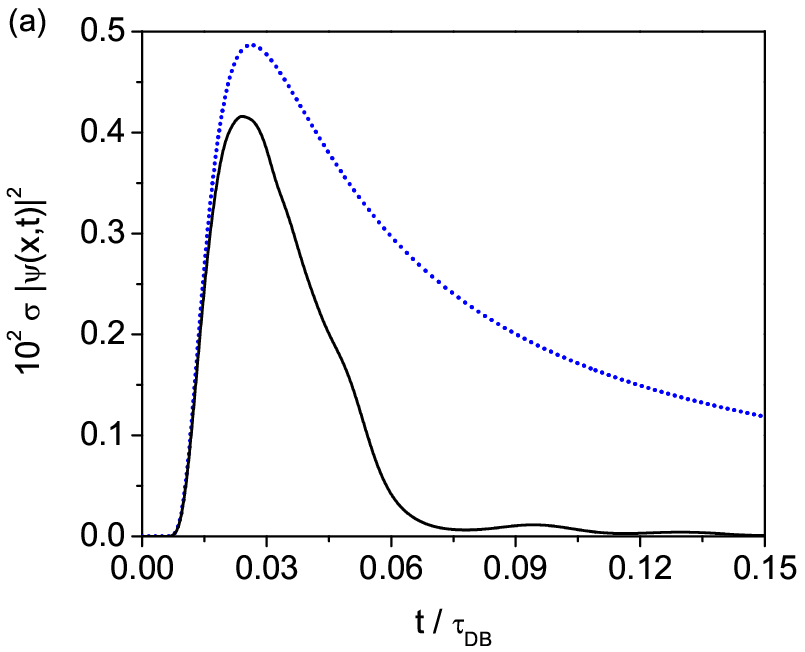}
\includegraphics[width = 2.6in]{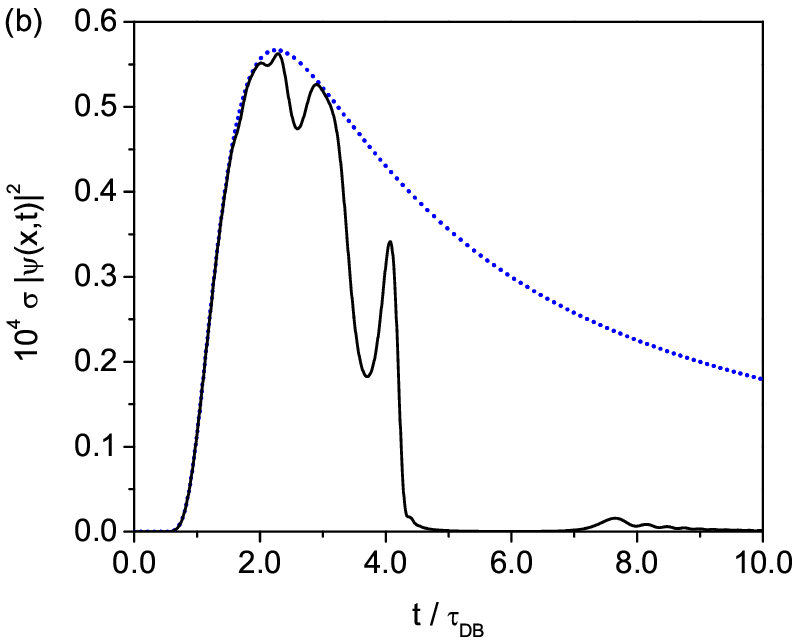}
\includegraphics[width = 2.6in]{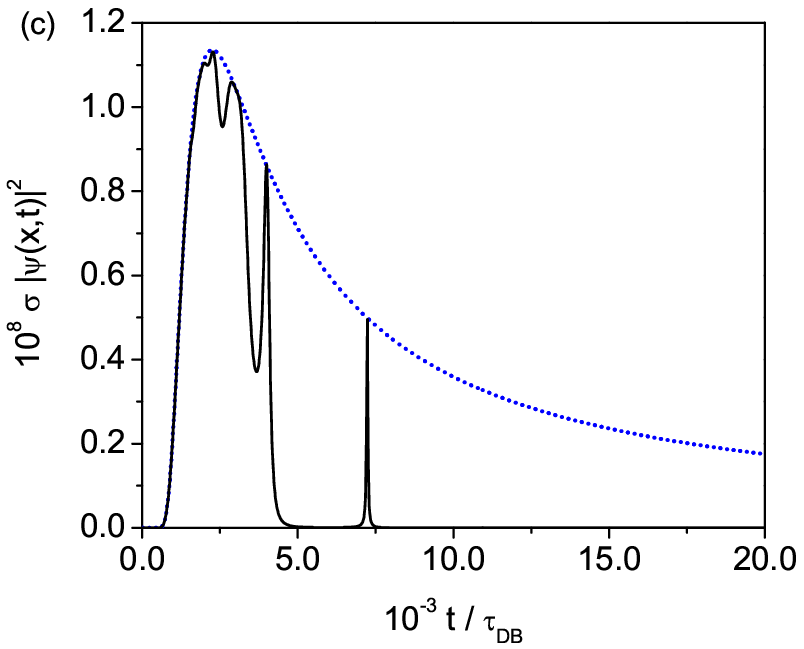}
\includegraphics[width = 2.6in]{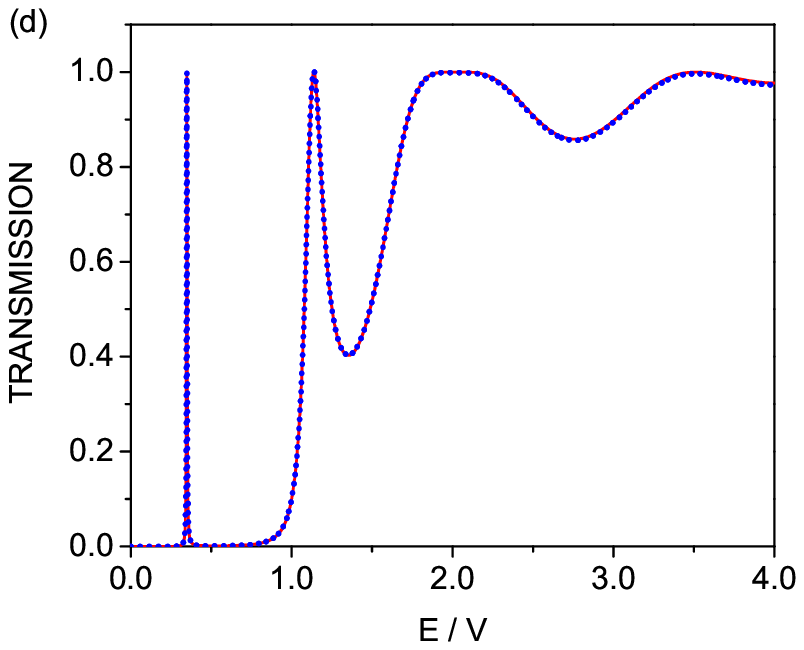}
\caption{ \footnotesize Figures (a), (b) and (c) display $\sigma |\psi(x,t)|^2$ as function of time in units of the lifetime $\tau_{DB}=\hbar/\Gamma_1$ (full lines) at different distances $x_d$ from the DB system, calculated using Eq. (\ref{18}): (a) $2L$, (b) $200L$ and (c) $2\times10^5L$, where $L$ is the length of the system. Also shown in these figures is the free evolving cutoff Gaussian wavepacket  (dotted line). Figure (d) exhibits the transmission spectra of the DB system as a function of energy, in units of the barrier height, using the exact numerical calculation (full line) and that obtained using the resonance expansion given by Eq. (\ref{e21}) (dot line). See text.}
\label{tef4}
\end{center}
\end{figure}

Recently, Cordero and Garc\'\i a-Calder\'on \cite{cgc08} have obtained a formal solution of the transmitted time-dependent function
for an arbitrary potential profile as an expansion in terms of resonance states and poles of the problem that
follows by substitution of Eq. (\ref{e21}) into Eq. (\ref{te9}), namely,
\begin{equation}
\psi(x,t)= \frac{1}{\sqrt{2\pi}}\sum_{n=-\infty}^{\infty} \varrho_ne^{-ik_nL}
\int_{-\infty}^{\infty}dk\,\frac{k}{k-k_n}\phi_0(k)e^{ikx-i\hbar k^2t/2m},
\label{tefs}
\end{equation}

where  $\varrho_n=u_n(0)u_n(L)/k_n$. For a cutoff Gaussian initial wavepacket, substitution of  Eq. (\ref{te17}) into Eq. (\ref{tefs}),
leads, in the  \textit{small truncation regime}, to the expression
\begin{eqnarray}
\psi(x,t)&=& T(k_0) \psi^f_a(x,t) \nonumber \\[.3cm]
&+ & i\sum_{n=-\infty}^{\infty} \varrho_nk_ne^{-ik_nL}\bigg[\frac{1}{k_n-k_0}\psi^f_a(x,t)  \nonumber \\ [.3truecm]
& +& D(z_0)e^{ik_0x - i\hbar k_0^2t/2m}  M(x',t';k_n') \bigg],
\label{18}
\end{eqnarray}
where $D(z_0)=-2i(2\pi)^{1/4}\sqrt{\sigma/\textrm{erfc}(z_0)}$ and the Moshinsky function $M(x',t';k_n')$, depends now on $k_n^{'}=k_n-k_0$, with $k_n$ the complex resonance pole. The other parameters remain as defined above.
Figures \ref{tef4} (a)-(c) exhibit  $\sigma |\psi(x,t)|^2$ in units of the lifetime, for a cutoff Gaussian initial  wavepacket
with parameters $x_0=-5.0$ nm,  $\sigma=0.5$ nm and an incident energy $E_0$  of half the potential barrier height $V$, tunneling through a double-barrier resonant system  (DB) with the same parameters as given in the discussion in \ref{quasim}.
The profile of the transmitted wavepacket exhibits a transient behavior that after a very large distance,
acquires a fixed shape that resembles, in time domain, the resonance spectra of the system. As a comparison, Fig. \ref{tef4} (d)
yields the transmission coefficient of the DB system \textit{vs} energy in units of the potential height, using an exact numerical calculation
(full line) and the resonance expansion given by Eq. (\ref{e21}).

\subsection{Relativistic effects in quantum transients}
An interesting feature of the solutions for cutoff initial waves, occurring both in the free case \cite{holland} and in the presence
of a potential interaction \cite{GR97}, is that, if initially there is a zero probability for the particle to be at $x > 0$, as soon as  $t \neq  0$, there is instantaneously, a finite, though very small, probability to find the particle at any point $x > 0$.
In Ref. \cite{GRV99}, Garc\'\i a-Calder\'on, Rubio and Villavicencio, studied the short time behavior of the solution of the quantum shutter in the presence of a finite range potential extending through the region $0 \leq x \leq L$ with the initial conditions given, respectively,  by Eqs. (\ref{te1p}) and (\ref{te6p}), which lead along the external region, $ x \geq L$ to the solutions given by Eqs.  (\ref{te2}) and
(\ref{te8}). The found that at short times the above solutions behave as
\begin{equation}
\psi(x,t) \sim \frac{A}{x}t^{1/2}; \,\,\, x \geq L,
\label{sb}
\end{equation}
where $A$ is a constant. The above expression tells that the probability density at any distance $x$ from the potential will rise instantaneously with time. Although this should not pose any conceptual difficulties because the treatment is not relativistic, the above authors ask
themselves whether the above nonlocal behavior arises because the initial condition is a cutoff wave. In order to solve this question
the authors considered the solution of the Klein-Gordon equation for a delta potential $V(x)=b_s\delta(x)$ by using Laplace transforming techniques and found an analytical solution of the problem that at short times becomes different from zero after a time $t_0=L/c$, with $c$ the velocity of light, thus restoring Einstein causality. Hence the non-local behavior of the Schr\"odinger solution is due to its nonrelativistic nature and not a result of the quantum shutter setup.

\section{Time scale of forerunners in quantum tunneling\label{tune}}

Quantum tunneling, the possibility that a particle traverses a classically forbidden region, constitutes one of the paradigms of quantum mechanics. Most textbooks usually address its stationary aspects only, and solve the stationary Schr\"odinger equation at a fixed energy $E$.
In the time domain the analysis of the transient behavior of the time evolution of the wave function along the internal and transmitted regions of a potential has been considered in the framework of the quantum shutter setup as a natural extension of the free case considered by Moshinsky.
Since the seminal work of B\"uttiker and Landauer \cite{BL82}, the time-dependent aspects of tunneling have attracted
a great deal of attention.
Much of this work has led to controversy, as several authors have proposed and defended different ``tunneling times''.
In fact each has its own virtues, weaknesses, physical content, and range of applicability. For reviews see \cite{HS89,LA90,LM94,Ghose99,MSE02}.

An analysis of the involved non-commuting observables (the projectors that
determine the final transmission and the probability to find the particle in the barrier region)
shows that, from a fundamental perspective, there is no unique tunneling time,
because several quantizations are possible due to different
operator orderings and defining criteria \cite{BSM94}.
It is thus necessary to specify precisely how to time the quantum
particle in the tunneling regime, since different procedures
lead to different relevant time scales.
For example, the traversal time of B\"uttiker and Landauer
(BL time) \cite{BL82}, $\tau_{BL}=L/v_{sc}$, given by the barrier length
$L$ divided by the ``semiclassical'' velocity $v_{sc}=[2(V-E)/m]^{1/2}$,
marks the transition from sudden to adiabatic regimes
for an oscillating barrier \cite{BL82}, and  determines
the rotation of the spin in a weak magnetic field in opaque conditions \cite{Buttiker83};
whereas the average over wavepacket components of the (monochromatic)
``phase times'' provides the mean arrival time of the transmitted wave
packet \cite{BSM94,ML00}.
These two time scales may be very different.
The B\"uttiker-Landauer time increases with
decreasing energies up to a finite value,
whereas $\tau^{Ph}(0,L)$ (the so called
extrapolated phase time) diverges as $E\to 0$, and
tends for increasing $L$ to a constant value,
$2\hbar/[v_{sc}(2mE)^{1/2}]$.
This latter property implies that the arrival of the transmitted wave
becomes independent of $L$ (Hartman effect \cite{Hartman62}), although
the independence only holds
until a certain critical length $L_c$ \cite{BSM94}
where above-the-barrier components start to dominate. For
$L>L_c$ the mean arrival time depends on $L$ linearly.
While $\tau_{BL}$ and $\tau^{Ph}$ are surely the most frequently
found tunneling times, they do not exhaust all
timing questions as we shall see.
\subsection{Transients for a square barrier\label{tsb}}
A simple and physically interesting complication of the elementary shutter
is the addition of a square barrier after the shutter edge.
Brouard and Muga studied this problem paying attention to the ``tunneling''
configuration in which the carrier energy is below the barrier top,
$E_0<V_0$ \cite{BM96}.
They used a complex momentum plane approach associated with asymptotic
expansions which has been applied later to different transient phenomena
\cite{PBM01,gcvdm02,DCM02,DMRGV,PBM03,DMR04,AP,APerr,PBM05}.
The aim of the approach is to describe the wave propagation with a minimum of elements and maximal efficiency.
It provides the wave function, as well as characteristic velocities and times,  and localizes
the origin of the main contributions
at critical points in the complex energy or momentum planes such as poles, saddle
points, and branch points, following the analysis by Sommerfeld
and Brillouin for the propagation of light in dispersive
media \cite{SB}.
This is useful conceptually,
and improves the efficiency of the numerical calculations.
Pioneering work in this direction was done by
Stevens \cite{Stevens}, who proposed a sequence of ``tunneling problems''
for the step potential barrier.
Using approximate arguments,
his main conclusion was that for energies below the
barrier height $E_0<V_0$, a wave front traveling with the semiclassical
velocity $v_{sc}=[2(V_0-E_0)/m]^{1/2}$ could be identified. He
related this wave front to the crossing of a pole by the
steepest-descent path. This conclusion has been later shown to be
generally unjustified, due to the simultaneous
effect of other critical points \cite{TKF87,JJ89,Ranfa90,Ranfa91}.
More on this in \ref{mbsources} below.
The study of finite width barriers is a natural extension of Stevens's tunneling problems for the
step potential. Along this line, Jauho and Jonson \cite{JJ89} examined
numerically the propagation of an initially sharp packet (a
cutoff plane wave) through a square barrier, and Moretti carried
out an approximate asymptotic analysis for this system
\cite{Mor92} as well.

Brouard and Muga \cite{BM96}
provided an exact solution of the dynamics that retains the
basic philosophy of working out the time-dependent wave
function by contour deformation in the complex momentum
plane and extracting contributions
from critical points, which in most cases reduce
to a $w$-function.

It is instructive to note the main features
of the analysis in \cite{BM96} by first studying 
Moshinsky's simplest shutter problem.
The general solution for free motion takes the form
\beq
\psi_0(x,t)=\int_{-\infty}^{\infty} \la x|k\ra e^{-iE_k t/\hbar}
\la k|\psi(t=0)\ra dk,
\label{superpo}
\eeq
where $E_k=k^2\hbar^2/2m$ and $\la k|k'\ra=\delta(k-k')$.
For an initial state corresponding to a cut-off plane wave $e^{ik_0 x}\Theta(-x)$,
\beq
\psi_0(x,t)=\frac{i}{2\pi} \int_{-\infty}^{\infty}
\frac{e^{i(kx-k^2\hbar t/2m)}}{k-k_0+i0}dk.
\eeq
The steepest descent path (SDP) is a $-45^o$ straight line cutting the real $k$ axis at the saddle point ($k=mx/t\hbar$)
of the exponent.
For a fixed position $x>0$, the saddle point moves from $\infty$ to 0 as $t$ grows
from 0 to $\infty$.
Deforming the contour along that path, completing the square,
and using the same $u$-variable of Eq. (\ref{mou}),
\beq
\label{uvar}
u=\frac{1+i}{2}\left(\frac{\hbar t}{m}\right)^{1/2}
\left(k-\frac{mx}{\hbar t}\right),
\eeq
which is zero at the saddle, and real along the SDP,
we get
\beq
\psi_0(x,t)=i\frac{e^{imx^2/2t\hbar}}{2\pi}\int_{\Gamma_u}\frac{e^{-u^2}}{u-u_0} du\;\;\;
(x,t>0),
\eeq
where $u_0=u(k=k_0)$
is the pole position in the $u$-plane. Note that it ``moves'' with $t$ and/or $x$;
by contrast, the pole in the $k$-plane is fixed at $k_0$.
The contour  $\Gamma_u$ goes along the real $u$ axis from
$-\infty$ to $\infty$ plus a circle around $u_0$ whenever its imaginary part is positive.
Finally,
\beq
\psi_0(x,t)=\frac{e^{imx^2/2t\hbar}}{2}w(-u_0)\;\;\; (x,t>0),
\eeq
which is Moshinsky's solution $M(x,k,t)$, Eq. (\ref{eqDIT}).
At fixed $x$, $u_0^2\sim ik_0^2 \hbar t/2m$ as $t\to\infty$ and a stationary regime with constant density is reached.
At small times the saddle at $u=0$ is far from the pole and it is the only important critical point.
Setting $u=0$ in the integrand,
\beq\label{shti}
|\psi(x,t)|^2\sim\frac{th}{4m\pi^2x^2}\;\;\; (x/t\to\infty).
\eeq
The treatment for the square barrier starts from an integral like
(\ref{superpo}), substituting the plane waves
by scattering states.
Thus, for the transmitted part, $x>a$,
\beq
\psi_T(x,t)=\frac{1}{(2\pi)^{1/2}}\int_{-\infty}^{\infty} T(k) e^{ikx}e^{-iE_k t/\hbar}
\la k|\psi(0)\ra dk,
\eeq
which is in fact a generic result for cut-off potentials and initial states.
The difference with the elementary treatment of the free motion case is the need to consider the complex poles
of $T(k)$ in the lower half-$k$-plane, $k_j$, which leads to a series in terms of $w$-functions for each of them.
The short time asymptotic behavior does not change with respect to the free
case, Eq. (\ref{shti}), because the poles are yet very far from the saddle and integration contour.
As time progresses the saddle gets near the poles, and
a velocity of propagation can be assigned to each resonant pole contribution from the
condition of the crossing of each pole by the contour. These contributions
are transient, fade away with time, their
residue terms being
proportional to $e^{-ik_j^2\hbar t/2m}$, and only the contribution of the structural pole $k_0$
remains eventually,\footnote{The ``structural poles'' are associated with the structure
of the wave-function and are to be distinguished from poles of the resolvent,
in particular from resonance poles \cite{MWS96,MWS96b}.} which sets in around the phase time.

The internal part, $0<x<L$, becomes  more complicated in this treatment
because of the interference between different terms, but the results
show clearly that a front moving with a semiclassical velocity
is not seen
in the exact solution.

\begin{figure}[!tbp]
\begin{center}
\rotatebox{0}{\includegraphics[width=3.3in]{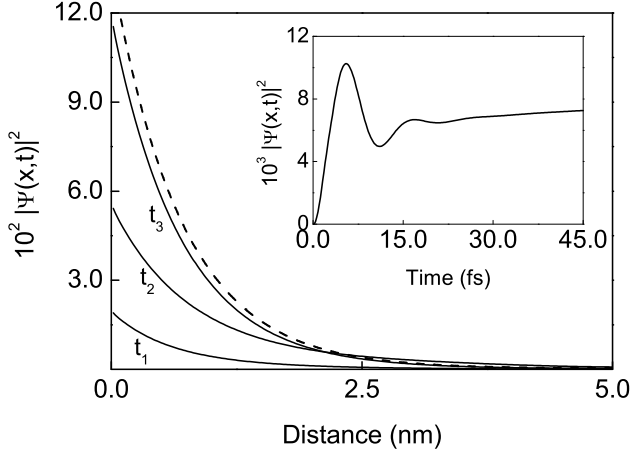}}
\caption{Plot of the probability density normalized to the incident flux along the internal region
of a rectangular potential, denoted by $|\psi(x,t)|^2$, for various times which exhibits that there is no a propagating wave
along the internal region: $t_{1}=1.0$ fs, $t_{2}=3.0$ fs, and $t_{3}=30.0 fs$. For comparison the stationary solution has been also
included (dashed line). The inset shows a replica \textit{time domain resonance} at the internal distance $x_0=2.0$ nm. See text.}
\label{ffig1}
\end{center}
\end{figure}

\begin{figure}[!tbp]
\begin{center}
\rotatebox{0}{\includegraphics[width=3.3in]{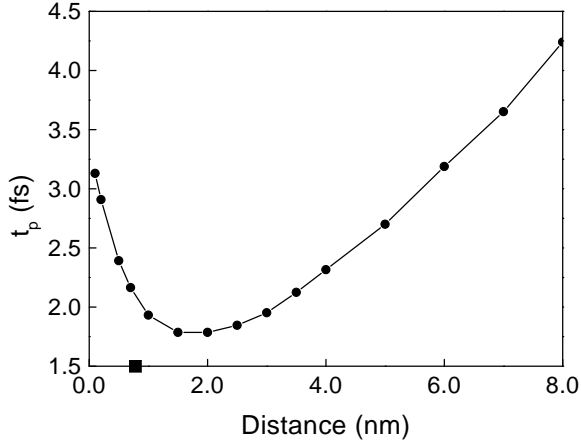}}
\caption{Exact calculation of $t_p$ (solid dots)
as a function of the position $x$. The full square indicates the value
of the penetration length $\kappa_0^{-1}$. See text.}
\label{basin}
\end{center}
\end{figure}

Alternatively, using the exact analytical approach discussed in the previous section,
Garc\'\i a-Calder\'on \cite{gc02}  examined the time evolution of the initially
cutoff wave given by Eq. (\ref{te6p}) with energy $E=0.01$ eV on a rectangular barrier.
The parameters of the barrier are typical of semiconductor quantum structures \cite{ferry}: Barrier height $V_0=0.23$ eV, barrier width $L=5.0$ nm and effective electronic mass $m=0.67m_e$, with $m_e$ the bare electron mass.
The results of the calculation exhibit an absence of a propagating wave along the internal region of the potential.
The absence of a propagating wave along the tunneling region may also be inferred from
the corresponding analytical solution (see Eq. (\ref{te7}) since the Moshinsky functions there are independent of $x$.  The inset to
Fig. \ref{ffig1} exhibits the probability density as a function of time at a fixed internal distance $x_0=2.0$ nm.
A peaked structure at $t \approx 5.6$ fs is clearly appreciated.
This is a replica of a similar structure that appears at $x=L$, named \textit{time domain resonance} \cite{GV01,gcv02,gcv03} which is further discussed in  \ref{tune3} and  \ref{tune4}.
Similar non-propagating structures with different peaked values $t_p$ appear at every internal point.  In Ref. \cite{gcvdm02} this was further
investigated. Figure \ref{basin} exhibits a plot of $t_p$ (solid dots) as a function of position along internal values of the distance $x_0$ of
a rectangular barrier with parameters: $V_0=1.0$ eV and $L=40.0$ nm. Figure \ref{basin} explores distances of the order of, or smaller than, the penetration length defined as $\kappa_0^{-1}$, where $\kappa_0=[2m(V_0-E_0)]^{1/2}/\hbar$, and $E_0=\hbar^2 k_0^2/2m$ stands for the incident energy.
One might clearly identify two regimes. The first of them corresponds to a basin,
where for small values of the position, $t_p$ decreases reaching a
minimum value as $x$ increases.
However, if $x$ is further increased, $t_p$
begins to grow as a function of the position.
Apparently this second regime corresponds to a situation where
$t_p$ increases linearly with $x$. The result depicted in Fig.
\ref{basin} appears also to be a replica of a similar behavior of the
probability density at the barrier edge $x=L$, reported in Ref. \cite{GV01}.
Before we further clarify and interpret these effects in \ref{tune3},
it will be useful to step back from the relative complexity
of the square barrier and analyze an even simpler case.
\subsection{Muga-B\"uttiker analysis of evanescent waves \label{mbsources}}
%
%
%
%
%
%

Here we shall follow \cite{MB00} to study the role of
the B\"uttiker-Landauer time
in quantum transients due to the
injection of particles with energy below the potential level,
so that an evanescent wave is finally formed.
For the Schr\"odinger equation
\beq
i\hbar\frac{\partial\psi}{\partial t}=-\frac{\hbar^2}{2m}
\frac{\partial^2\psi}{\partial x^2}+V\psi,
\eeq
with $V=$constant,
%
%
%
%
we define wavenumber $k$
and frequency $\omega$ as
\beqa
\omega&=&E/\hbar=V/\hbar+k^2\hbar/2m,
\\
k&=&[2m(\omega-V/\hbar)/\hbar]^{1/2}.
\eeqa
Particle injection is modeled by fixing the wavefunction at
the origin for all times,
\beq
\label{source0}
\psi(x=0,t)=\Theta(t)\, e^{-i\om_0 t},
\eeq
for a frequency $\om_0=\hbar k_0^2/2m$, and assuming $\psi(x,t<0)=0$.


The inverse Fourier transform of the source
$\psi_0(x=0,k_0,t)=e^{-i\om_0 t}\Theta(t)$ is given by
\beq
\widehat{\psi}(x=0,k_0,\omega)=\frac{1}{\sqrt{2\pi}}\int_{-\infty}^{\infty}\psi_0(x=0,k_0,t)e^{i\om t}dt
=\frac{i}{\sqrt{2\pi }}\,
\frac{1}{\omega-\omega_{0}+i0}.
\eeq
For $x,t>0$, the wave function evolves according to
\begin{displaymath}
\psi_0(x,k_0,t)=\frac{i}{2\pi }\int_{-\infty}^{\infty}
d\omega\,\frac{e^{i kx-i\omega t}}{\omega-\omega_{0}+i0},
\end{displaymath}
where ${\rm{Im}} k\ge 0$.
If we are interested in carrier frequencies below threshold,
$\omega_0<V/\hbar$, it is also useful to define $\kappa_0=[2m(V/\hbar-\omega_0)/\hbar]^{1/2}$.
The integral can be written in the complex $k$-plane by deforming the contour of
integration to $\Gamma_{+}$, which goes from
$-\infty$ to $\infty$ passing above the poles,
\begin{eqnarray}
\psi_0(x,k_0,t)&=&\frac{i}{2\pi }
\int_{\Gamma_+}dk\,2 k\,\frac{e^{i kx -i(k^{2}\hbar/2m+V/\hbar)t}}{k^{2}+\kappa_0^2}\nonumber\\
&=&
\frac{i}{2\pi }\int_{\Gamma_+}\!\!\!dk\,
\left(\frac{1}{k+i\kappa_{0}}\!+\!\frac{1}{k-i\kappa_{0}}\right)
e^{i kx-i k^{2}\hbar t/2m-iVt/\hbar}.
\nonumber
\end{eqnarray}
%
The contour can be deformed further along the steepest descent path
from the saddle at $k_s=xm/t\hbar$, the straight line
\beq
k_I=-k_R+xm/t\hbar
\eeq
($k=k_R+ik_I$), plus a small circle around the pole at $k_0=i\kappa_0$
after it has been crossed by the steepest descent path, for fixed $x$, at the critical time
\beq
\tau_c\equiv xm/\hbar[{\rm Re}(k_{0})+{\rm Im}(k_{0})]
\label{tauc}.
\eeq
For tunneling ${\rm Re}(k_0)=0$ so that $\tau_c=\tau_{BL}$, where
now
\beq
\tau_{BL}=\frac{xm}{\kappa_0\hbar}
\eeq
%
is a slight generalization of the B\"uttiker-Landauer
time, with $x$ playing the role of $L$.
The solution, using the $u$-variable (\ref{uvar}),  
real along the steepest descent path, is
\beq\label{soluci}
\psi(x,t)=\frac{e^{-iVt/\hbar}e^{ix^2m/2t\hbar}}{2}
\left[w(-u_0)+w(-u_0')\right],
\eeq
%
where $u_0=u(k=k_0)$, and $u_0'=u(k=-k_0)$.\footnote{This formal result
holds for all carrier frequencies $\omega_0$,
above or below threshold. Putting $V=0$ it reads
simply $\psi(x,t)=M(x,k_0,t)+M(x,-k_0,t)$ in terms of Moshinsky functions, i.e., the source
boundary condition imposed at $x=0$ gives the same wavefunction as
the shutter with reflection amplitude $R=1$ opened at $t=0$.
The case $R=-1$
and its relation to inhomogeneous sources was examined by Moshinsky \cite{MoshinskyRMF54} and is reviewed in Appendix \ref{apis}.}
%
%
Eq. (\ref{soluci}) can be written for $x\kappa_0 \gg 1$ (opaque conditions)
as the sum of contributions
from the saddle point
and the pole,
\beqa
\psi(x,t)&=&\psi_p(x,t)+\psi_s(x,t),
\label{apro}\\
\psi_p(x,t)&=&e^{-i\omega_0 t+ik_0 x}
\Theta\left(t-\tau_c\right),
\\
\label{saddle}
\psi_s(x,t)
&=&\frac{e^{-iVt/\hbar}e^{ik_s^2\hbar t/2m}}
{2i\pi^{1/2}}\left(\frac{1}{u_0}+\frac{1}{u_0'}\right).
\eeqa
Asymptotically, only the ``monochromatic''
pole contribution remains and a stationary evanescent wave
is formed, but,
before that regime,
a forerunner (bump in the density with a well defined maximum) is observed at fixed $x$.
For carrier frequencies below the cut-off
$\tau_{BL}$ plays clearly a role in the exact solution
since
\beqa\label{u0u0}
u_0&=&\frac{1+i}{2}\sqrt{\frac{\hbar t}{m}}\kappa_0 (i-\tau_{BL}/t),
\\
u_0'&=&\frac{1+i}{2}\sqrt{\frac{\hbar t}{m}}\kappa_0 (-i-\tau_{BL}/t).
\eeqa
However, it cannot be identified with the transition to the asymptotic
regime since the saddle contribution dominates up to an exponentially large time
$t_{tr}=t_{tr}(x)>\tau_{BL}$
that can be identified from the condition $|\psi_p|=|\psi_s|$
\cite{MB00}. (By contrast above threshold, $\tau_c$ in Eq. (\ref{tauc})
is a good scale for the arrival of
the main peak.)
A different instant is the time of arrival, $t_p=t_p(x)$,
of the temporal peak
of the forerunner at a point $x$.
It can be obtained from from $\partial|\psi_s|^2/\partial t=0$ and,
for opaque conditions,
$t_p$ turned out to be (surprisingly)
proportional to the BL time $\tau_{BL}$,
\beq
\label{tf}
t_p=\tau_{BL}/3^{1/2},
\eeq
even though the time-frequency analysis of the forerunner \cite{Cohen95,MB00}
confirmed that it was composed by frequencies above
threshold so it was not tunneling.\footnote{The simplest time-frequency
information is contained in the local instantaneous frequency.
Writing $\psi(x,t)=|\psi(x,t)|e^{i\phi(x,t)}$ it is defined as
$\bar{\omega}(x,t)=-\partial\phi(x,t)/\partial t$. More sophisticated treatments
such as spectrograms, Wigner functions or other time-frequency distributions
provide further moments and information \cite{Cohen95,MB00}.}
The frequency of the saddle is $\omega_s=E_s/\hbar$,
where $E_s=V+x^2m/(2t^2)=V+k_s^2\hbar^2/(2m)$, coincides with
the energy of a classical particle traveling from the source to
$x$ in a time $t$.
It is very remarkable that in spite of its frequency content, the forerunner's peak
``travels'' with a velocity proportional to $v_{sc}$, $v_p=3^{1/2}v_{sc}$,
which increases with decreasing energies, and its intensity
diminishes exponentially as it progresses along the coordinate $x$.
Other works had already pointed out the dominance of non-tunneling
components in the forerunner \cite{RMFP90,RMA91,TKF87,JJ89,BM96}
but had not characterized its time dependence.
\subsection{Forerunners which do tunnel\label{tune3}}
B\"uttiker and Thomas \cite{BT98} proposed to enhance
the importance of the monochromatic front associated with $\omega_0$
compared to the forerunners by limiting the frequency band of the
source or of the detector; it was shown later in \cite{MB00}
that the monochromatic front could not be seen in
opaque conditions even with the frequency band limitation.
However, a clear separation of the amplitude into two terms, one
associated with saddle and forerunner and the other with the pole
and the ``monochromatic front'', is only possible for opaque conditions.
Non-semiclassical conditions have been much less investigated \cite{Nimtz},
even though these are actually easier to observe because a
stronger signal may be obtained.
Ref. \cite{gcvdm02} and this subsection are mainly concerned with them.
\begin{figure}
\begin{center}
\rotatebox{0}{\includegraphics[width=3.3in]{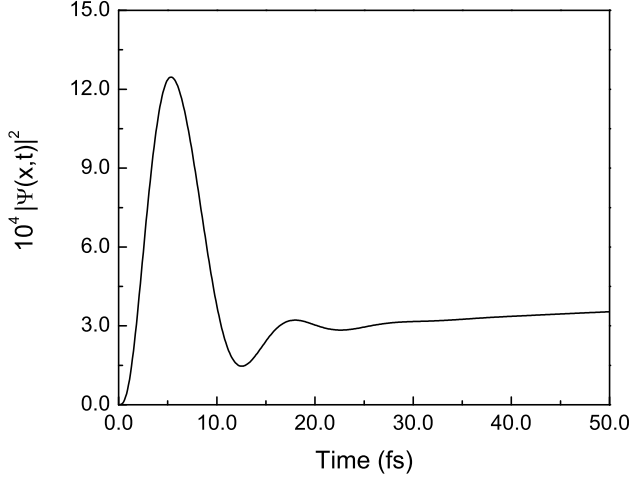}}
\caption{Time evolution of $|\psi(x,t)|^2$  for a rectangular barrier of height $V=0.3$ eV and width $L=5.0$ nm, evaluated at  $x=L$, exhibiting a \textit{time domain resonance} peaked at $t_p=5.3$ fs. The incident energy is $E=0.01$ eV
and the effective electron mass is $m=0.067m_e$, where  $m_e$ stands for the electron mass.}
\label{tdr}
\end{center}
\end{figure}

\begin{figure}[!tbp]
\begin{center}
\rotatebox{0}{\includegraphics[width=3.3in]{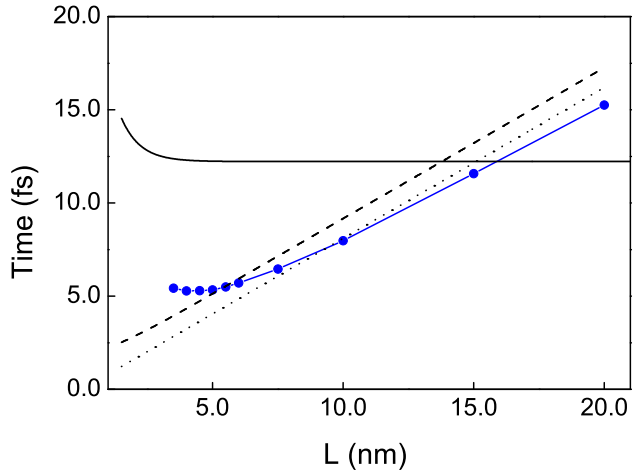}}
\caption{Maximum of the {\it time-domain resonance} $t_{p}$ (solid dots)
as a function of the barrier width $L$ with fixed barrier height $V=0.3$ eV and  incidence energy $E=0.001$ eV.
For comparison, are also plotted the B\"uttiker traversal time $t_B$, the B\"uttiker-Landauer time $t_{BL}$ and
the phase-delay time $t_D$.}
\label{baslin_new}
\end{center}
\end{figure}

Garc\'{\i}a-Calder\'on and Villavicencio \cite{GV01} examined the time evolution of an initial cutoff wave truncated at the
left edge of a rectangular barrier potential (quantum shutter setup).
There it was found that the probability density at the barrier edge $x=L$,
exhibits at short times a transient structure named {\it time domain resonance}, see Fig. \ref{tdr}.
The behavior of the maximum, $t_p$, of the {\it time domain resonance}
showed a plateau region for small $L$, or more accurately  a shallow basin,
followed by a linear dependence for larger $L$; this behavior is reminiscent
of the Hartman effect, but the time of the plateau did not coincide
with the phase-time estimate. It was found that for a broad range of parameters,
$t_p$ in the basin may be written approximately as
\begin{equation}
t_p^B=\frac{\hbar \pi}{\epsilon_1-E_0},
\label{tB}
\end{equation}
where $\epsilon_1$ and $E_0$, correspond to the energy of the first top-barrier resonance
and the incidence energy, respectively.
On the other hand, along the linear regime, at larger values of $L$, $t_p$ is described by
\begin{equation}
t_p^L=\frac{L}{v_1},
\label{tL}
\end{equation}
where $v_1=\hbar a_1/m$, with $a_1$ the real part of the first top-barrier pole $k_1=a_1-i b_1$.
The above considerations are illustrated in Fig. \ref{baslin_new}. This figure also shows a comparison
with exact calculations for the B\"uttiker traversal time $t_B$, the semiclassical B\"uttiker-Landauer time $t_{BL}$
and the  phase-delay time $t_D$, see Eqs. (3.12), (1.7) and (3.2) of [196]. Notice that $t_B$ and $t_{BL}$ do not describe the basin regime and that they are close to the values of $t_p$  along the regime that increases linearly with $L$.

It was shown in \cite{gcvdm02} that this basin dependence, and the corresponding time scale, is not only present at
the transmission edge of the square barrier studied in \cite{GV01}. It may also be found, mutatis mutandis, in the
sharp onset source model with constant $V$ examined in \cite{MB00} for small $x\kappa_0$. As discussed above,
the basin is also found for the time of the forerunner
versus position in the internal region of the rectangular barrier, and for a step potential barrier.
There are no resonances in the sharp onset source model, or for the step potential, so the time scale
of the forerunner at the basin minimum is in these cases inversely proportional to $\kappa_0^2$,
namely to the difference between the ``potential level'' $V$ and the source
main energy $E_0$. This is to be contrasted with the dependence on $\kappa_0^{-1}$ of the traversal time $\tau_{BL}$.
Apart from certain peculiarities, all models show a small length region of the order of the penetration
length $\kappa_0^{-1}$ where the forerunner is dominated by tunneling components and
arrives at a time proportional to $\kappa_0^{-2}$.

The time-frequency analysis showed that the peak of the \textit{forerunner}
at small lengths, in the evanescent region, is composed predominantly
by under-the-barrier components, so that indeed there is a genuine tunneling time scale different from
the phase or BL times \cite{gcvdm02}.

In Ref. \cite{gcv02}  the analytic solution to the Schr\"odinger equation for cutoff wave initial conditions given by  Eq. (\ref{te8}) was used to investigate the time evolution of the transmitted probability density for a rectangular potential.
For a broad range of values of the \textit{opacity} defined as
\begin{equation}
\alpha=[2mV]^{1/2}L/\hbar,
\label{teopa}
\end{equation}
it was found that the
transmitted probability exhibits two evolving structures. One refers to a \textit{forerunner} which follows from the propagation of  the \textit{time-domain resonance} whereas the other structure consists of a semiclassical propagating wavefront, that resembles the time evolution of the free term modulated by the value of the transmission coefficient. A regime is also found for values of the opacity $\alpha$ below a critical value $\alpha_c$ where there are no \textit{time domain resonances} and hence no \textit{forerunners}. Moreover, it is also found that this situation is related to a \textit{time advance}, i.e. a positive \textit{delay time}. This occurs for very thin or very shallow potential barriers. For the effective electron mass $m=0.067m_e$, typical of $GaAs$ semiconductor materials,  $\alpha_c=2.0653$.

\begin{figure}[!tbp]
\begin{center}
\rotatebox{0}{\includegraphics[width=3.3in]{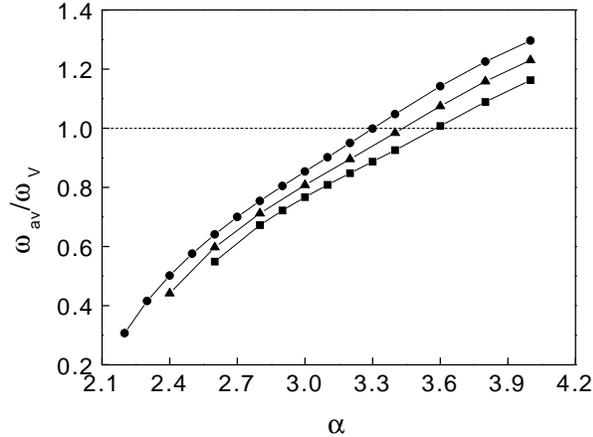}}
\caption{Relative frequency $\omega_{av}/\omega_V$ measured
at the barrier edge $x=L$, as a function of the opacity
$\alpha$. The parameters: $u=300$ (solid dot), $u=10$
(solid triangle), and $u=5$ (solid square). Note that for
values of the opacity smaller than $\alpha \simeq 3.3$, the relative
frequencies for all values of $u$ are below the cutoff-frequency
$\omega_{av}/\omega_V=1$ (dashed line). See text.}
\label{omegalfa}
\end{center}
\end{figure}
\subsection{Barrier opacity and time scale for tunneling as a transient effect\label{tune4}}
The short-length tunneling time scale discussed before is
further investigated in Ref. \cite{gcv03}:
Garc\'{\i}a-Calder\'on and Villavicencio
look at the
peak value of the probability density at the barrier edge $x=L$
in the shutter problem
and
characterize the regime that corresponds to under-the-barrier tunneling by the \textit{opacity} of the system, \textit{i.e.}, Eq. (\ref{teopa}).
A time-frequency analysis  establishes the existence of a range of values of the \textit{opacity}, independent of the incidence energy, where the peak value of the \textit{time domain resonance} is governed by a single frequency, which suggests that the system acts as a frequency filter. Figure \ref{omegalfa} exhibits the relative frequency $\omega_{av}/\omega_V$ as a function of the \textit{opacity} for three different
values of the parameter $u=V/E$. Here $\omega_{av}={\rm Im}[(d\psi/dt)/\psi]$, and the cutoff frequency $\omega_V=V/\hbar$. The time-dependent solution is given by Eq. (\ref{te7}).  Notice the existence of an \textit{opacity} ``window'', in the range of values $2.065 \leq \alpha \leq 3.3$, where the average frequencies are always below the cutoff frequency, \textit{i.e.}. In this example the
value of $V=0.3$ eV and the mass $m=0.067m_e$, with $m_e$ the bare electron mass.

\subsection{Time scales for relativistic equations\label{relativi}}
In \cite{DMRGV} the authors investigated
the role played by $\tau_{BL}$, and the characteristics of the
forerunners at large
and small distance from the source for under-cut-off, unit-step modulated excitations
with the (dimensionless) relativistic wave equation
\beq
\left[\frac{\partial^2}{\partial T^2}-\frac{\partial^2}{\partial Y^2}
+1\right]\Psi=0,
\label{KGdl}
\eeq
which turns out to be more accessible experimentally (with
waveguides in evanescent conditions \cite{Nimtz,RM04})
than the corresponding Schr\"odinger equation
\beq
i\partial\Psi/\partial t=-\partial^2\Psi/\partial Y^2+\Psi.
\label{Sch}
\eeq
The stationary
equations corresponding to Eqs. (\ref{KGdl}) and (\ref{Sch}) are
equal and have equal solutions, but the two
cases are not equivalent in the time domain, because of the
dispersion relations between frequency and wavenumber.
%
%
In fact a wealth of
qualitative changes with respect to the Schr\"odinger scenario
were found \cite{DMRGV}.

Eq. (\ref{KGdl})
leads to the Klein-Gordon equation for spin-0 particles by means of
the substitutions
\beqa
Y&=&x(m_0c)/\hbar,
\\
T&=&tm_0c^2/\hbar,
\eeqa
where $x$ and $t$ denote dimensional position and time,
and $m_0$ and $c$ are the particle's rest mass and the velocity of light
in vacuum.
Another interesting connection, much more promising to
implement the analysis experimentally \cite{RM04},
and free from the conceptual puzzles
of the former,
may be established
with the equation that governs the electromagnetic field
components in waveguides
of general constant cross section with perfectly
conducting walls.
This requires the substitutions
\beqa
Y&=&x\lambda,
\\
T&=&t\lambda c,
\eeqa
where $\lambda>0$ is one of the eigenvalues
of the waveguide.
The main differences
between the relativistic and non-relativistic cases are: (a)
the relativistic solutions are not
simply related to each other by time and position scaling as in the
non-relativistic case, so that qualitatively different
``shallow'' and ``deep''
tunneling regimes may be distinguished;  (b) tunneling is more robust
relativistically, both in the precursors (more tunneling frequencies)
and asymptotically (more
``density'');
(c) the ``first'' relativistic precursor, right after the
limit imposed by causality does not have a non-relativistic counterpart,
and does tunnel; (d) the ``second'' precursor,
which tends to the non-relativistic one for excitation
frequencies near cut-off, has an oscillating
structure that may be characterized by its envelopes. The traversal time
$\tau_{BL}$ is not an exact measure of their arrival except in the
non-relativistic limit.
\section{Transient phenomena caused by a sudden perturbation\label{sape}}
We now turn our attention to the dynamics following a general sudden perturbation in
 which the potential of the system changes from $V_- (x)$ to $V_+ (x)$ abruptly.
\subsection{Potential steps, resonant tunneling}
One of the most important tunneling processes
is resonant tunneling, which
has been employed in
several semiconductor devices since
the original double-barrier diode of Tsu and Esaki \cite{TE73},
and more recently
to control the energy levels, transport through, and state
occupancy of semiconductor quantum dots and wells \cite{TAR96}
and nanowire resonant tunneling diodes \cite{LUND1}. Quantum dots are
candidates as elements in quantum circuits, and
basic quantum manipulation is possible by applying sharp
voltage pulses or microwave signals to
the contact leads surrounding the dot region
\cite{Oosterkamp98,FATHT02N,FATHT02P}.
The response of quantum dots to these stimuli in the sub-nano-second
to almost milli-second time domain is of much current interest.
Two basic manipulations that can significantly affect 
charge transport across the quantum well (dot)
are the change {\it (i)} in the source-drain
voltage applied across the quantum well (dot),
which tilts the whole potential structure, and {\it (ii)} in the
gate electrode voltage,
which may modify essentially (and electrostatically)
the quantum well (dot) depth \cite{Fuji1,Fuji2}.
At resonance, full transmission occurs,
whereas off-resonance, there is essentially full reflection.
Understanding the transitions that fill or
deplete a resonance or a bound state following a change in potential
and determining the speed of the change is clearly of fundamental interest.

For scattering systems, an elementary transient process
is the transition between eigenstates of the original and final Hamiltonian in the continuum.
A stationary scattering wave function
responds to an abrupt change
in the potential shape by forming a new stationary state at any finite
position $x$.
A physical realization would be an
abrupt change in the bias voltage of an electronic device.
The characteristic times of the transient
are of practical interest to determine
the transport properties
of small mesoscopic structures, but modelling the
process by means of a grid discretization of
space in a ``finite box'' is far from simple
\cite{Frensley90}, since
the boundary conditions at the box edges are not known a priori and involve
simultaneous injection to and absorption from the
simulation (box) region. Some approximate ways to deal with the
transients have been proposed
\cite{MH88,Frensley90,RRH91,YE95}. An exact numerical solution generally requires
non-local in time ``transparent'' boundary conditions \cite{Arnold}.
Some cases are solvable analytically in terms of known functions:
Delgado {\it et al}. \cite{DCM02} worked out an explicit
and exact solution of the transition between stationary monochromatic
waves due to an abrupt potential switch
for
a step potential $-V_0\Theta(x)$ that changes
the step height suddenly to $-V_0'\Theta(x)$, $V_0'>V_0>0$,
and other potential profiles, e.g., containing
square single or double barriers can be treated similarly.
The basic trick to find the exact solution is to implement
the action of the evolution operator of the
new Hamiltonian on the initial state
using an integral expression in the complex momentum plane obtained by
Hammer, Weber and Zidell \cite{HWZ77}.
For the step barrier that changes its height abruptly one starts with a stationary
wave function of the original Hamiltonian with incident (momentum $q_0$),
reflected and transmitted waves,
$\psi(t=0)=\psi_1+\psi_2+\psi_3$.
Each part is
treated separately as a modified Moshinsky shutter problem and combined later.
Let us analyze, for example, the incident case: The eigenstates of the final Hamiltonian
can be written as the incident plane wave (for all $x$) plus a rest. The time evolution
for $\psi_1$ is then given by
\beq
\psi_1(x,t)=\int_{C_1} dq\, \psi_q(x)\phi_1(q,t=0)e^{-iE_qt/\hbar},
\eeq
where $q$ is the momentum from the left potential, $C_1$ goes from $-\infty$ to $\infty$
above all singularities and $\phi_1(q,t=0)$ is the momentum representation of $\psi_1(x,t=0)$.
From here the integral is solved by deforming the contour along the steepest descent path.
This provides $w$-function terms and branch-cut contributions, but the latter
can be neglected in many cases. The characteristic time scales for the transitions in the step potential admit simple
classical analogs.
Essentially the same method can be applied to the outer parts of the wave for more complex cut-off potentials such as a double barrier with a well
that can be adjusted to make the wave on or off-resonance \cite{AP,APerr}.
The contour deformation speeds up convergence and provides physical insight
as discussed in \ref{tsb}.
However the method is difficult to apply for the evolution of the wave component initially inside
the quantum well region which was calculated with the  systematic pole expansion
of Garc\'\i a-Calder\'on and coworkers \cite{AP}, see Section \ref{epo}.
In summary, the build-up process is dominated
by the interference between the incident and resonance poles, leading to
a non-exponential growth in which
the resonance lifetime is a relevant characteristic time;
the decay process is dominated by the displaced resonance; and the trapping
process involves very long time transients due to interferences between
a non-exponential resonance contribution and a bound-virtual state pair.
Although a real quantum dot structure is much more complicated, a simple
model potential, with realistic parameters, shows the importance of transient
behavior over a wide (on the pico-second to nano-second and even longer)
time scale. This work was extended to the case of interacting electrons
at the mean field level by coupling Poisson and Schr\"odinger equations \cite{DMCLA05}.
A critical coupling was found (determined by the electron sheet density in the well)
above which permanent trapping does not occur.

Julve and Urr\'\i es have similarly described the transient from a pure plane wave
on all space and the corresponding state for a square barrier using Laplace transform techniques \cite{JU08},
and Moshinsky and Sadurn\'i have studied the dynamics of a deuteron suddenly subjected to an electrostatic field \cite{MS05}.
Finally, Abdallah and Pinaud \cite{Pinaud} have determined non-local in time ``transparent boundary conditions''
necessary to solve the coupled Schr\"odinger-Poisson equation
for the transient evolution of a resonant tunneling diode when the incoming
amplitude representing the electron injection is prescribed, and Antoine {\it et al}.
have recently reviewed different techniques to solve numerically
the time-dependent linear and non-linear Schr\"odinger equation
on unbounded domains \cite{Arnold}.
\section{Other transient phenomena\label{otp}}
\subsection{Ultrafast propagation and
simultaneous arrival of information in absorbing media}
Delgado, Ruschhaupt and Muga have shown that the temporal peak of
a quantum wave  may arrive at different locations simultaneously in an absorbing medium \cite{DMR04,RM04}.
The arrival occurs at the lifetime of the particle in the medium
from the instant when a point source with a sharp onset
is turned on, or when a Moshinsky shutter is opened.
(The effect remains for smoothed pulses \cite{RM04}.)
Other characteristic times may be identified.
In particular, the ``traversal'' or ``B\"uttiker-Landauer'' time
(which grows linearly with the distance to the source)
for the Hermitian, non-absorbing case is substituted by
several characteristic quantities in the absorbing case.
The simultaneous arrival due to absorption, unlike the Hartman
effect, occurs for carrier frequencies under or above the
cut-off, and for arbitrarily large distances. It holds also in a
relativistic generalization but limited by causality, i.e.,
limited to distances where the time of arrival of the peak is larger than the time of the  very first front.
The ubiquitous peak is dominated by the saddle-point contributions above the cutoff frequency.
A proposal for an experimental verification is described in \cite{DMR04} making use
of a quantum optical measurement of atomic densities
in a metastable state by fluorescence. Another interesting possibility is the use of absorbing
wave guides and electromagnetic waves.
One advantage of the waveguide realization with respect to quantum particles is
that the effect could be measured in a non-invasive way.
It also makes it possible to trigger several devices at the same time
or transmit information so that it arrives simultaneously
at unknown locations.
This cannot be achieved by standard transmission methods because
a receiver could not resend an information bit to the closest one
faster than the velocity of light in vacuum $c$;
nor can we design the timing of a series of signals
from the source so that they arrive simultaneously at different receivers
if their locations are unknown.

The simplest case is that of a source with a sharp onset, and
the treatment is very similar to the one used in \ref{mbsources},
substituting the real potential by a complex one,
\beq
i\hbar\frac{\partial \psi}{\partial t}=
-\frac{\hbar^2}{2m}\frac{\partial^2 \psi}{\partial x^2}+(V_R-iV_I)\psi.
\label{wdim}
\eeq
$k_0$ becomes also a complex number so that the role of $\tau_{BL}$
in the formal solutions (see Eq. (\ref{u0u0})) 
is now played by the complex number
\beq
\tau_{cBL}=mx/(-ik_0\hbar),
\eeq
and the steepest descent path in the $k$-plane crosses the pole in $k_0$
at a time $\tau_c$, Eq. (\ref{tauc}).
For carrier frequencies below the cut-off the saddle contribution
dominates up to exponentially large times as in
the Hermitian case \cite{MB00}, so that $\tau_{c}$ is not of much
significance whereas, above threshold, $\tau_{c}$ gives
approximately the arrival of the main peak.
From $\partial|\psi(x,t)|^2/\partial x=0$
we may obtain the position $x(t)$ of the ``spatial'' maximum at time $t$.
This also defines (by inverting $x(t)$) a function $\tau_S(x)$,
namely, the time when this spatial maximum arrives at $x$.
One finds that the spatial maximum in the large-$x$ region
is given by $\tau_S=|\tau_{cBL}|$,
a role played by the real $\tau_{BL}$ in the absence of
absorption \cite{vrspra02}.

Thus a single real quantity $\tau_{BL}$ in the Hermitian case
has been substituted, for a non-zero $V_I$, by
three different quantities: the time for pole-cutting $\tau_c$,
a complex $\tau_{cBL}$, and its modulus $|\tau_{cBL}|$, all of which tend to
the ``B\"uttiker-Landauer'' traversal time $\tau_{BL}\equiv\tau_{cBL}(V_I=0)$ without
absorption \cite{BL82}.

Similarly, we may fix $x$ and solve
$\partial |\psi(x,t)|^2/\partial t=0$ to get $\tau_t(x)$, namely,
the time of the first ``temporal'' maximum.
At variance with the Hermitian case,
the time of arrival of the maximum is {\it not} proportional to
$\tau_{BL}$ ($\tau_t=\tau_{BL}/\sqrt{3}$ when $V_I=0$).
The equation $\partial|\psi_s|^2/\partial t=0$ cannot be solved analytically
for $V_I\ne 0$ in a generic case, so there is no explicit formula for $\tau_t(x)$.
Nevertheless, if $x\gg 2^{-1/2}|k_0|\hbar^2/mV_I$, one finds
\beq
\tau_t\approx \frac{\hbar}{2V_1},
\eeq
i.e., the temporal maximum coincides with the mean ``life time''
$\hbar/2V_1$ of a particle immersed in the absorbing potential, and
it is independent of $x$ and $\omega_0$, which is the
most important result of \cite{DMR04}. A time-frequency analysis \cite{Cohen95,MB00}
shows that the maximum is not tunneling but it is dominated by frequencies above the cut-off,
in particular by the saddle frequency
corresponding to the ``classical''
velocity required to arrive at $x$ at time $\tau_t(x)$.

%
%
\subsection{Breakdown of classical conservation of energy in quantum collisions}
In most quantum interferometers the phase differential between
paths that meet in a coordinate-space point or region at a given time
lead to constructive or destructive wave combinations and thus
to fringes, but the paths can also interfere in momentum space and produce
momentum-space fringes. An example
of interference in momentum representation is the collisional,
transitory enhancement of the high momentum components
of a wave packet \cite{BM98a,BM98b,PBM01,PBM05}. This enhancement leads to a
violation of the classical conservation of energy since the
probability of finding the particle with a momentum larger than
a given value may exceed the classical bound during the
collision. The effect was shown to be quantitatively significant for incident energies
well above the barrier top when a transient suppression of the
main incident momentum components ``pushes'' the momentum
distribution towards lower and higher values \cite{PBM01,PBM05}. This
was explained as an interference in momentum representation between components associated with incidence and transmission.
A simple picture emerges by assuming that in the midst of the collision the wavepacket density is unaltered with respect
to free motion, but the transmitted part acquires the phase of the transmission
amplitude $e^{i\phi_T}$. Adjusting this phase with the incident or barrier energies
constructive or destructive interferences appear in momentum space.
For Gaussian states impinging on
square barriers, an analytical
approximate expression for the wave
function is \cite{PBM01}
\beq
\psi(k,t) \propto w[u_I(k,t)]+T(k) w[-u_T(k,t)],
\label{equ1}
\eeq
where $u_I(k, t)$ and $u_T(k, t)$ are known functions of momentum
and time \cite{PBM01}, and $T(k)$ is the transmission amplitude.
If, for the central momentum components, the
modulus $|T|$ is close to unity and its phase is an odd
multiple of $\pi$, the two contributions in Eq. (\ref{equ1}) interfere destructively
to cause the transitory suppression of the main
momentum components of the incident wave packet. This is
rather surprising from a classical perspective since only a
negligible part of the wave packet is located at the potential
barrier or, equivalently, most of the wave packet is on regions
with zero potential energy. Both transmitted and incident
components are dominated by the central momentum of
the distribution, but instead of adding up classically, they
interfere destructively.

By sending the wave packet towards the appropriate barrier
potential and then switching the potential off at the time
when the maximum interference occurs, this effect may be
used to convert a Gaussian-shaped wave packet into a bimodal
one in momentum representation, since the free evolution
afterwards will not modify the momentum distribution.
However, the transitory character of the effect could hinder
its observation or its use as a preparation tool for controlling
the atomic state of motion in atom optics or quantum information
applications. To remedy this problem an extra potential barrier
was added in \cite{PBM05}. The interference effect
becomes more complex but it is possible to increase its duration
significantly making it more robust and easier to observe.

For an alternative to the scattering configuration using
trapped atoms and ``phase imprinting'' by detuned lasers
which illuminate only half the atomic cloud see \cite{quantph}.
If a $\pi$ phase is imprinted, the interference dip formed in momentum space is centered
at $k=0$, but the minimum is displaced linearly if the phase deviates
slightly from $\pi$. Turning off the external trap thus creates a dark notch
in coordinate space moving with controllable velocity, similar to the
``dark solitons'' observed experimentally \cite{Bongs}. In \cite{quantph} the
perturbation of imprinting imperfections and many body effects are investigated
as a necessary step to determine the potential applicability of momentum space
interferometry. The imprinted phase carries information
about the laser interaction (time, laser intensity,
frequency) that can be obtained from the observable notch.
\subsection{Effect of classically forbidden momenta}
Suppose that a classical ensemble of independent particles in one
dimension is initially confined (at $t=0$) in the spatial interval $a<x<b\le 0$, and
allowed to move freely after $t=0$.
Only particles with positive momenta may arrive at positive positions for
$t>0$. In contrast, the quantum wave function
involves negative-momentum contributions as well,
\begin{equation}
\psi(x,t)= (2\pi)^{-1/2} \int_{-\infty}^\infty dk\,e^{i k x}
\widetilde\psi(k) e^{-iE_kt/\hbar},
\end{equation}
where $E_k=k^2\hbar^2/2m$, and
\begin{equation}
\widetilde\psi(k)=(2\pi)^{-1/2}\int_{-\infty}^\infty dx\,e^{-i k x} \psi(x,0)
\end{equation}
is the wavenumber representation of the initial state.
The effect of negative momenta for $x,t>0$ is however a transient one; the
total final probability to find the particle at $x>0$ is given only by positive
momentum components,
\begin{equation}
P_T(\infty)
\equiv\lim_{t\to\infty}\int_0^\infty dx\,|\psi(x,t)|^2=\int_0^\infty dk\,
|\widetilde\psi(k)|^2,
\end{equation}
see e.g. \cite{JW} or \cite{Allcock69}, since there are no bound
states. This negative-momentum effect is also
present in collisions, where it may be combined with other
classically forbidden effects \cite{Bauteandco}.
Consider the family of cut-off potentials of the form
\begin{equation}
\label{pot}
V(x)=\cases{0 &if $\,\, x<c$\,,\cr
U(x), &if $\,\, c\le x\le d$\,,\cr
V_0,&if $\,\, x>d$,\cr}
\end{equation}
where $V_0\ge 0$ and $V(x)$ are real.
%
%
Let us suppose that the maximum
value of the potential is $V_M$.
For a classical ensemble confined between $a$ and $b$ (such that
$a< b\leq c\leq d\le 0$)
only particles with {\it initial} momentum above the barrier, $k>k_M$,
$k_M\equiv(2mV_M)^{1/2}/\hbar$, may pass to the right of the potential region.
In the quantum case however, there are contributions
from all the components of the initial wave packet:
(a) $k>k_M\equiv(2mV_M)^{1/2}/\hbar$
(above-the-barrier transmission); (b) $k_0<k<k_M$
(``asymptotic'' tunneling, these momenta contribute to
the transmission probability $P_T(\infty)$); (c) $0<k<k_0$ (``transient tunneling'',
these momenta do not contribute to $P_T(\infty)$);
(d) $-\infty<k<0$ (transient negative-momentum effect);
(e) $k_j=i\gamma_j$, $\gamma_j>0$ (tunnel effect associated with bound states, which do contribute to $P_T(\infty)$).

The negative momentum effect is the only one that survives for free motion, and it has been frequently overlooked,
exceptions being \cite{HWZ77,WH,MBS,Bauteandco,BEM01}.
%
%
%
%
%
%
%
\subsection{Time of arrival and Zeno effect}
An elementary model for the time of arrival measurement at a point in space, $x=0$,
consists in performing instantaneous
and frequent projections \cite{Allcock69}
$
\psi(x,{t_j}_+)=\psi(x,{t_j}_-)\Theta(-x),
$
with the rate of norm lost playing the role of an operational time-of-arrival distribution.
After each chopping, the time evolution can be expressed as a combination of Moshinsky
functions \cite{ML00}, and from their short-time asymptotics it follows that
when the projections are performed very frequently the wave packet is totally reflected, according to the quantum Zeno effect.
The first discussions of the Zeno effect, understood as
the hindered passage of the system between orthogonal subspaces because of
frequent instantaneous measurements,
emphasized its problematic status and regarded it as a failure
to simulate or define quantum passage-time distributions \cite{Allcock69, MS77}.
Muga {\it et al.} however \cite{ECM08},
have recently showed that in fact there is a ``bright side'' of the effect:
by normalizing the little bits of norm $N$ removed at each projection
step,
\beq
\Pi_{Zeno}=\lim_{\delta t\to 0}\frac{-\delta N/\delta t}{1-N(\infty)}\;,
\eeq
a physical time distribution defined for the freely moving system
emerges, which turns out to be in correspondence with a kinetic energy density \cite{ked05}, rather than with a time-of-arrival distribution.

\section{Experiments \label{experiments}}
Compared to the abundant theoretical work on elementary quantum transients, the pace of experimental observations has been relatively slow, but fortunately 
the trend is changing very rapidly. 
One reason for the difficulties is the feebleness
of diffraction in time in some systems.   
Dissipation \cite{MS01}, environmental noise \cite{DMM07},
and repulsive interatomic interactions \cite{DM05b} tend to suppress DIT-like phenomena.
Moreover, the smoothing of the aperture function of the shutter \cite{DM05} leads to apodization in time, which delays the manifestation of quantum transients 
until the revival time \cite{DMM07} discussed in Section \ref{onepulse}.
Similarly, DIT may be suppressed by superposition of carrier frequencies. A finite width $\Delta v$ of the velocity distribution induces a spreading $\Delta vt$ in coordinate space which should be negligible compared to the width of the main fringes in (\ref{fringewidth}). Hence, DIT weakens with the width-broadening of the velocity distribution \cite{DM05b,MB00},  which often arises as a direct consequence of the uncertainty principle. In particular, it comes into play in the dynamics of beams of finite-size and velocity selection processes \cite{GN57,GK76} as well as
localized states under confinement \cite{Godoy02,Godoy03,DM05b}.

The analysis of Goldemberg and Nussenzveig \cite{GN57} rightly pointed out that the experimental observation of quantum transients
would be simpler for the time-energy uncertainty principle than for  
interference effects on the density profile, though the latter can be enhanced in matter-waves with attractive interactions \cite{KGAK03}
and using constructive interference with reflected components from moving mirrors \cite{DMK08}.

Despite the hindrance, a number of remarkable experiments have focused on quantum transients and time-dependent optics of different matter waves.
A series of experiments \cite{SSDD95,ASDS96,SGAD96} culminated in
the confirmation of the time-energy uncertainty relation derived by Moshinsky for pulse formation \cite{Moshinsky76}. The matter wave source consisted of caesium atoms in a magneto-optical trap (MOT)
which are released and fall under gravity to bounce off of a mirror, made of an evanescent laser wave.
A selection of atoms with a given total energy was achieved in a first bounce, while the double-slit in time was implemented during the second bounce, after which the resulting pulse was probed. In such a way the energy distribution of the beam when chopped in pulses was measured in agreement with the time-energy uncertainty relation \cite{MoshinskyRMF52b,Moshinsky76}. Moreover, it was possible to register the interference pattern in the arrival-time distribution of the cold atoms after reflection at the double temporal slit. A similar setup was used by Colombe {\it et al.} \cite{CMPL05} in which a Bose-Einstein condensate, with negligible non-linear interactions, was dropped over a harmonically vibrating mirror which can act as a phase modulator \cite{SSDD95}. After the bounce the atomic carrier and sidebands were directly measured by absorption imaging of the density profile \cite{CMPL05}.

Cold atoms imply long wavelengths and time scales favorable to the observation of transients. 
The recent development of guided atom lasers \cite{1datlaser1,1datlaser2}, may pave the way for the direct observation of the DIT fringes of a suddenly turned on matter wave beam \cite{DMM07,DLPMM08}, allowing for a close implementation of the initial setup discussed by Moshinsky \cite{Moshinsky52}.

Historically, ultracold neutrons were initially the favorite matter waves to study such phenomena.
In this case, the interaction with a potential periodically modulated in time, was shown  to induce frequency side-bands. The resulting energy spread of the matter wave could be detected using a high-resolution time-of flight instrument \cite{HFGGGW98,BAKSZ98,BAKOSZ00} and was shown to be in agreement with previous theoretical insight \cite{GKZ81,GZ91,RWCKW92,Frank1,Frank2}. The interaction with a grating moving across a beam of ultracold neutrons similarly induces an energy spreading measured in \cite{Frank3}.

Further studies on the time-energy uncertainly relation have been performed at the attosecond scale during the photoionization of Argon atoms. A double slit in time was implemented using few-cycle ionizing fields, where the momentum distribution of the resulting photoelectron carries the information of the time at which the ionization occurs. The interference in the energy distribution was then measured as a function of the time interval between the generation of different isoenergetic photoelectrons \cite{LSWBGKMBBP05}.

A boost in the experimental progress on quantum transients has recently taken place \cite{chinos} exploiting the analogy between paraxial optics and the time-dependent Schr\"odinger equation \cite{QCA}. 
By means of this analogy, the bifurcation in the density profile of freely expanding excited states of a hard-wall trap predicted in \cite{DM05b} (see section \ref{free_expansion}) has been reported. The free propagation of eigenstates and coherent states of square quantum billiards (essentially a 2D hard-wall potential), was {\it simulated} using the transverse modes of oxide-confined vertical-cavity surface-emitting lasers and recorded with a CCD camera. The analogy with matter-wave dynamics holds under the replacements $t\rightarrow z$, where the $z$-axis is the direction of vertical emission, and $m/\hbar\rightarrow 2\pi/\lambda$,  $\lambda$ being the lasing wavelength. Moreover, for a laser cavity chaotically shaped a random branching interference of the coherent lasing modes was observed.

One more recent new route for experimental access to quantum transients and DIT phenomena is based on the time evolution of the population of an atomic level, i.e., transients do not only occur for translational degrees of freedom but also 
for internal degrees of freedom after or during laser excitations.  
Specifically, linearly chirped pulses where the instantaneous frequency changes in time passing trough the resonance condition for a particular transition of rubidium atoms have been shown to produce DIT oscillations (``coherent transients'') of the excited state in the low field regime \cite{ct2001,ct2006}. 
They are probed with pulses by a second laser on the subpicosecond time scale and can be attributed to the interference between resonant and off-resonant excitations.   

These later developments and, 
in addition, the fact that 
techniques of ultra-short laser pulses in the attosecond time scale (without the need for high-intensity X-rays) 
are a gateway to a plethora 
of time-dependent electronic processes that were too fast to be observed, \cite{science2007}
make the prospects for quantum transient experiments rather brilliant.  
%
%
%
%
%
%
\section{Final comments}
More than fifty years after Moshinsky's pioneering investigation on the
quantum shutter, quantum transients keep posing
theoretical and experimental challenges
and fascinating open questions. The initial interest in nuclear physics has
shifted to applications in cold atoms, quantum optics, and semiconductor structures. Recent experimental capabilities
to prepare and control light pulses, atoms, electrons and their interactions
make possible a whole new world of quantum transients that has been barely explored
and will provide insight into  
many aspects of fundamental physics that had been until now difficult to probe or change. 
Few and many-body physics will offer, for example, ample opportunities
for discovering and exploiting interesting phenomena without
single-particle counterparts. At the single particle level, new 
techniques, ideas, and physical realizations of quantum transients 
set a sound ground for realizing many effects described in the theoretical studies, and will itself motivate further theoretical work.    
A challenge is the direct observation of matter-wave diffraction-in-time oscillations, but promising and versatile substitutes are the electromagnetic
analogy that has been experimentally realized near the completion of this work, 
and ``coherent transients'' of atomic populations under controlled field excitation.  
Applications such as the control and understanding
of quantum dot dynamics, quantum gates operation,
tunneling, atom lasers, momentum interferometry, coherent control, 
or simultaneous transmission of information hold promise of a vigorous
development of research on quantum transients, technological advances
and exploration of new physics. 
\ack
The authors pay homage to Marcos Moshinsky, who unfortunately has passed away recently, and acknowledge discussions
over the years with him;  also  with
D. G. Austing, S. Brouard, H. Buljan, M. B\"uttiker, S. Cordero, F. Delgado, I. L. Egusquiza, M. D. Girardeau, S. Godoy, D. Gu\'ery-Odelin, C. Grosche, M. Kleber, V. V. Konotop, V. I. Man'ko, J. P. Palao, M. G. Raizen, R. Romo, A. Ruschhaupt, E. Sadurn\'{\i}, R. F. Snider, E. Torr\'ontegui and J. Villavicencio and S. Weber. Support by the Ministerio de {\-E\-du\-ca\-ci\'on} y Ciencia (FIS2006-10268-C03-01), the Basque Country University (UPV-EHU, GIU07/40), the EU Integrated Project QAP, and the EPSRC Project QIP-IRC (GR/S82176/0) is acknowledged.

\appendix

\vskip1truecm

%
%
%
%
%
\section{Propagators in quantum transients\label{propa}}
Many relevant quantum transients can be formulated as an initial value problem.
A rather general setting considers a given eigenstate $|\psi(t=0)\ra$ of a Hamiltonian $H_0$
which is suddenly perturbed to a new Hamiltonian $H$.
The initial state undergoes a non-trivial evolution $|\psi(t)\ra$, which for non-relativistic particles
is governed by the time-dependent Schr\"odinger equation.
%
%
Formally, the time-dependent state vector becomes $|\psi(t)\ra={U}(t,0)|\psi(0)\ra$
where the evolution operator for Hermitian Hamiltonians satisfies unitarity,
${U}^{\dagger}{U}={U}{U}^{\dagger}=1$, and the composition property
${U}(t,t'){U}(t',0)={U}(t,0)$.

The most general form of the time-evolution operator is given by the Dyson series,
\beqa
{U}(t,t')=1+\sum_{n=0}^{\infty}\left(\frac{1}{i\hbar}\right)^n\int_{t'}^tdt_1 \int_{t'}^{t_1}dt_2\cdots
\int_{t'}^{t_{n-1}}dt_n {H}(t_1){H}(t_2)\dots {H}(t_n),
\eeqa
though as it is generally the case specific problems admit simpler forms.
Indeed, for a time-independent Hamiltonian, ${U}=\exp(-i{H}t/\hbar)$; while, if $\partial_t{H}\neq0$ but the Hamiltonian at different times commute ($[{H}(t),{H}(t')]=0$ for any $t$ and $t'$),
${U}(t,0)=\exp(-i\int_{0}^{t}dt'{H}(t')/\hbar)$.
In coordinate representation, and using the resolution of the identity,
\beqa
\la x|\psi(t)\ra=\int dx' \la x|U(t,t')|x'\ra\la x'|\psi(t')\ra,
\label{propagatoreq}
\eeqa
one naturally arrives at the concept of the propagator, which is the kernel $K(x,t|x',t')=\la x|{U}(t,t')|x'\ra$, this is, the coordinate representation of ${U}$.
In the general case in which the Hamiltonian has both a discrete set of eigenstates $\{\phi_n\}$ labelled by the quantum number $n$, and scattering states $\{\varphi_E\}$ of energy $E$, the {\it spectral decomposition} of the propagator reads
\beqa
K(x,t|x',t')&=&\sum_n\phi_n(x)\phi_n(x')^{*}e^{-iE_n(t-t')/\hbar}
\nonumber\\
& & +\int_{0}^{\infty}dE \varphi_E(x)\varphi_{E}(x')^{*}e^{-iE(t-t')/\hbar},
\eeqa
where $^*$ denotes complex conjugation.
A path integral representation of the propagator is possible,
\beqa
K(x,t|x',t')=\int\mathcal{D}x[t]\exp\left(\frac{i}{\hbar}S[x]\right),
\eeqa
in terms of the classical action $S[x]=\int_{t'}^{t}d\tau(m\dot{x}^ 2/2-V(x))$.
The propagator is thus an integral over all paths $x[\tau]$ satisfying the boundary conditions $x[t']=x'$ and $x[t]=x$.
Without intention of being exhaustive,
we next point out some specific features relevant to the study of quantum transients:
\begin{itemize}
\item
The propagator for quadratic Hamiltonians acquires the simple form \cite{FH65}
\beqa
K(x,t|x',t')=\mathcal{A}(t)\exp\left(\frac{i}{\hbar}S_{cl}[x]\right),
\eeqa
where $S_{cl}[x]$ is the action corresponding to the classical path, as it follows from the stationary phase approximation.
Here $\mathcal{A}(t)$ is just a time-dependent function independent of the position variables.
For bounded regions, other propagators are known to collapse into a countable sum of classical paths \cite{Crandall93}.
\item Hamiltonians with an arbitrary time-dependent linear potential can be mapped to the free particle problem \cite{Song03}.
\item Perturbations of the delta function type can be taken into
account via path integral summation of perturbation series \cite{GS98} and the Laplace method \cite{EK88,DM06b}.
\item There are important classes of time-dependent problems which can be mapped to time-independent ones making use of integrals of motion \cite{DMN92} or the so-called Duru's method \cite{Duru89}.
\item The propagator for a problem with Dirichlet, von Neumann and mixed boundary conditions
can be obtained from the unconstrained one using the method of images \cite{FH65,Schulman81,Kleinert90,GS98,AS97}.
Simple examples involve a hard-wall at the origin,
$K_{wall}(x,t|x',t')=K(x,t|x',t')-K(-x,t|x',t')$, where $K$ is for instance the propagator for the free Hamiltonian or harmonic oscillator. Similarly, the dynamics of a particle in a box can be described by taking into account infinite images,
even in the presence of moving boundaries \cite{LuzCheng92,GS98}.
%
\end{itemize}
The general theory for the path integral approach to quantum mechanics can be found in \cite{FH65,Schulman81,Kleinert90}.
In particular, explicit expressions for propagators are collected in \cite{GS98},
an excellent account of exactly solvable path integral problems. A recent extension of some of these results
exploits methods of supersymmetric quantum mechanics \cite{PS05,SP06,PSG07}. Finally, one should mention  representations of
the retarded time-dependent Green function as expansions involving resonant states \cite{gcp76,gcmm95}, which are summarized in
Appendix \ref{ressta}.
\section{The Moshinsky function}
\label{app_moshfunction}
The Moshinsky function arises in most of the problems where 1D quantum dynamics
involves sharp boundaries well in time or space domains
\cite{Moshinsky51,Moshinsky52,Moshinsky76,Kleber94}.
Indeed, it can be considered ``the basic propagator of the Schr\"odinger transient mode'' \cite{Nuss92}.
Similarly, it is found when considering free propagators perturbed with point-interactions.
Its standard definition reads
\beqa
\label{A1moshw}
M(x,k,t)&:=&\frac{e^{i\frac{mx^{2}}{2\hbar t}}}{2}w(-z),
\eeqa
where
\beq
\label{z}
z=\frac{1+i}{2}\sqrt{\frac{\hbar t}{m}}\left(k-\frac{mx}{\hbar t}\right).
\eeq
An equivalent expression is
\beq
M(x,k,t):=\frac{e^{-i\frac{\hbar k^{2}t}{2m}+ikx}}{2}{\rm erfc}\left[\frac{x-\hbar k t/m}{\sqrt{2i\hbar t/m}}\right].
\eeq
Its absolute square value, $|M(x,k,t)|^2$, admits a simple geometric interpretation
in terms of the Cornu spiral or clothoid, which is the curve that
results from a parametric representation of the Fresnel integrals. Indeed, another frequent representation of the Moshinsky makes use of the complex Fresnel function $\mathcal{F}$,
\beqa
\label{mosh_fresnel}
M(x,k,t)=\frac{1}{\sqrt{2i}}
e^{-i\frac{\hbar k^{2}t}{2m}+ikx}
\bigg[
\sqrt{\frac{i}{2}}
-\mathcal{F}
\left(\theta\right)
\bigg],
\eeqa
where $\mathcal{F}(\theta)=\int_0^{\theta}{\rm exp}(i\pi u^2/2)du=C(\theta)+iS(\theta)$ is an odd function of its argument
\beq
\theta=\sqrt{\frac{\hbar t}{m\pi}}\left(k-\frac{mx}{\hbar t}\right).
\eeq
Directly from Eq.(\ref{mosh_fresnel}), or using Eqs. (\ref{A1fresnel}), (\ref{A1fresnelw}), and (\ref{A1moshw}),
it follows that
\beq
\vert M(x, k,t)\vert^{2}=\frac{1}{2}
\Bigg\{\left[\frac{1}{2}+C(\theta)\right]^{2}
+\left[\frac{1}{2}+S(\theta)\right]^{2}\Bigg\},
\eeq
which, up to the definition of $\theta$, is identical
to the result from Fresnel diffraction from a straight edge \cite{Fowles68,BW99}.
It is possible to understand physically
the Moshinsky function as a freely evolved cut-off plane wave.
Using the free propagator (\ref{freepropagator}),
\beq
M(x,k,t)=\int_{-\infty}^{\infty}dx'K_0(x,t|x',0)e^{ikx'}\Theta(-x')=\int_{-\infty}^{0}dx'
\sqrt{\frac{m}{2 \pi i \hbar t}}e^{i\frac{m(x-x')^2}{2\hbar t}+ikx'}.
\eeq
An integral representation in the $k$-variable, follows from the same physical problem,
\beq
M(x,k,t)=\frac{i}{2\pi}\int_{-\infty}^{\infty}dk'
\frac{e^{-i\hbar k'^{2}t/2m+ik'x}}{k'-k+i0}
=\frac{i}{2\pi}\int_{\Gamma_{+}}dk'\frac{e^{-i\frac{\hbar k'^{2}t}{2m}+ik'x}}{k'-k},
\eeq
where $\Gamma_{+}$ is a contour in the complex $z$-plane which goes from
$-\infty$ to $\infty$ passing above the pole.

Many properties of the Moshinsky function directly follow from its relation to the Faddeyeva $w(z)$ function.
In particular, under inversion of both $x$ and $k$, it satisfies
\beqa
M(x,k,t)+M(-x,-k,t)=e^{ikx-i\frac{\hbar k^2}{2m}t}.
\eeqa

Similarly, the asymptotic behavior of the Moshinsky functions for $|x-\hbar kt/m|\rightarrow\infty$ can be found
from those of $w(z)$ when $z\rightarrow\infty$ \cite{MB00}.
For $x\leq \hbar kt/m$, and $z$ as in Eq. (\ref{z}), one finds in terms of the Gamma function
$\Gamma(y)$ that
\beqa
M(x,k,t)&\sim&
e^{ikx-i\frac{\hbar k^2 t}{2m}}
+\frac{e^{i\frac{mx^2}{2\hbar t}}}{2i\sqrt{\pi}z}\bigg[1+\sum_{n=1}^{\infty}\frac{(2n-1)!!}{(2z^2)^n}\bigg]\nonumber\\
&=&
e^{ikx-i\frac{\hbar k^2 t}{2m}}
+\frac{e^{i\frac{mx^2}{2\hbar t}}}{2\pi i}
\sum_{n=0}^{\infty}\frac{\Gamma\left(n+\frac{1}{2}\right)}{z^{2n+1}},
\eeqa
whereas in the complementary region $x>\hbar kt/m$ only the series survives.
\subsection{Integral and differential equations}
The following relations are known \cite{EK88}
\beqa
\partial_x M(x,k,t)&=&ikM(x,k,t)-K_0(x,t|0,0),\\
\partial_k M(x,k,t)&=&ikM(x,k,t)-\frac{\hbar t}{m}\partial_x M(x,k,t),\\
\partial_t M(x,k,t)&=&\frac{i\hbar }{2m}\partial^{2}_x M(x,k,t),\\
&=&\frac{x}{2t}K_0(x,t|0,0)-\frac{\hbar k}{2m}\partial_x M(x,k,t),
\eeqa
where $K_0(x,t|0,0)=\sqrt{\frac{m}{2\pi i\hbar  t}}\e^{i\frac{mx^{2}}{2\hbar t}}$, this is, the free propagator from the origin.
In addition, the indefinite integral (which arises in tunneling problem through short-range potentials \cite{EK88,DM06b}) holds
\beqa
\int^x M(ax'+b,c,t)e^{iqx'}dx=\frac{e^{iqx}}{i(q+ca)}
\big[M(ax+b,c,t)-M(ax+b,-q,t)\big],\nonumber\\
\eeqa
but for $q=-ac$, when l'Hopital's rule is to be applied.
\section{The $w$-function}
\label{app_wfunction}

The Faddeyeva function is also known as the complex error or plasma function \cite{FT61}.
It is an an entire function related to the complementary error function \cite{AS65},
\beq
w(z):= e^{-z^{2}}{\rm{erfc}}(-i z).
\eeq
Moreover, it admits the integral representation
\beqa
\label{wz}
w(z)=\frac{1}{i\pi}\int_{\Gamma_{-}}du\frac{e^{-u^{2}}}{u-z},
\eeqa
where $\Gamma_{-}$ is a contour in the complex $z$-plane which goes from
$-\infty$ to $\infty$ passing below the pole.
Actually, it satisfies the following equation,
\beqa
\int_{-\infty}^{\infty}du\frac{e^{-u^{2}}}{u-z}=i\pi{\rm sign}(\Im z)w[{\rm sign}(\Im z)z],
\eeqa
and it obeys the so-called Faddeyeva identity,
which follows from Cauchy's theorem
\beq
w(z)+w(-z)=2e^{-z^{2}}.
\eeq
Moreover,
\beq
w(z^{*})=w(-z)^{*}.
\eeq
We further notice that the Fresnel integrals
\beqa
\label{A1fresnel}
S(\theta)&=&\int_{0}^{\theta} \cos\frac{\pi t^2}{2}dt,\nonumber\\
C(\theta)&=&\int_{0}^{\theta} \cos\frac{\pi t^2}{2}dt,
\eeqa
are related to the $w$-function by
\cite{AS65},
\beq
\label{A1fresnelw}
C(\theta)+iS(\theta)=\frac{1+i}{2}
\Bigg[1-e^{i\pi\theta^2/2}w\left(\frac{1+i}{2}\pi^{1/2}\theta\right)\Bigg].
\eeq
It has a series expansion
\beq
w(z)=\sum_{n=0}^{\infty}
\frac{(iz)^{n}}{\Gamma\left(\frac{n}{2}+1\right)},
\eeq
whereas, its asymptotic expansion reads (expanding $(u-z)^{-1}$ around the
origin in Eq. (\ref{wz}) and
integrating term by term)
\beq
w(z)\thicksim\frac{i}{\sqrt{\pi}z}
\left[1+\sum_{m=1}^{\infty}\frac{(2m-1)!!}{(2z^{2})^{m}}\right]
+2e^{-z^2}\Theta(-\Im z),
\eeq
for $z\rightarrow\infty$.
We also note that its derivatives can be evaluated in a recursive way, using the relation
\beq
w'(z)=-2zw(z)+\frac{2i}{\sqrt{\pi}}.
\eeq
In the upper half-plane, $0 \le |w(z)| < 1$, and its efficient computation is described in \cite{w1,w2}.
The real part of the $w$-function is proportional
to the Voigt spectral-line profile, a convolution of
a Lorentz and a Gaussian profile.

Apart from its close connection with the Moshinsky function, $w$-functions arise also naturally in the dynamics of some initial states which are not
truncated in position space: for Gaussian-like examples in free motion see \cite{MPL99}, and for Lorentzian states in momentum representation \cite{EK88} impinging on a separable
interaction see \cite{MP98}.
\section{Inhomogeneous quantum sources}
\label{apis}
The  time-dependent Schr\"odinger equation for a freely moving particle with an inhomogeneous
source term switched on at $t=0$ reads
\beqa
L\psi(x,t)&=&\Theta(t)\sigma(x,t),\nonumber\\
L&=&i\hbar\frac{\partial}{\partial t}
+\frac{\hbar^2}{2m}\frac{\partial^2}{\partial x^2}.
\eeqa
Using the free Green's function,
\beq
G_0(x,t\vert x',t')=\Theta(t-t')K_0(x,t\vert x't'),
\eeq
which obeys
\beq
LG_0(x,t\vert x',t')=i\hbar \delta(x-x')\delta(t-t'),
\eeq
the solution for $\psi(x,t\ge 0)$, assuming $\psi(x,t<0)=0$, is given by
\beq
\label{F4}
\psi(x,t)=\frac{1}{i\hbar}\int_{0}^{t^+} dt'\int dx' G_0(x,t\vert x',t')\sigma(x',t'),
\eeq
as can be verified by substitution, note the upper limit $t^+$
in the time integral.

In what follows we shall be interested in the dynamics of an inhomogeneous source
of the form $\sigma(x,t)=e^{-i\om t} \varrho(x)$,
where $\hbar\om=(\hbar k)^2/2m$.
This can be related to the Moshinsky shutter problem with the initial condition $\psi(x,0)=\sin(kx)\Theta(-x)$,
which evolves into $\psi(x,t)=[M(x,k,t)-M(x,-k,t)]/(2i)$.
We note that
\beq
\psi(x,t)=\frac{1}{\sqrt{2\pi}}\int g(q)e^{iq x-i\frac{\hbar q^2 t}{2m}}dq,
\eeq
where
\beq
g(q)=\frac{1}{\sqrt{2\pi}}\int \psi(x,0)e^{-iq x}dx=\frac{k}{\sqrt{2\pi} [(q+i\epsilon)^2-k^2]},
\eeq
has two poles in the lower half-plane, close to the real axis.
The integral can be deformed along the contour $\Gamma_{+}$ which goes from $-\infty$ to $\infty$ passing above the poles at $q=\pm k$,
\beqa
\psi(x,t)=\frac{k}{{2\pi}}\int_{\Gamma_{+}} \frac{e^{-i\frac{\hbar q^2t}{2m}}}{q^2-k^2} e^{iqx} dq.
\eeqa
Rewriting
\beqa
\frac{e^{-i\frac{\hbar q^2t}{2m}}}{q^2-k^2}=\frac{\hbar e^{-i\frac{\hbar\kappa^2t}{2m}}}{2im}
\int_{0}^{t^+} e^{-i\frac{\hbar(\kappa^2-k^2)t'}{2m}}dt'+\frac{e^{-i\frac{\hbar k^2t}{2m}}}{\kappa^2-k^2},
\eeqa
only the first term contributes along the contour $\Gamma_{+}$, and therefore
\beqa
\label{eqmoshishutter}
\psi(x,t)=\frac{\hbar k}{2im}\int_{0}^{t^+} e^{-i\frac{\hbar k^2t'}{2m}}
\Big[\frac{1}{2\pi}\int_{\Gamma_{+}}e^{i qx-i\frac{\hbar q^2(t-t')}{2m}}dq\Big]dt'.
\eeqa
%
Identifying the term in brackets with the free Green's function
and comparing (\ref{eqmoshishutter}) with (\ref{F4}),
it follows that
\beqa
\rho(x)=\frac{\hbar^2k}{2m}\delta(x).
\eeqa
In other words, the DIT setup with a totally reflecting shutter
(reflection amplitude $R=-1$) is equivalent to an inhomogeneous  point-like source suddenly turned on.
This result was first obtained by Moshinsky \cite{MoshinskyRMF54}.

\section{Resonant states\label{ressta}}

The first ideas for a theory of resonant states  originated at the end of the twenties of the last century with the work of Gamow on nuclear radioactive decay \cite{gamow}. He considered a situation where a particle, initially confined inside a three dimensional region by a potential barrier, goes out of it by tunneling. In order to describe the above process, Gamow  restricted the discussion to spherically symmetric systems and looked for solutions to the Schr\" odinger equation which at large distances consist of purely outgoing waves. He realized that the absence of incoming waves in the solution is fulfilled if the energy eigenvalues are complex. In 1939, Siegert used  these purely  outgoing  boundary conditions to derive a dispersion formula for elastic scattering by a potential of finite range \cite{siegert}. The same definition of resonant states was also used by Humblet and Rosenfeld to formulate a theory of nuclear reactions \cite{rosenfeld}. One sees that resonant states provide a unified description of decay and scattering problems. Following an idea by Peierls \cite{peierls},  subsequent developments involved a consideration of the analytical properties of the outgoing Green function to the problem \cite{more,gcp76}. These led to expansions of this function and  to a consistent normalization condition for resonant states \cite{gcp76}. The extension of resonant states to one dimension (i.e. to the full line)  was made in Ref. \cite{gc87}.  Along the full line, the fact that there are two edges to the potential leads to a modification of expression for the normalization condition \cite{gcrr93,GR97}. For the half-line the resonance formalism is similar to the case of zero angular momentum in three dimensions.

Here we present some properties of resonant states relevant for the discussion of sections \ref{epo} and \ref{tune}.

Most of the work on resonant states in one dimension refers to interactions of finite range, \textit{i.e.}, extending along a finite region of space. A given resonant state satisfies the Schr\"odinger equation of the problem with complex energy eigenvalues \cite{GR97},
\begin{equation}
\frac{d^{\,2}}{dx^2}u_n(x)+\left [k_n^2-U(x)\right ]u_n(x)=0
\label{e1}
\end{equation}
where $k_n^2=2mE_n/\hbar^2$, with the complex energy $E_n=\mathcal{E}_n-i\Gamma_n/2$, $U(x)=[2mV(x)]^{1/2}/\hbar$ where $V(x)$ vanishes beyond the interval $0 \leq x \leq L$. The solutions to the above equation satisfy outgoing boundary conditions at $x=0$ and $x=L$, given respectively by
\begin{equation}
\left [\frac{d}{dx}u_n \right ]_{x=0}=-ik_nu_n(0),
\label{e2}
\end{equation}
and
\begin{equation}
\left [\frac{d}{dx}u_n\right ]_{x=L}=ik_n u_n(L).
\label{e3}
\end{equation}
The amplitude of a resonant state increases exponentially with distance beyond the interaction region, \textit{i.e.}, since $k_n=\alpha_n-i \beta_n$, say for $x > L$, $u_n(x)\sim \exp(i\alpha x)\exp(\beta x)$, and therefore the usual rules concerning normalization, orthogonality and completeness do not apply for resonant states. For that reason it was generally believed that these functions were not very useful for calculations.
As mentioned above, developments involving the analytical properties of the outgoing Green function to the problem have led to a consistent normalization condition and completeness condition for resonant states.
Notice also that at distances beyond  the interaction region, say $x > L$, the resonant state may be written as
\begin{equation}
u_n(x,t) \sim e^{i(\alpha_nx-\mathcal{E}_nt/\hbar)} e^{\beta_nx}e^{-\Gamma_nt/\hbar},
\label{e3a}
\end{equation}
which represents an outgoing wave of wave number $\alpha_n$ associated with a state of positive energy $\mathcal{E}_n$. The second factor
shows that this state decays at the rate $\Gamma_n/\hbar$, in agreement with the usual interpretation of the width of a resonance level.
Since the velocity of the outgoing particle is $v_n=\hbar\alpha_n/m$, one sees that the exponential increase with $x$ of the last factor
in Eq. (\ref{e3a}), may be interpreted as due to the fact that at a distance $x$ one finds those particles which left the interaction region
at a time $t-x/v_n$ when the amplitude there was larger by the factor $\exp[\Gamma_nx/(2\hbar v_n)]=\exp(\beta_nx)$.

It is well known that for finite range interactions the outgoing Green  function $G^+(r,r';k)$, as a function of $k$, can be extended analytically to the whole complex $k$-plane where it possesses an infinite number of poles distributed in a well known manner \cite{taylor,newton}. Purely imaginary poles, situated on the upper half $k$-plane, correspond to bound states of the problem whereas purely imaginary poles seated on the lower half $k$-plane are related to antibound  states, also called virtual states by some authors. On the other hand, complex poles are only found on the lower half $k$-plane and are associated with resonant states. To each complex pole at $k_n=\alpha_n-i\beta_n$ with $\alpha_n, \beta_n >0$ there corresponds, from time reversal invariance, a complex pole $k_{-n}$ situated symmetrically with respect to the imaginary axis, i.e., $k_{-n}=-k_n^*$.
Also $u_{-n}(x)=u_n^*( x)$.

\subsection{Time-independent resonance expansions}

The outgoing Green function to the problem satisfies the equation
\begin{equation}
{ \partial^2 \over \partial x^2} G^+(x,x^{\prime};k) + \left [ k^2-U(x)\right ] G^+(x,x^{\prime};k) = \left(\frac{2m}{\hbar^2}\right )\delta(x-x^{\prime}),
\label{e4}
\end{equation}
where $k^2=2mE/\hbar^2$, with $E$ the energy. The solution  to Eq. (\ref{e4}) satisfies  outgoing boundary conditions at $x=0$ and $x=L$ given respectively by
\begin{equation}
\left[{\partial \over \partial x }G^+(x,x^{\prime};k)\right]_{x=0_-}=
-ikG^+(0,x^{\prime};k)
\label{e5}
\end{equation}
and
\begin{equation}
\left[{\partial \over \partial x }G^+(x,x^{\prime};k)\right]_{x=L_-}=
ikG^+(L,x^{\prime};k).
\label{e6}
\end{equation}
It may be shown that the residue $C_n(x,x^{\prime})$ of $G^+(x,x^{\prime};k)$ at a pole $k_n$ seated on the complex $k$-plane is given
by \cite{GR97}
\begin{equation}
C_n(x,x^{\prime})= \left(\frac{2m}{\hbar^2}\right )\frac{u_n(x) u_n(x^{\prime})}{2k_n},
\label{e7}
\end{equation}
provided the resonant states are normalized according to the condition
\begin{equation}
\int_0^L u_n^2(x)dx +i {u_n^2(0)+u_n^2(L) \over 2k_n} =1.
\label{e8}
\end{equation}
Notice that for a bound state (i.e., $k_n=i\gamma_n$) the contribution of the two terms on right-hand side of Eq. (\ref{e7})
corresponds exactly to two integral terms that allow to express the normalization condition for the bound state in the usual form
$\int_{-\infty}^{\infty} u_n^2(x)dx=1$.

An interesting expression follows by considering the Green theorem between Eq. (\ref{e1}) for $u_n(x)$ and a similar equation for $u_m(x)$
and then using the corresponding boundary conditions for  $u_n(x)$, \textit{i.e.}, Eqs. (\ref{e2}) and (\ref{e3}),  and similarly  for $u_m(x)$,
to obtain
\begin{equation}
\int_0^L u_n(x)u_m(x)dx +i \frac{u_n(0)u_m(0)+u_n(L)u_m(L)} {k_n+k_m}=0.
\label{e8a}
\end{equation}

The above results suggest to look for an expansion of the outgoing Green function  of the problem in terms of the set of resonant states.
This may be achieved by considering the following integral,
\begin{equation}
I={1 \over 2\pi i} \int_C {G^+(x,x^{\prime};k) \over k^{\prime}-k}\,dk^{\prime},
\label{e8b}
\end{equation}
where $C$ represents a large closed contour of radius $L$ in the $k$-plane about the
origin in the $clockwise$ direction which excludes all the poles, $k_n$
and the value at $k^{\prime}=k$. Using Cauchy theorem it follows that $I=0$ and
hence one may write
\begin{equation}
I= {1 \over 2\pi i}\left [ -\,\int_{C_R} {G^+(k^{\prime}) \over k^{\prime}-k}dk^{\prime} +
\sum_n \int_{C_n} {G^+(k^{\prime}) \over k^{\prime}-k}dk^{\prime}
+\int_{C_k} {G^+(k^{\prime}) \over k^{\prime}-k}dk^{\prime} \right]=0,
\label{e9}
\end{equation}
where for simplicity of notation $G^+(x,x^{\prime};k)$ is denoted by
$G^+(k^{\prime})$ and $C_R$ represents a large circle centered at the origin;
the contours $C_n$ encircle each of the poles $k_n$ that are enclosed
by $C_R$, and the contour $C_k$ encloses the value $k^{\prime}=k$. All these
contours are in the \textit{counterclockwise} direction.

Using the theorem of residues and Eq. (\ref{e7}) one may write Eq.  (\ref{e9}) as
\begin{equation}
G^+(x,x^{\prime};k)= \left(\frac{2m}{\hbar^2}\right )\sum_{n=-N}^N{u_n(x)u_n(x^{\prime}) \over 2k_n(k-k_n)} +
{1 \over 2\pi i}\int_{C_R}{G^+(x,x^{\prime};k) \over k^{\prime}-k}dk^{\prime}.
\label{e9a}
\end{equation}
The above equation is defined for all $x,x^{\prime} \geq 0$. However, for many applications 
we need to know $G^+(x,x^{\prime};k)$ only along the internal region of the
interaction. In this case it has been  proved rigorously along the half-line that
the outgoing Green function goes exponentially to zero along all directions in the complex $k$-plane
for all values of $x$ and $x'$ along the internal interaction region,  except at the points $x=x'=0=L$ \cite{gcb79,romo80}.
This is denoted as $ 0 \preceq (x,\,x^{\prime}) \preceq\,L$  \textit{i.e.},
\begin{equation}
G^+(x,x^{\prime};k) \to 0 \,\,\,\,as\,\,\,  |k| \to \infty\, , \hskip1.5truecm
 0 \preceq (x,\,x^{\prime}) \preceq\,L.
\label{e10}
\end{equation}
Therefore by extending the radius $R$ of the contour $C_R$
up to infinity one then obtains that the integral term along $C_R$ in  Eq. (\ref{e9}) vanishes exactly.
Hence, one is left with an infinite sum over the full set of resonant terms,
\begin{equation}
G^+(x,x^{\prime};k) =  \left(\frac{2m}{\hbar^2}\right )\sum_{n=-\infty}^{\infty} {u_n(x)u(x^{\prime}_n) \over 2k_n(k-k_n)}\,,
\hskip1.5truecm   0 \preceq (x,\,x^{\prime}) \preceq\,L.
\label{e11}
\end{equation}
Substitution of Eq. (\ref{e11}) into Eq. (\ref{e4}) implies the fulfillment of the relations
\begin{equation}
\sum_{n=-\infty}^{\infty} \frac{u_n(x)u_n(x')}{k_n}=0, \hskip1.5truecm   0 \preceq (x,\,x^{\prime}) \preceq\,L,
\label{e12}
\end{equation}
and
\begin{equation}
\frac{1}{2}\sum_{n=-\infty}^{\infty} u_n(x)u_n(x')=\delta(x-x'), \hskip1.5truecm   0 \preceq (x,\,x^{\prime}) \preceq\,L.
\label{e13}
\end{equation}
The first of the above two expressions indicates that resonance states are not independent of each other and the second one may
be seen as a modified closure relationship.

Let us now consider a relationship between the outgoing Green function and the continuum wave function of the problem.
The continuum wave function satisfies the equation
\begin{equation}
\frac{d^2}{dx^2}\psi(k,x) +[k^2 -U(x)]\psi(k,x)=0.
\label{e14}
\end{equation}
For a particle approaching the potential from the left ($x<0$), the solutions to the above equation read, respectively
for $x <0$ and $x> L$,
\begin{equation}
\psi(k,x)=e^{ikx} + R(k)e^{-ikx},
\label{e15}
\end{equation}
and
\begin{equation}
\psi(k,x)=T(k)e^{ikx},
\label{e16}
\end{equation}
where $R(k)$ and $T(k)$  are, respectively, the reflection and transmission coefficients to the problem.
It then follows  by applying Green theorem between Eqs. (\ref{e14}) and (\ref{e4}), and using the  boundary relations given
by Eqs.  (\ref{e2}) and (\ref{e3}), that
\begin{equation}
\psi(k,x)= \left(\frac{\hbar^2}{2m}\right )2ikG^+(0,x;k),  \hskip1.5truecm  0 < x \leq L.
\label{e17}
\end{equation}
By considering the solution of the wave function at $x=L$, given by Eq. (\ref{e16}), into the above expression yields
for the transmission amplitude
\begin{equation}
T(k)= \left(\frac{\hbar^2}{2m}\right )2ikG^+(0,L;k)e^{-ikL},
\label{e18}
\end{equation}
and similarly, using Eq. (\ref{e15}), yields for the reflection amplitude
\begin{equation}
R(k)= \left(\frac{\hbar^2}{2m}\right )2ikG^+(0,0;k)-1.
\label{e19}
\end{equation}
Using Eq. (\ref{e11}) into Eqs. (\ref{e17}) and (\ref{e18}) then leads, respectively,  to resonance expansions
for the wave solution and the transmission amplitude, namely,
\begin{equation}
\psi(k,x)=ik \sum_{n=-\infty}^{\infty}C_n u_n(x)  \hskip1.5truecm  0 < x \leq L,
\label{e17a}
\end{equation}
where $C_n=u_n(0)/[k_n(k-k_n)]$, and
\begin{equation}
T(k)=ik\sum_{n=-\infty}^{\infty} \frac{u_n(0)u_n(L)}{k_n(k-k_n)}e^{-ikL}.
\label{e20}
\end{equation}
The expansion for the reflection amplitude requires of  subtraction terms.
For the transmission amplitude one may alternatively expand  $G^+(x,x';k)\exp(-ikL)$, to obtain
\begin{equation}
T(k)=ik\sum_{n=-\infty}^{\infty} \frac{u_n(0)u_n(L)}{k_n(k-k_n)}e^{-ik_nL}.
\label{e21}
\end{equation}
\subsection{Resonance expansion of the retarded time-dependent  Green function}
The solution $|\psi(t)\rangle$ to the time-dependent Schr\"odinger equation
may be written in terms of the retarded time evolution operator
of the problem $\exp(-iHt)$, where $ t \geq 0$,  and the known arbitrary initial state $|\psi(0)\rangle$ as
\begin{equation}
|\psi(t)\rangle=e^{-iHt}|\psi(0)\rangle.
\label{e22}
\end{equation}
In coordinate representation $g(x,x';t)=\langle x|\exp(-iHt)|x'\rangle$ is referred to as the  retarded Green function.
It may be written, using Laplace transform techniques, in terms of the outgoing Green function to the problem,
\begin{equation}
g(x,x';t)= \left(\frac{\hbar^2}{2m}\right ){i \over 2 \pi} \int_{C_{\circ}} G^+(x,x';k) {\rm e}^{-i\hbar k^2t/2m} \,2kdk,
\label{e23}
\end{equation}
where $C_{\circ}$ stands for the Bromwich contour and corresponds to an hyperbolic contour along the first quadrant on the
complex $k$ plane.

One may write a representation of the time-dependent
Green function involving resonant states plus an integral
contribution in a similar fashion as discussed for the half-line in \cite{gcp76}. This may be done by deforming appropriately
the integration contour $C_{\circ}$ on the complex $k$-plane.
Since the variation of $G^+(x,x';k)$ with $k$ is at most exponential, the behavior
of the integrand with $k$  in Eq.\,(\ref{e23}) is dominated by $\exp(-i\hbar k^2t/2m)$.
A convenient choice is to deform it to a contour involving two semi-circles $C_s$ along the second and fourth quadrants of the
$k$-plane plus a straight line $C_L$ that passes through the origin at $45^{\circ}$ off the real $k$-axis.
In doing that one passes over some poles of the outgoing Green function. In general these include  bound states and the subset
of complex poles associated with the so called proper resonant states, \textit{i.e.}, ${\rm Re}\, k_p > {\rm Im} \,k_p$ .
By extending the integration contour up to infinity one
obtains that the semi-circles $C_s$ yield
a vanishing contribution so that one is left with a infinite sum of terms
plus an integral contribution \cite{gcp76}, namely,
\begin{eqnarray}
g(x,x';t)= &&\sum_{n=1}^{\infty} u_n(x)u_n(x')e^{-iE_nt/\hbar}\nonumber\\[.5cm]
&&+\left(\frac{\hbar^2}{2m}\right ){i \over 2\pi} \int_{C_L} G^+(x,x';k)e^{-i\hbar k^2t/2m}\,2kdk.
\label{e24}
\end{eqnarray}
Here the summation includes all bound states $u_b$, and all proper resonant states, $u_p$. In this last case, the energies are
complex, \textit{i.e.}, $E_p=\mathcal{E}_p-i\Gamma_p/2$. In order to derive the above expression, one makes use of Eq. (\ref{e7}) for the residue of a complex pole, with the resonant functions  normalized according to the condition given by Eq. (\ref{e8}). One may manipulate the integrand in the integral term of Eq. (\ref{e24}), as in Ref. \cite{gcp76} for the half-line, to write the retarded Green function as a sum over bound and resonance terms plus a continuum integral term along the negative imaginary energy axis, a form that resembles the usual expansion as a sum of bound states and a continuum contribution along the real energy axis.

The time-dependent expansion given by Eq. (\ref{e24}) holds for arbitrary values of $x$ and $x'$. However, by restricting
these values according to the condition $0 \preceq (x,\,x^{\prime}) \preceq\,L$, discussed above,
one may obtain a time-dependent expansion that includes the full set of bound, antibound and resonant states (associated with both the third
and fourth quadrant poles of the complex $k$-plane).
Noticing that
\begin{equation}
\frac{1}{2k_n(k-k_n)} \equiv \frac{1}{2k} \left [ \frac{1}{k-k_n} + \frac{1}{k_n} \right ],
\label{e25}
\end{equation}
allows to write Eq. (\ref{e11}), using Eq. (\ref{e12}), as
\begin{equation}
G^+(x,x^{\prime};k) = \left(\frac{2m}{\hbar^2}\right )\frac{1}{2k} \sum_n^{\infty} \frac{u_n(x)u(x^{\prime}_n)}{(k-k_n)}; \,\,\, 0 \preceq (x,\,x^{\prime}) \preceq\,L.
\label{e26}
\end{equation}
The above representation is very convenient because substituting it into  the integral term in Eq. (\ref{e24})
leads to an expansion over the full set of bound, antibound and resonance states of the system, similar to that obtained for the half-line
\cite{gcmm95,nueva},
\begin{equation}
g(x,x';t)= \sum_{n=-\infty}^{\infty} u_n(x)u_n(x')M(k_n,t),
\label{e27}
\end{equation}
where $M(k_n,t)\equiv M(0,k_n,t)$ stands for the Moshinsky function.

\end{document}